\documentclass[aps,prd,USenglish,reprint,nofootinbib,showpacs]{revtex4-1}

\pdfoutput=1

\usepackage[preprint]{atlascover}

\usepackage[subfigure=true]{atlaspackage}
\usepackage[process=true]{atlasphysics}
\usepackage{atlasbsm}
\usepackage{atlasmisc}
\usepackage{multirow}
\usepackage{rotating}

\usepackage{jetetmisssymbols}
\usepackage{boostedsymbols}
\usepackage{susysymbols}
\usepackage{edittools}

\newcommand{\fatjetetacut}       {2.5}

\newcommand{\intlumi}            {$20.3 \pm 0.6$\invfb}

\newcommand{\ExpLimitLF}         {853 GeV}
\newcommand{\ObsLimitLF}         {917 GeV}

\newcommand{\ExpLimitB}          {921 GeV}
\newcommand{\ObsLimitB}          {929 GeV}

\newcommand{\ExpLimitBT}         {938 GeV}
\newcommand{\ObsLimitBT}         {874 GeV}

\graphicspath{{logos/}{figures/}{../INT/Merged/figures/}{../INT/Resolved/figures/}{forPub/}}
\newcommand{\papertitle}{
Search for massive supersymmetric particles decaying to many jets using the ATLAS detector in $pp$ collisions at $\sqrt{s} = 8$~TeV
}

\AtlasAbstract{%
Results of a search for decays of massive particles to fully hadronic final states are presented. This search uses 20.3 fb$^{-1}$ of data collected by the ATLAS detector in $\sqrt{s} = 8$~TeV proton--proton collisions at the LHC. Signatures based on high jet multiplicities without requirements on the missing transverse momentum are used to search for $R$-parity-violating supersymmetric gluino pair production with subsequent decays to quarks. The analysis is performed using a requirement on the number of jets, in combination with separate requirements on the number of $b$-tagged jets, as well as a topological observable formed from the scalar sum of the mass values of large-radius jets in the event. Results are interpreted in the context of all possible branching ratios of direct gluino decays to various quark flavors. No significant deviation is observed from the expected Standard Model backgrounds estimated using jet-counting as well as data-driven templates of the total-jet-mass spectra. Gluino pair decays to ten or more quarks via intermediate neutralinos are excluded for a gluino with mass $m_{\tilde{g}} < 1$~TeV for a neutralino mass $m_{\tilde{\chi}^0_1} = 500$~GeV. Direct gluino decays to six quarks are excluded for $m_{\tilde{g}} < 917$~GeV for light-flavor final states, and results for various flavor hypotheses are presented.
}

\AtlasTitle{\papertitle}
\AtlasRefCode{SUSY-2013-07}             
\AtlasVersion{2.4}                      
\AtlasJournal{PRD}                      

\PreprintIdNumber{CERN-PH-EP-2015-020}  

\AtlasCoverEditor{D. W. Miller}{David.W.Miller@cern.ch}
\AtlasCoverEditor{M. Swiatlowski}{maximilian.j.swiatlowski@cern.ch}

\AtlasCoverSupportingNote{Jet-counting analysis supporting note}{https://cds.cern.ch/record/1544938}
\AtlasCoverSupportingNote{Total-jet-mass analysis supporting note}{https://cds.cern.ch/record/1646768}

\AtlasCoverCommentsDeadline{February 3, 2015}

\AtlasCoverAnalysisTeam{Tobias Golling, Lawrence Lee~(*), David W. Miller~(*), Joakim Olsson, Ariel Schwartzman, Maximilian Swiatlowski~(*), Lian-Tao Wang}

\AtlasCoverEgroupEditors{atlas-susy-2013-07-editors@cern.ch}
\AtlasCoverEgroupEdBoard{atlas-susy-2013-07-editorial-board@cern.ch}

\AtlasCoverEdBoardMember{Jonathan Butterworth}
\AtlasCoverEdBoardMember{Davide Costanzo~(*)}
\AtlasCoverEdBoardMember{Gregor Herten}
\AtlasCoverEdBoardMember{Juan Terron}

\HepDataRecord{ZZZZZZZZ}
\arXivId{15XX.YYYYY}

\hypersetup{pdftitle={rpvmultijet-prd},pdfauthor={The ATLAS Collaboration},pageanchor=false}
                             
\begin{document}

\title{\AtlasTitleText}


\author{The ATLAS Collaboration}

\begin{abstract}
	\AtlasAbstractText
\end{abstract}

\maketitle

\clearpage

\section{Introduction}
\label{sec:introduction}

Supersymmetry (\SUSY)~\cite{Miyazawa:1966,Ramond:1971gb,Golfand:1971iw,Neveu:1971rx,Neveu:1971iv,Gervais:1971ji,Volkov:1973ix,Wess:1973kz,Wess:1974tw} is a theoretical extension of the Standard Model (SM) which fundamentally relates fermions and bosons. It is an alluring theoretical possibility given its potential to solve the naturalness problem~\cite{Dimopoulos:1981zb,Witten:1981nf,Dine:1981za,Dimopoulos:1981au,Sakai:1981gr,Kaul:1981hi} and to provide a dark-matter candidate~\cite{Goldberg:1983nd,Ellis:1983ew}. Partially as a result of the latter possibility, most searches for \SUSY focus on scenarios such as a minimal supersymmetric standard model (MSSM) in which $R$-parity is conserved (RPC)~\cite{SUSYZeroLep2011, SUSYOneLep2011, SUSYTwoLep2011, SUSYJetMult2011}. In these models, \SUSY particles must be produced in pairs and must decay to a stable lightest supersymmetric particle (LSP). With strong constraints now placed on standard RPC \SUSY scenarios by the experiments at the Large Hadron Collider (LHC), it is important to expand the scope of the \SUSY search program and explore models where $R$-parity may be violated and the LSP may decay to SM particles, particularly as these variations can alleviate to some degree the fine-tuning many \SUSY models currently exhibit~\cite{Arvanitaki:2013yja}.

In $R$-parity-violating (RPV) scenarios, many of the constraints placed on the MSSM in terms of the allowed parameter space of gluino (\gluino) and squark (\squark) masses are relaxed. The reduced sensitivity of standard \SUSY searches to RPV scenarios is due primarily to the high missing transverse momentum (\Etmiss) requirements used in the event selection common to many of those searches. This choice is motivated by the assumed presence of two weakly interacting and therefore undetected LSPs. Consequently, the primary challenge in searches for RPV \SUSY final states is to identify suitable substitutes for the canonical large \Etmiss\ signature of RPC SUSY used to distinguish signals from background processes. Common signatures used for RPV searches include resonant lepton pair production~\cite{SUSYRPVemu2010, SUSYRPVemu20111fb, SUSYRPVemu20112fb}, exotic decays of long-lived particles, and displaced vertices~\cite{SUSYRhadron2010, SUSYRPVLL2010, SUSYRPVDV2010, SUSYStopGluino2010}.

  \begin{figure}[!ht]
    \centering
    \includegraphics[width=0.47\columnwidth]{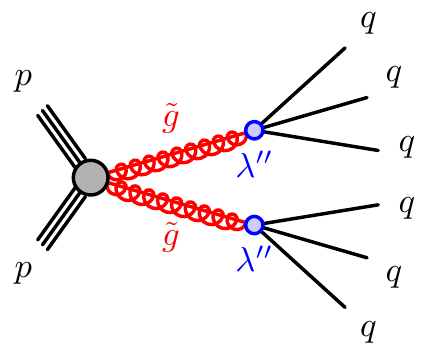}
    \includegraphics[width=0.47\columnwidth]{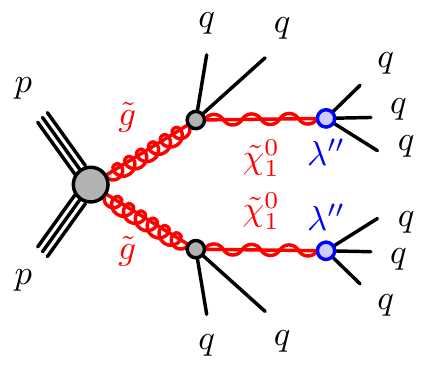}
    \caption{Diagrams for the benchmark processes considered for this analysis. The solid black lines represent Standard Model particles, the solid red lines represent SUSY partners, the gray shaded circles represent effective vertices that include off-shell propagators (e.g. heavy squarks coupling to a \ninoone neutralino and a quark), and the blue shaded circles represent effective RPV vertices allowed by the baryon-number-violating \lampp\ couplings with off-shell propagators (e.g. heavy squarks coupling to two quarks).}
    \label{fig:intro:diagram}
  \end{figure}

New analyses that do not rely on \met are required in order to search for fully hadronic final states involving RPV gluino decays directly to quarks or via \ninoone neutralinos as shown in the diagrams in \figref{intro:diagram}. Cases in which pair-produced massive new particles decay directly to a total of six quarks, as well as cascade decays with at least ten quarks, are considered. Three-body decays of the type shown in \figref{intro:diagram} are given by effective RPV vertices allowed by the baryon-number-violating \lampp\ couplings as described in \secref{rpvandudd} with off-shell squark propagators. This analysis is an extension of the search conducted at \sqsseven for the pair production of massive gluinos, each decaying directly into three quarks~\cite{RPVGluino7TeV}.

The diagrams shown in \figref{intro:diagram} represent the benchmark processes used in the optimization and design of the search presented in this paper. The extension to considering cascade decays of massive particles creates the potential for significantly higher hadronic final-state multiplicities and motivates a shift in technique with respect to previous searches. Therefore, the analysis is extended to look for events characterized by much higher reconstructed jet multiplicities as well as with event topologies representative of these complex final states. Two complementary search strategies are thus adopted: a jet-counting analysis that searches for an excess of $\geq$6-jet or $\geq$7-jet events, and a data-driven template-based analysis that uses a topological observable called the \textit{total-jet-mass} of large-radius (\largeR) jets. The former exploits the predictable scaling of the number of $n$-jet events ($n=6,7$) as a function of the transverse momentum (\pt) requirement placed on the $n^{\rm th}$ leading jet in \pt for background processes. This analysis is sensitive to the models presented here because this scaling relation differs significantly between the signal and the background. The latter analysis uses templates of the event-level observable formed by the scalar sum of the four leading \largeR\ jet masses in the event, which is significantly larger for the signal than for the SM backgrounds.

This paper is organized as follows: \secref{rpvandudd} describes the motivation and theoretical underpinnings of the benchmark processes used in this analysis. \Secref{detector} and \secref{data-mc} present details of the detector, the data collection and selection procedures, and the Monte Carlo (MC) simulation samples used for this search. The physics object definitions used to identify and discriminate between signal and background are described in \secref{event-selection}. The details of the methods are separated for the two analyses employed. The jet-counting analysis is presented in \secref{Resolved}, while the total-jet-mass analysis using more advanced observables is presented in \secref{Merged}. The combined results of this search and the final sensitivity to the benchmark processes are then described in \secref{results}. The results using the total-jet-mass analysis are presented first, in \secref{results:merged}, as they only apply to the ten-quark final states. The jet-counting analysis additionally yields interpretations across the flavor structure allowed by the \lampp\ couplings. This comprehensive set of results is presented in \secref{results:resolved}. Comparisons between the two analyses are then made in \secref{results:combined}.

\section{$R$-parity-Violating Supersymmetry and Baryon-Number Violation}
\label{sec:rpvandudd}

The benchmark model used to interpret the results of the search for high multiplicity hadronic final states is the baryon-number-violating RPV SUSY scenario. The RPV component of the generic supersymmetry superpotential can be written as~\cite{Dreiner:1998wm, Allanach:2003eb}:
%
  \begin{eqnarray} 
    \Wrpv &= \frac{1}{2} \lamijk L_{i}L_{j}\Ebar_{k} + \lampijk L_{i}Q_{j}\Dbar_{k} +  \nonumber \\
       & \quad \frac{1}{2}\lamppijk\Ubar_{i}\Dbar_{j}\Dbar_{k} + \kappa_iL_{i}H_{2},
    \label{eq:rpvandudd:wrpv}
  \end{eqnarray} 

\noindent where $i,j,k=1,2,3$ are generation indices. The generation indices are sometimes omitted in the discussions that follow if the statement being made is not specific to any generation. The first three terms in \equref{rpvandudd:wrpv} are often referred to as the trilinear couplings, whereas the last term is bilinear. The $L_i$, $Q_i$ represent the lepton and quark $SU(2)_{\rm L}$ doublet superfields, whereas $H_2$ is the Higgs superfield. The $\Ebar_j$, $\Dbar_j$, and $\Ubar_j$ are the charged lepton, down-type quark, and up-type quark $SU(2)_{\rm L}$ singlet superfields, respectively. The Yukawa couplings for each term are given by \lam, \lamp, and \lampp, and $\kappa$ is a dimensionful mass parameter. In general, the particle content of the RPV MSSM is identical to that of the RPC MSSM but with the additional interactions given by $\Wrpv$.

Generically, the addition of \Wrpv into the overall SUSY superpotential 
allows for the possibility of rapid proton decay. The simultaneous presence of 
lepton-number-violating (e.g. $\lamp \neq 0$) and baryon-number-violating 
operators ($\lampp \neq 0$) leads to proton decay rates larger than allowed by the experimental limit on the proton lifetime unless, for example,~\cite{PhysRevD.47.279}
%
\begin{eqnarray}
  \lamp_{11k}\cdot\lampp_{11k}\; \lsim\; 10^{-23} \left( \frac{{\msquark}} {100~\gev} \right)^2,
  \label{eq:rpvandudd:protondecay}
\end{eqnarray}
%
\noindent where \msquark is the typical squark mass. As a result, even when considering this more generic form of the \SUSY superpotential by including \Wrpv, it is still necessary to impose an \textit{ad hoc}, albeit experimentally motivated, symmetry to protect the proton from decay. It is generally necessary that at least one of \lam, \lamp, \lampp be exactly equal to zero. Consequently, it is common to consider each term in \equref{rpvandudd:wrpv} independently. In the case of nonzero \lam and \lamp, the typical signature involves leptons in the final state. However, for $\lamppijk \neq 0$, the final state is characterized by jets, either from direct gluino decay or from the cascade decay of the gluino to the lightest neutralino (\ninoone), as also considered here. Because of the structure of \equref{rpvandudd:wrpv}, scenarios in which only $\lamppijk \neq 0$ are often referred to as UDD scenarios.

Current indirect experimental constraints~\cite{Allanach:1999ic} on the sizes of each of the UDD couplings \lamppijk from sources other than proton decay are valid primarily for low squark masses, as suggested by \equref{rpvandudd:protondecay}. Those limits are driven by double nucleon decay~\cite{Sher:1994sp} (for $\lampp_{112}$), neutron oscillations~\cite{Zwirner1983103} (for $\lampp_{113}$), and $Z$ boson branching ratios~\cite{Bhattacharyya:1997vv}.

Hadron collider searches are hindered in the search for an all-hadronic decay of new particles by the fact that the SM background from multi-jet production is very high. Nonetheless, searches have been carried out by several collider experiments. The CDF Collaboration~\cite{Aaltonen:2011sg} excluded gluino masses up to 240 GeV for light-flavor models. The CMS Collaboration~\cite{Chatrchyan:2013gia} excludes such gluinos up to a mass of 650 GeV and additionally sets limits on some heavy-flavor UDD models. The ATLAS Collaboration~\cite{RPVGluino7TeV} has also previously set limits in a search for anomalous six-quark production, excluding gluino masses up to 666 GeV for light-flavor models. The search presented here uniquely probes the flavor structure of the UDD couplings and employs new techniques both in analysis and theoretical interpretation.

\section{The ATLAS detector}
\label{sec:detector}

The ATLAS detector~\cite{detPaper} provides nearly full solid angle coverage around the collision point with an inner tracking system covering the pseudorapidity\footnote{ATLAS uses a right-handed coordinate system with its origin at the nominal interaction point (IP) in the center of the detector and the $z$-axis along the beam pipe. The $x$-axis points from the IP to the center of the LHC ring, and the $y$-axis points upward. Cylindrical coordinates $(r,\phi)$ are used in the transverse plane, $\phi$ being the azimuthal angle around the beam pipe. The pseudorapidity is defined in terms of the polar angle $\theta$ as $\eta=-\ln\tan(\theta/2)$.} range $|\eta|<2.5$,  electromagnetic (EM) and hadronic calorimeters covering $|\eta|<4.9$, and a muon spectrometer covering $|\eta|<2.7$.  

The ATLAS tracking system is composed of a silicon pixel tracker closest to the beamline, a microstrip silicon tracker, and a straw-tube transition radiation tracker. These systems are layered radially around each other in the central region. A thin solenoid surrounding the tracker provides an axial 2 T field enabling measurement of charged-particle momenta.

The calorimeter, which spans the pseudorapidity range up to $|\eta|= 4.9$, is comprised of multiple subdetectors with different designs. The high granularity liquid argon (LAr) electromagnetic calorimeter system includes separate barrel ($|\eta|<1.475$), endcap ($1.375<|\eta|<3.2$), and forward subsystems ($3.1<|\eta|<4.9$). The tile hadronic calorimeter ($|\eta|<1.7$) is composed of scintillator tiles and iron absorbers. As described below, jets used in the analyses presented here are typically required to have $|\eta|<2.8$ such that they are fully contained within the barrel and endcap calorimeter systems.

A three-level trigger system is used to select events to record for offline analysis.  The level-1 trigger is implemented in hardware and uses a subset of detector information to reduce the event rate to a design value of at most 75~kHz during 2012. This is followed by two software-based triggers, level-2 and the event filter (collectively called the \textit{high-level trigger}), which together reduce the event rate to a few hundred Hz. The primary triggers used in this analysis collected the full integrated luminosity of the \eighttev dataset with good efficiency for the event selections described in this paper.

\section{Data and Monte Carlo samples}
\label{sec:data-mc}

The data used in this analysis correspond to \intlumi~\cite{Aad:2011dr,ATLAS-CONF-2011-116} of integrated luminosity taken during periods in which the data satisfied baseline quality criteria. Further details of the event selections applied, including the ATLAS data quality criteria and trigger strategy, are given in \secref{event-selection}. The primary systems of interest in these studies are the electromagnetic and hadronic calorimeters and the inner tracking detector. The data were collected with triggers based on either single-jet or multi-jet signatures. The single-jet trigger selection has a transverse momentum threshold of 360~GeV using a \largeR\ \antikt\ jet definition~\cite{Cacciari:2008gp} with a nominal radius of $R=1.0$ within the high-level jet trigger. The multi-jet trigger selection requires at least six \antikt\ $R=0.4$ jets with a nominal \pt threshold of 45~GeV in the high-level trigger. Data collected using several additional multi-jet requirements (from three to five jets) are also used for background estimation studies.

Multiple simultaneous proton--proton (\pp) interactions, or \pileup, occur in each bunch crossing at the LHC. The additional collisions occurring in the same and neighboring bunch crossings with respect to the event of interest are referred to as in-time and out-of-time \pileup, respectively, and are uncorrelated with the hard-scattering process.

The benchmark RPV SUSY signal processes of both the six-quark and ten-quark models (see \secref{introduction}) were simulated using \Herwigpp 6.520~\cite{Herwigpp} for several gluino and neutralino mass hypotheses using the parton distribution function (PDF) set CTEQ6L1~\cite{PDF-CTEQ, cteq6l1}. For both models, all squark masses are set to 5~TeV and thus gluinos decay directly to three quarks or to two quarks and a neutralino through standard RPC couplings. In the ten-quark cascade decay model, the neutralinos each decay to three quarks via an off-shell squark and the RPV UDD decay vertex with coupling \lampp. In this model, the neutralino is the lightest supersymmetric particle.

Samples are produced covering a wide range of both $\mgluino$ and $\mninoone$. In the six-quark direct gluino decay model, the gluino mass is varied from 500 to 1200~GeV. In the case of the cascade decays, for each gluino mass (400~GeV to 1.4~TeV), separate samples are generated with multiple neutralino masses ranging from 50~GeV to 1.3~TeV. In each case, $\mninoone < \mgluino$. In order to ensure the result has minimal sensitivity to the effects of initial state radiation (ISR), which could be poorly modeled in the signal samples,\footnote{\Herwigpp, which is used for signal simulation, is not expected to model additional energetic jets from ISR well because the leading-order evaluation of the matrix element is only performed for the $2\ra2$ particle scattering process.} the region with $(\mgluino-\mninoone)<100$~GeV is not considered. Due to the potentially large theoretical uncertainty on the non-SM colorflow given by UDD couplings, results are presented for a single model of radiation and no systematic uncertainty is assigned for this effect, further justifying the unevaluated region described above. All possible $\lamppijk$ flavor combinations given by the structure of \equref{rpvandudd:wrpv} are allowed to proceed with equal probability. As discussed in \secref{results}, the analysis maintains approximately equal sensitivity to all flavor modes.  All samples are produced assuming that the gluino and neutralino widths are narrow and that their decays are prompt. Cross-section calculations are performed at next-to-leading order in the strong coupling constant, adding the resummation of soft gluon emission at next-to-leading-logarithmic accuracy (NLO+NLL)~\cite{Beenakker:1996ch,Kulesza:2008jb,Kulesza:2009kq,Beenakker:2009ha,Beenakker:2011fu}.

Dijet and multi-jet events, as well as top quark pair production processes, were simulated in order to study the SM contributions and background estimation techniques. In the case of the vastly dominant background from SM jet production, several MC simulations were compared with data for the suitability of their descriptions of jet and multi-jet kinematic observables and topologies. For signal region selections that use $b$-tagging (the identification of jets containing $B$-hadrons), other backgrounds such as $t\bar{t}$, single top, and $W/Z$+jets become significant as well. These other backgrounds are estimated directly from the simulation.

In order to develop the data-driven background estimation techniques for multi-jet events from QCD processes, comparisons are made among various generators and tunes. In the case of the jet-counting analysis, the ATLAS tune AUET2B LO**~\cite{MC11c} of \Pythia 6.426~\cite{pythia} is used in estimating the rate of $n$-jet events (where $n=6,7$) as a function of the jet \pt\ requirement on the $n^{\rm th}$ jet. For the total-jet-mass analysis, \Sherpa 1.4.0~\cite{Gleisberg:2008ta} is used to develop and test the method. For the \Sherpa multi-jet samples, up to three partons are included in the matrix-element calculation and no electroweak processes are included. Heavy ($c$ and $b$) quarks are treated as massive. The next largest background after multi-jets is fully hadronic \ttbar production, which is also simulated with \Sherpa 1.4.0 and is used to estimate any background contamination in the control and signal regions defined in the analysis.

The jet-counting and total-jet-mass analyses use different multi-jet generators because of the different approaches to the background estimation employed by each analysis. The low-to-high jet-multiplicity extrapolation of the jet-counting analysis, described in \secref{resolved:backgrounds}, favors a generator that treats the production of an additional jet in a consistent manner, such as \Pythia, rather than a generator that treats the multileg matrix element separately from the additional radiation given by a separate parton shower model. In contrast, the total-jet-mass analysis uses the multi-jet simulation only to test the background estimation method and optimize the analysis as described in \secref{merged:backgrounds} and \secref{merged:SRCR}, and uses \Sherpa as it provides a better description of jet substructure variables, such as the jet mass used in this analysis.

The ATLAS simulation framework~\cite{simulation} is used to process both the signal and background events, including a full \geant~\cite{Geant4} description of the detector system. The simulation includes the effect of both in-time and out-of-time \pileup and is weighted to reproduce the observed distribution of the average number of collisions per bunch crossing in the data.

\section{Physics objects and event preselection}
\label{sec:event-selection}

\subsection{Data quality criteria}
\label{sec:event-selection:DQ}

The data are required to have met criteria designed to reject events with significant contamination from detector noise, noncollision beam backgrounds, cosmic rays, and other spurious effects. The selection related to these quality criteria is based upon individual assessments for each subdetector, usually separated into barrel, forward and endcap regions, as well as for the trigger and for each type of reconstructed physics object (i.e. jets).

To reject noncollision beam backgrounds and cosmic rays, events are required to contain a primary vertex consistent with the LHC beamspot, reconstructed from at least two tracks with transverse momenta $\pttrk>400$~MeV. Jet-specific requirements are also applied. All jets reconstructed with the \akt algorithm using a radius parameter of $R=0.4$ and a measured $\ptjet>20$~GeV are required to satisfy the ``looser'' requirements discussed in detail in Ref.~\cite{JetCleaning2011}. This selection requires that jets deposit at least 5\% of their measured total energy in the electromagnetic (EM) calorimeter as well as no more than 99\% of their energy in a single calorimeter layer.

The above quality criteria selections for jets are extended to prevent contamination from detector noise through several detector-region-specific requirements. Jets with spurious energy deposits in the forward hadronic endcap calorimeter are rejected and jets in the central region ($|\eta|<2.0$) that are at least 95\% contained within the EM calorimeter are required not to exhibit any electronic pulse shape anomalies~\cite{1748-0221-9-07-P07024}. Any event with a jet that fails the above requirements is removed from the analysis.

\subsection{Object definitions}
\label{sec:event-selection:objects}

Jets are reconstructed using the \akt algorithm with radius parameters of both $R=0.4$ and $R=1.0$. The former are referred to as \textit{standard} jets and the latter as \textit{\largeR} jets. 
The inputs to the jet reconstruction are three-dimensional topological clusters~\cite{TopoClusters}. This method first clusters together topologically connected calorimeter cells and classifies these clusters as either electromagnetic or hadronic. The classification uses a local cluster weighting (\LCW) calibration scheme based on cell-energy density and longitudinal depth within the calorimeter~\cite{Aad:2014bia}.
Based on this classification, energy corrections derived from single-pion MC simulations are applied. Dedicated corrections are derived for the effects of noncompensation, signal losses due to noise-suppression threshold effects, and energy lost in noninstrumented regions. 
An additional jet energy calibration is derived from MC simulation as a correction relating the calorimeter response to the true jet energy. In order to determine these corrections, the identical jet definition used in the reconstruction is applied to particles with lifetimes greater than 10~ps output by MC generators, excluding muons and neutrinos. Finally, the standard jets are further calibrated with additional correction factors derived \textit{in-situ} from a combination of $\gamma$+jet, $Z$+jet, and dijet balance methods~\cite{Aad:2014bia}.

No explicit veto is applied to events with leptons or \Etmiss. This renders the analysis as inclusive as possible and leaves open the possibility for additional interpretations of the results. There is no explicit requirement removing identified leptons from the  jets considered in an event. Calorimeter deposits from leptons may be considered as jets in this analysis given that the data quality criteria described in \secref{event-selection:DQ} are satisfied. A further consequence of these requirements is that events containing hard isolated photons, which are not separately identified and distinguished from jets, have a high probability of failing to satisfy the signal event selection criteria. For the signals considered, typically $1\%$ of events fail these quality requirements.

The standard jet \pT requirement is always chosen to be at least 60 GeV in order to reside in the fully efficient region of the multi-jet trigger. For the jet-counting analysis selection (\secref{Resolved}), a requirement of $\ptjet>80$~GeV is imposed for each jet in most of the background control regions, and a higher requirement is used for the majority of the signal regions of the analysis. All jets used in this analysis are required to have $|\eta| < 2.8$. The effect of \pileup\ on jets is negligible for the kinematic range considered, and no selection to reduce \pileup\ sensitivity is included.

In order to constrain specific UDD couplings to heavy flavor quarks, $b$-tagging requirements are also applied to some signal regions. In these cases, one or two standard jets are required to satisfy $b$-tagging criteria based on track transverse impact parameters and secondary vertex identification~\cite{mv1}. In simulated \ttbar\ events, this algorithm yields a 70\% (20\%) tagging efficiency for real $b$-($c$-)jets and an efficiency of 0.7\% for selecting light quark and gluon jets. The $b$-tagging efficiency and misidentification are corrected by scale factors derived in data~\cite{mv1}. These jets are additionally required to lie within the range $|\eta| < 2.5$.

The topological selection based on the total mass of \largeR\ jets (\secref{Merged}) employs the \textit{trimming} algorithm~\cite{Krohn2010}. This algorithm takes advantage of the fact that contamination from the underlying event and \pileup in the reconstructed jet is often much softer than the outgoing partons from the hard-scatter. The ratio of the \pT\ of small subjets (jets composed of the constituents of the original jet) to that of the jet is used as a selection criterion.  The procedure uses a \kt algorithm~\cite{Ellis1993, Catani1993} to create subjets with a radius $\drsub=0.3$. Any subjets with $\pti/\ptjet < \fcut$ are removed, where \pti\ is the transverse momentum of the $i^{\rm th}$ subjet, and $\fcut=0.05$ is determined to be an optimal setting~\cite{Aad:2013gja}. The remaining constituents form the trimmed jet, and the mass of the jet is the invariant mass of the remaining subjets (which in turn is the invariant mass of the massless topological clusters that compose the subjet). Using these trimming parameters, the full mass spectrum is insensitive to \pileup.

The total-jet-mass analysis uses a sample from the high-\ptjet\ single-jet triggers. A requirement that the leading \largeR\ jet have $\ptjet>500$~GeV is applied to ensure that these triggers are fully efficient.

\section{Jet-counting analysis}
\label{sec:Resolved}

\subsection{Method and techniques}
\label{sec:resolved:backgrounds}

The jet-counting analysis searches for an excess of events with $\geq$ 6 or $\geq$ 7 high \pt jets jets (with at least 80 GeV), with $\geq$ 0, $\geq$ 1, or $\geq$ 2 $b$-jet requirements added to enhance the sensitivity to couplings that favor decays to heavy-flavor quarks. The number of jets, the \pt requirement that is used to select jets, and the number of $b$-tagged jets are optimized separately for each signal model taking into account experimental and theoretical uncertainties.

The background yield in each signal region is estimated by starting with a signal-depleted control region in data and extrapolating its yield into the signal region using a factor that is determined from a multi-jet simulation, with corrections applied to account for additional minor background processes. This can be expressed as:
%
\begin{eqnarray}   \label{eq:ProjectionFormula}
N_{n\text{-jet}} &= \left( N^{\rm data}_{m\text{-jet}} - N^{\rm MC}_{m\text{-jet, Other BGs}} \right) \times \left( \frac{ N^{\rm MC}_{n\text{-jet}} }{ N^{\rm MC}_{m\text{-jet}} } \right)  \nonumber \\
&\quad + N^{\rm MC}_{n\text{-jet, Other BGs} }
\end{eqnarray}
%
%
where the number of predicted background events with $n$ jets ($N_{n\text{-jet}}$) is determined starting from the number of events in the data with $m$ jets ($N^{\rm data}_{m\text{-jet}}$). The extrapolation factor, $\frac{ N^{\rm MC}_{n\text{-jet}} }{ N^{\rm MC}_{m\text{-jet}} }$, is determined from multi-jet simulation and validated in the data. This procedure is performed in exclusive bins of jet multiplicity. Since the simulation is not guaranteed to predict this scaling perfectly, cross-checks in the data and a data-driven determination of systematic uncertainties are performed as described in \secref{reolved:systematics}. It is assumed here that the simulation used for this extrapolation given by \Pythia 6.426 predicts the relative rate of events with one additional order in the strong coupling constant in a consistent way across jet-multiplicity regions. This assumption comes from the behavior of the parton shower model used by \Pythia to obtain configurations with more than two partons and is shown to be consistent with data in the measurement of multi-jet cross-sections~\cite{Aad:2011tqa}. Other models were studied and are discussed in \secref{reolved:systematics}.

Small corrections from other backgrounds (\ttbar, single top, and $W/Z$+jet events) are applied based on estimates from the simulation. Without $b$-tagging, the contribution of events from these other backgrounds is less than 1\%. Including two $b$-tagged jets increases this relative contribution to as much as 10\%.

\begin{figure*}[!htb]
    \centering
    \subfigure[$\ge0$ $b$-tagged jets required] {
        \includegraphics[width=0.45\textwidth]{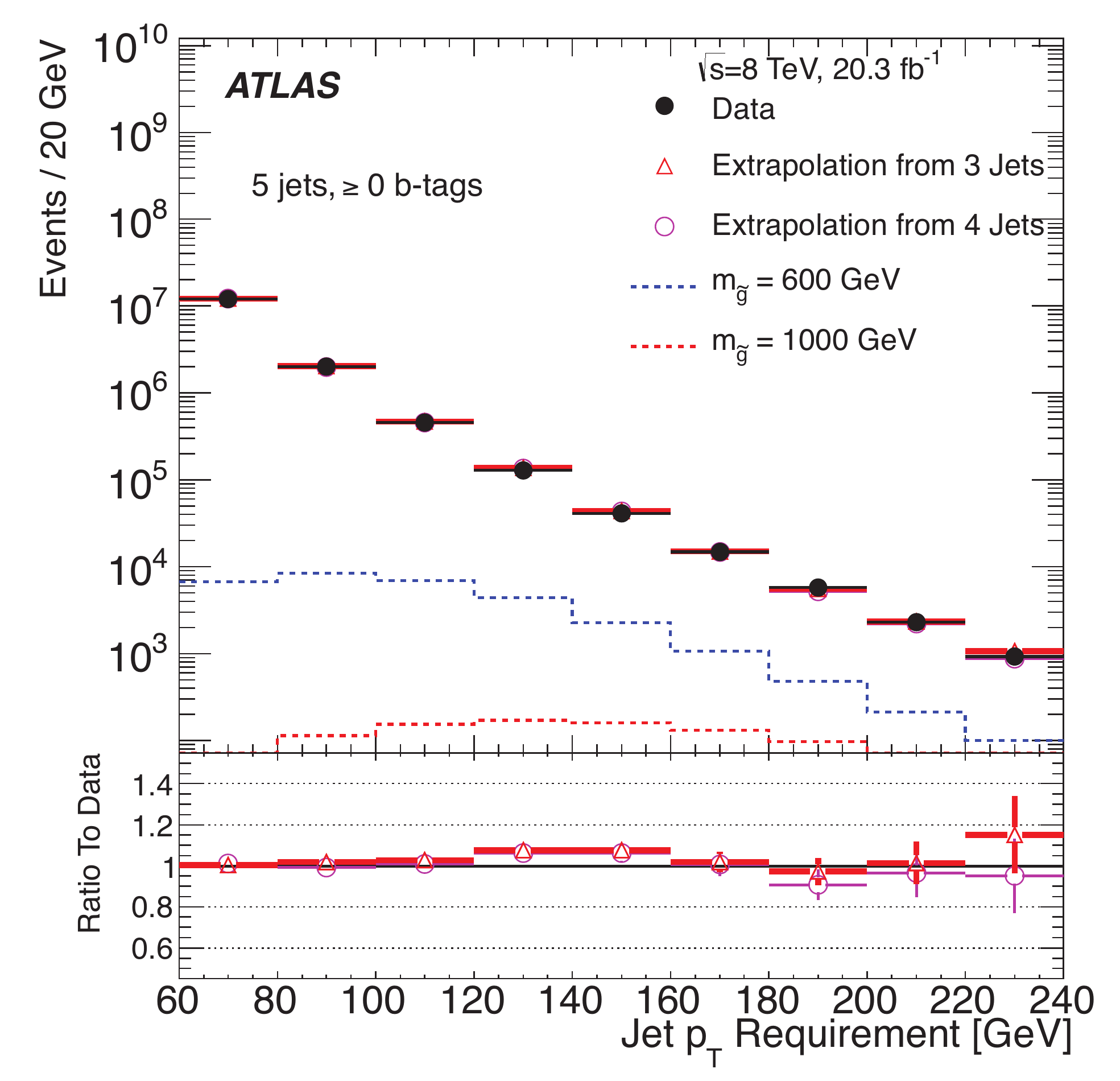}
    }\label{fig:5JetPythiaValidationPretag}
    \subfigure[$\ge1$ $b$-tagged jets required] {
        \includegraphics[width=0.45\textwidth]{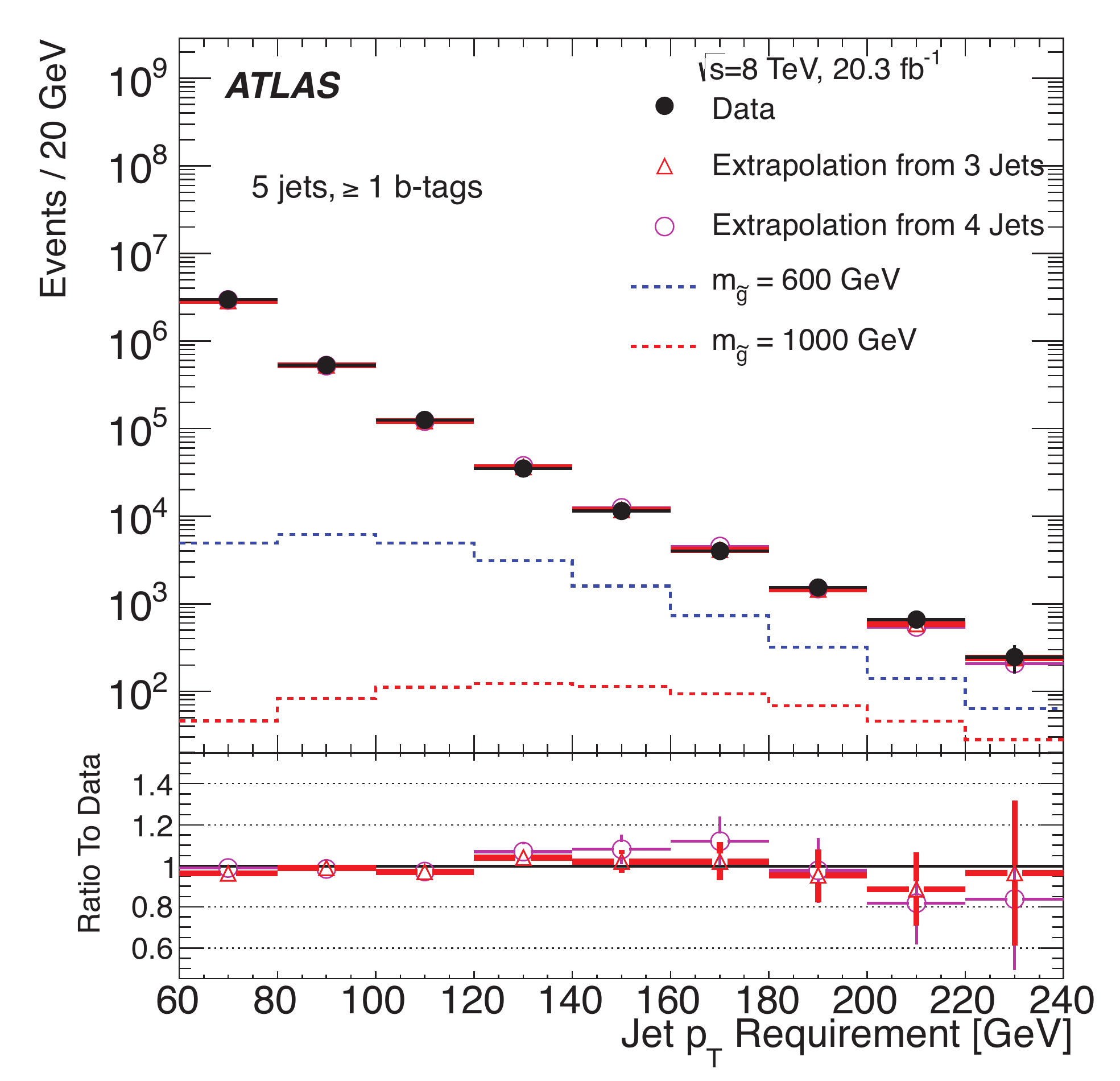}
    }\label{fig:5JetPythiaValidationOneBTag}
    \subfigure[$\ge2$ $b$-tagged jets required] {
        \includegraphics[width=0.45\textwidth]{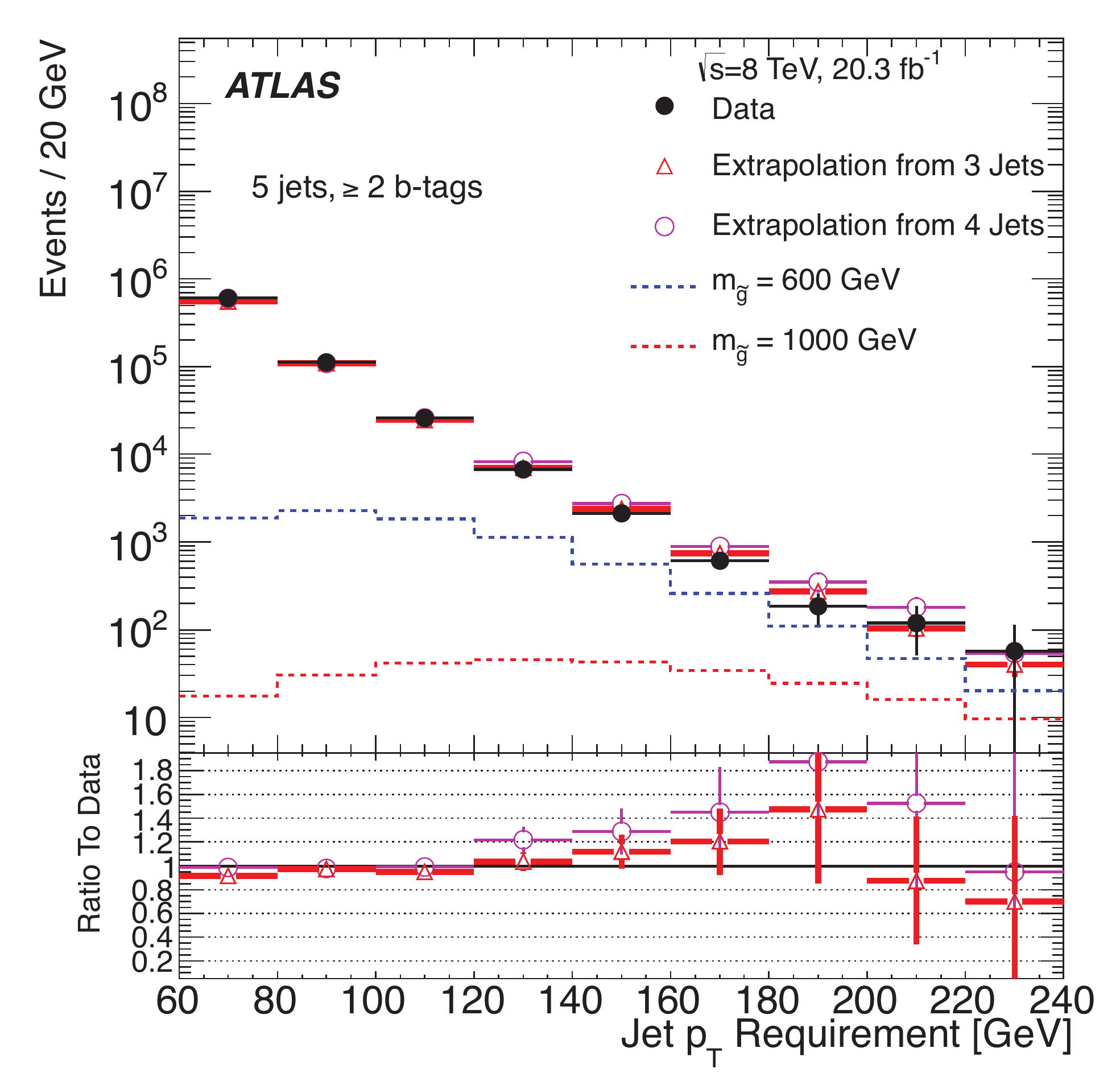}
    }\label{fig:5JetPythiaValidationTwoBTags}
    \caption{The number of observed events in the 5-jet bin is compared to the background expectation that is determined by using \Pythia to extrapolate the number of events in data from the low jet-multiplicity control regions. The contents of the bins represent the number of events with 5-jets passing a given jet \pt requirement. These bins are inclusive in jet \pt. Results with various $b$-tagging requirements are shown.}
    \label{fig:5JetPythiaValidation}
\end{figure*}

\subsection{Signal and control region definitions}
\label{sec:resolved:SRCR}

Control regions are defined with $m \leq 5$, for which the background contribution is much larger than the expected signal contributions from the benchmark signal processes. Extrapolation factors with $n,m \leq 5$ are used to validate the background model and to assign systematic uncertainties. For $n>5$, the expected signal contributions can become significant and an optimization is performed to choose the best signal region definitions for a given model. Signal regions are chosen with simultaneous optimizations of the jet-multiplicity requirement ($\ge6$ or $\ge7$ jets), the associated transverse momentum requirement (80--220 GeV in 20 GeV steps), and the minimum number of $b$-tagged jets ($\ge0$, $\ge1$, or $\ge2$) for a total of 48 possible signal regions. Alternative control regions are constructed from some $n>5$ regions when the signal significance is expected to be low as described in \secref{reolved:systematics}. Such regions are then excluded from the list of allowed signal regions. For a given signal model, the signal region deemed most effective by this optimization procedure is used for the final interpretations. The signal regions chosen by the optimization procedure tend to pick regions with signal acceptances as low as 0.5\% and as high as roughly 20\%.

Although other choices are also studied to determine background yield systematic uncertainties from the data, the background contributions are estimated in the final signal regions using extrapolations across two jet-multiplicity bins ($n = m+2$). This choice leads to negligible signal contamination in the control regions used for this nominal prediction.

\begin{figure}[!htb]
\centering

  \subfigure[Extrapolation in $4\ra6$-jet bin] {
    \includegraphics[width=0.45\textwidth]{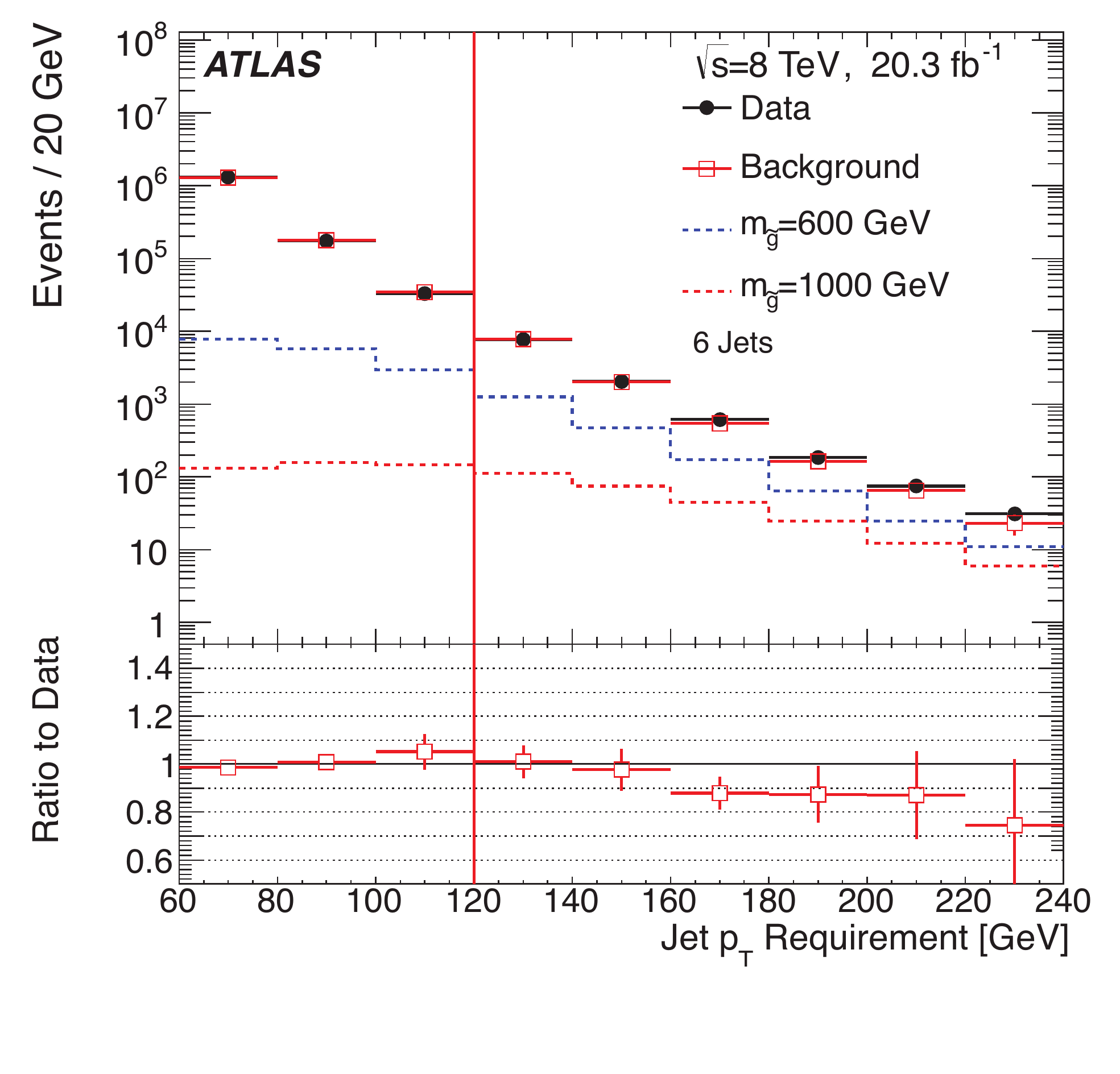}
    \label{fig:validation:lowpt}
  }
  \subfigure[Extrapolation in $5\ra7$-jet bin with $\langle|\eta|\rangle > 1.0$] {
    \includegraphics[width=0.45\textwidth]{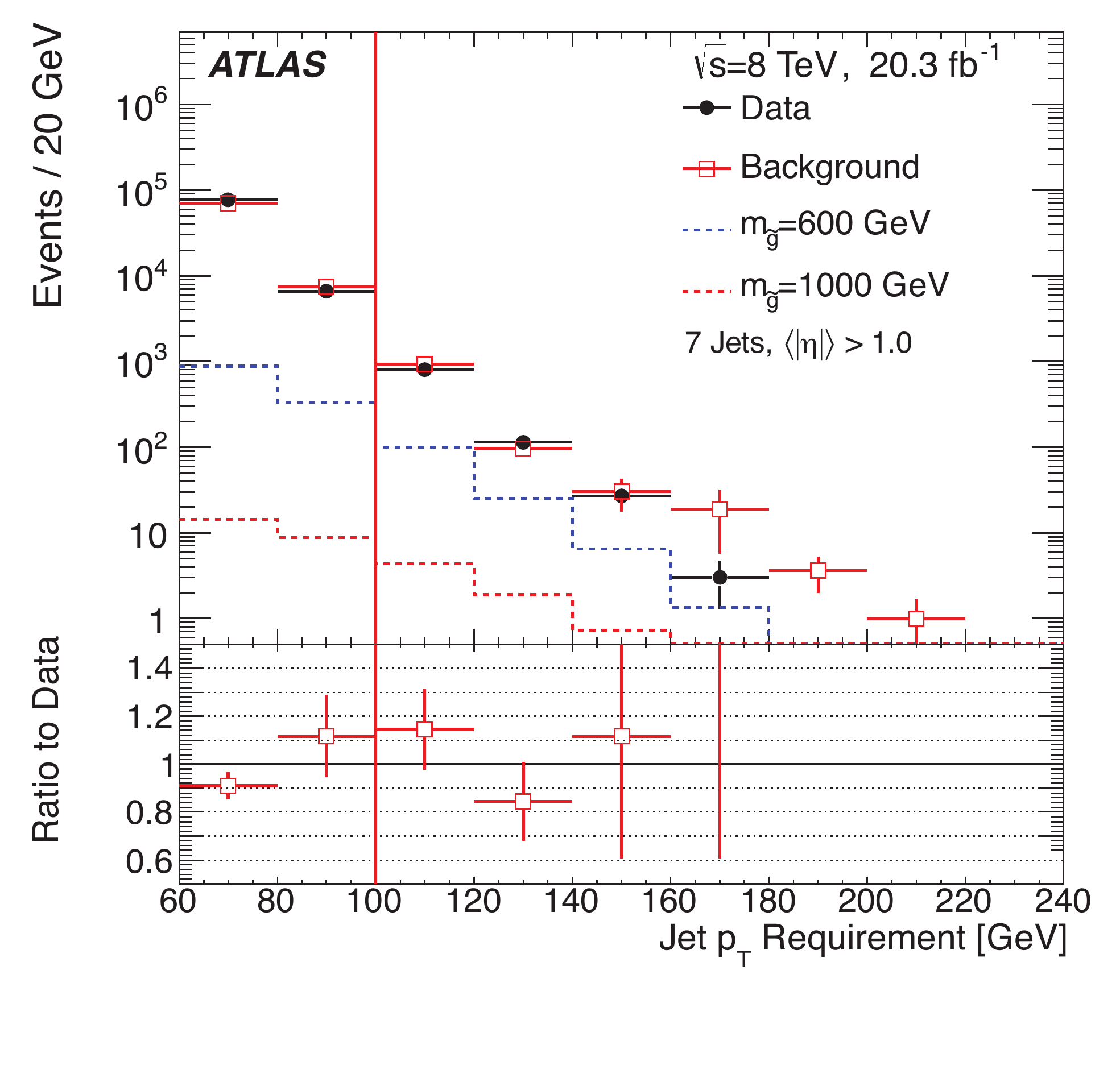}
    \label{fig:validation:deta}
  }
  
  \caption{The data are compared with the expected background shapes in the exclusive 6- and 7-jet bins before $b$-tagging. The contents of the bins represent the number of events with the given number of jets passing a given jet \pt requirement. The bins with less than 10\% expected signal contamination are control regions that are considered when assigning systematic uncertainties to the background yield. These control regions are the bins to the left of the vertical red lines in the plots. \subref{fig:validation:lowpt} shows the 6-jet region, and \subref{fig:validation:deta} shows the 7-jet region with $\langle|\eta|\rangle > 1.0$.}
  
  \label{fig:validation}
\end{figure}  

\begin{figure}[htb]
  \centering
    \subfigure[$\ge6$ jets, $\ge0$ $b$-tagged jets] {
        \includegraphics[width=0.4\textwidth]{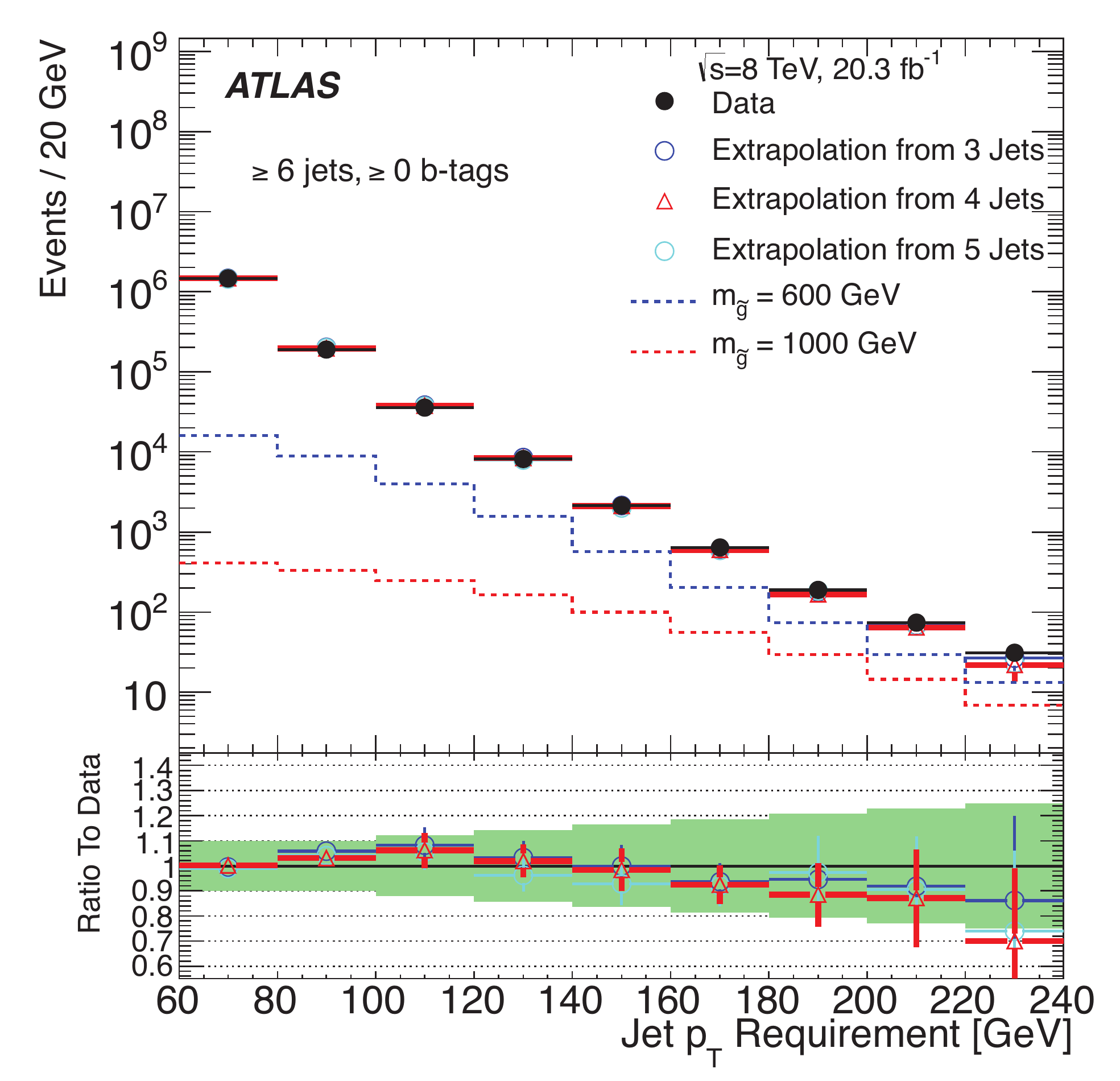}
    } \\
    \subfigure[$\ge7$ jets, $\ge0$ $b$-tagged jets] {
        \includegraphics[width=0.4\textwidth]{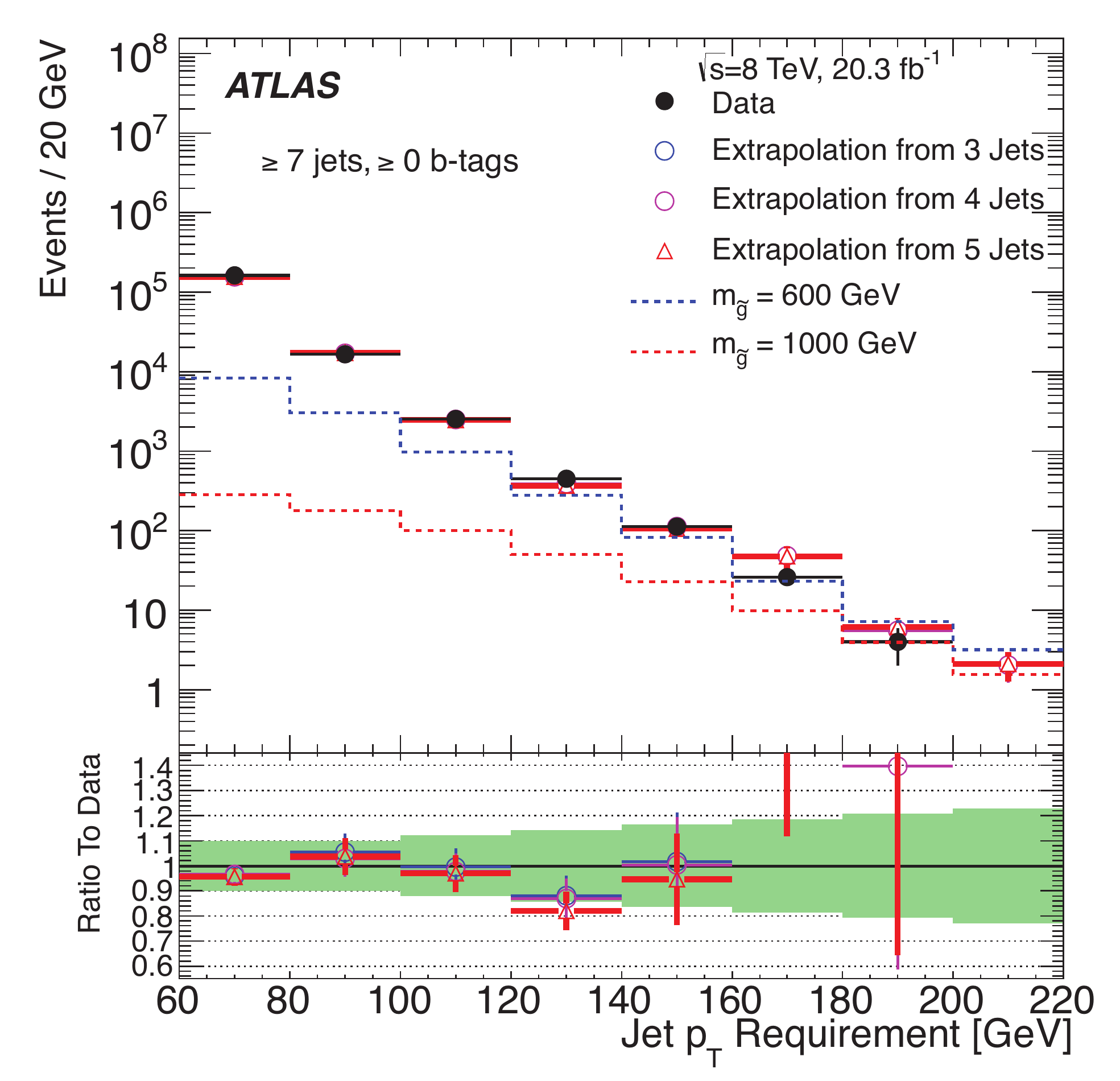}
    }
    \caption{The number of observed events in the inclusive $\ge6$-jet (top) and $\ge7$-jet (bottom) signal regions compared with expectations using the \Pythia extrapolations from low jet-multiplicity control regions, as a function of the jet \pT requirement. The distributions representing the extrapolations across two units in jet multiplicity (red triangles) are used as the final background prediction in each case, while the other extrapolations are treated as cross-checks. $\ge0$ $b$-tagged jets are required. In the ratio plots the green shaded regions represent the background systematic uncertainties.}
    
    \label{fig:projections0b}
\end{figure}

\begin{figure}[htb]
  \centering
    \subfigure[$\ge6$ jets, $\ge1$ $b$-tagged jet] {
        \includegraphics[width=0.4\textwidth]{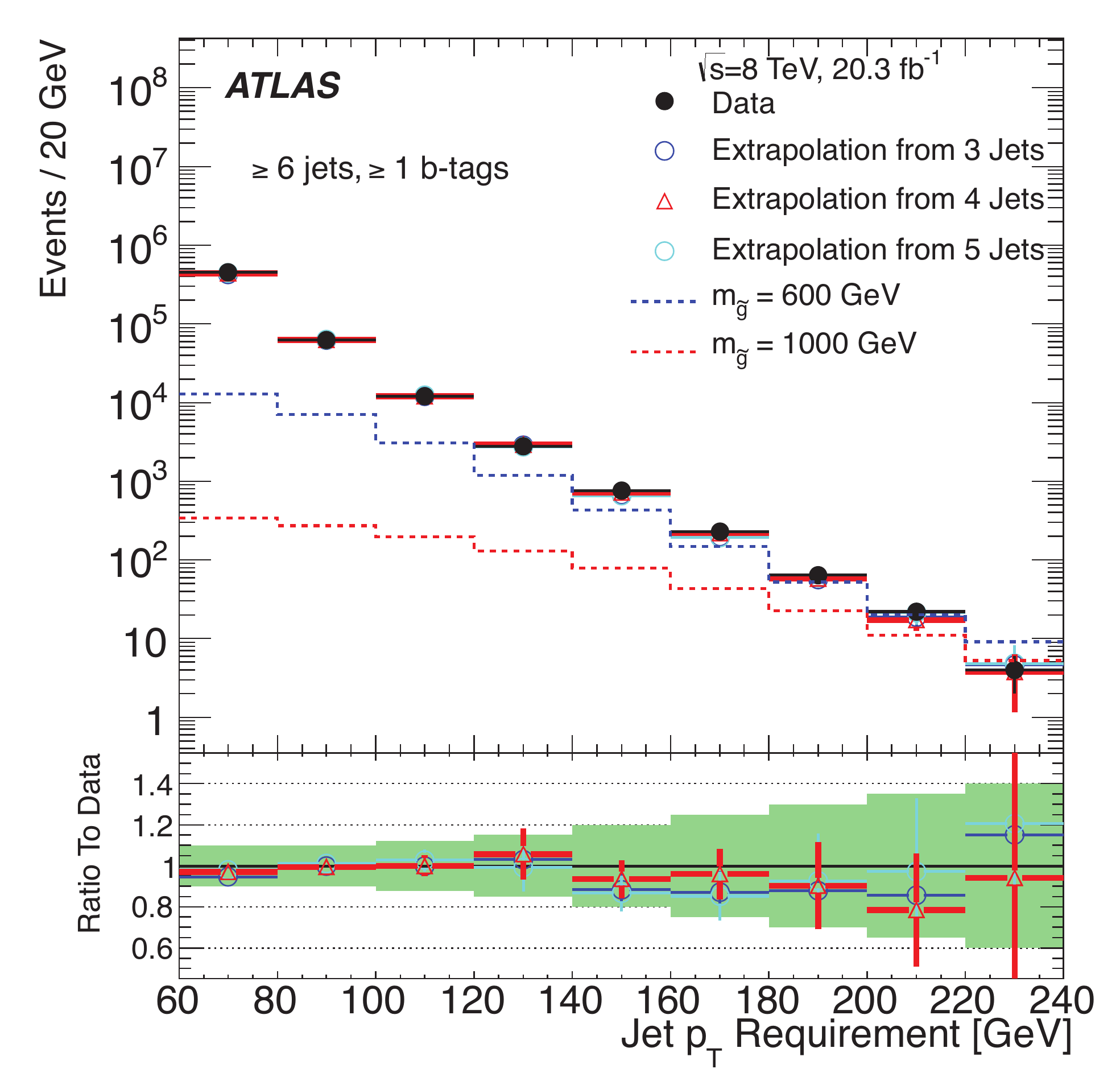}
    }\\
    \subfigure[$\ge7$ jets, $\ge1$ $b$-tagged jet] {
        \includegraphics[width=0.4\textwidth]{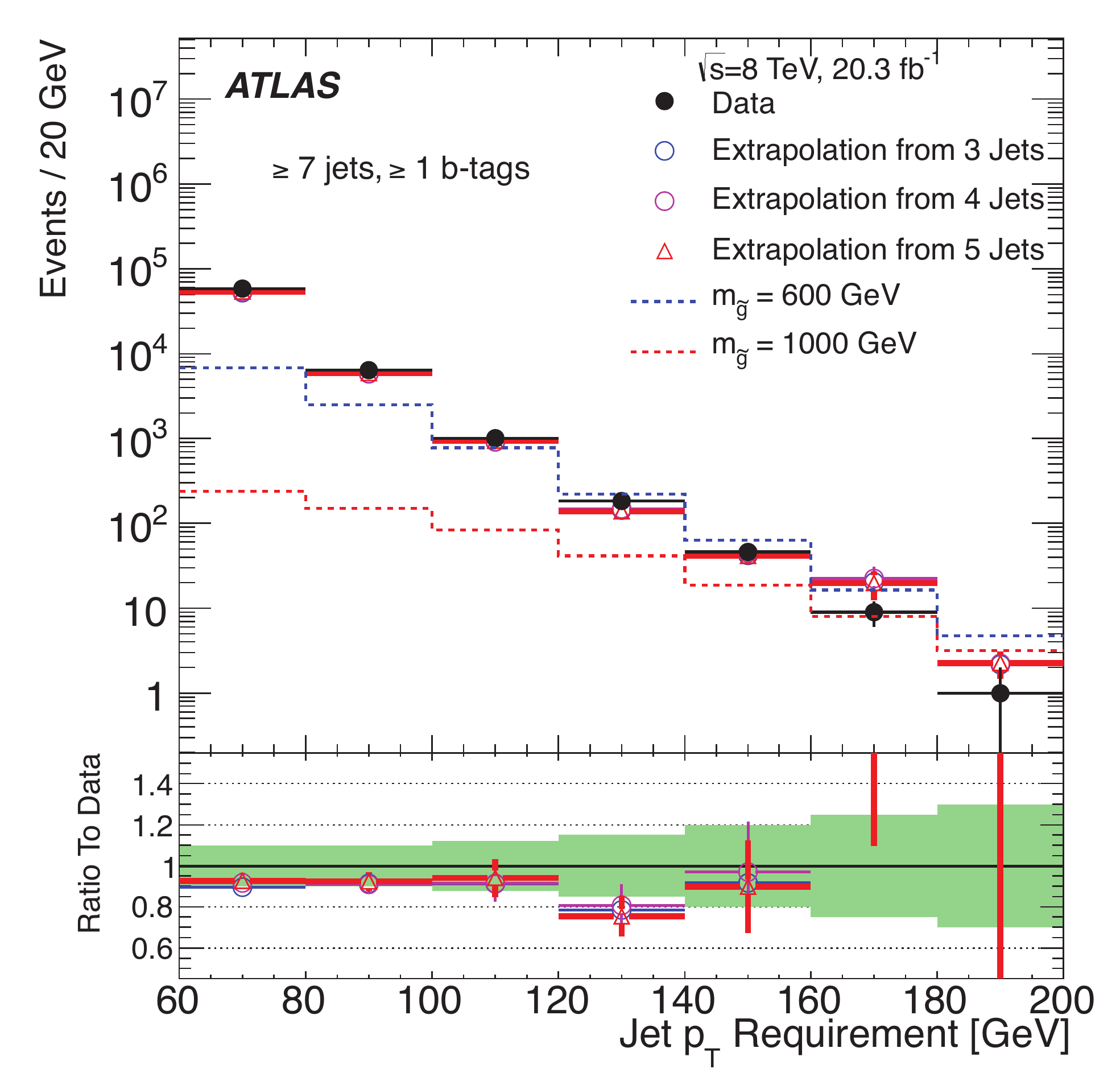}
    }
    \caption{Distributions shown here are as in \figref{projections0b} but with $\ge1$ $b$-tagged jets required.}
    
    \label{fig:projections1b}
\end{figure}

\begin{figure}[htb]
  \centering
    \subfigure[$\ge6$ jets, $\ge2$ $b$-tagged jets] {
        \includegraphics[width=0.4\textwidth]{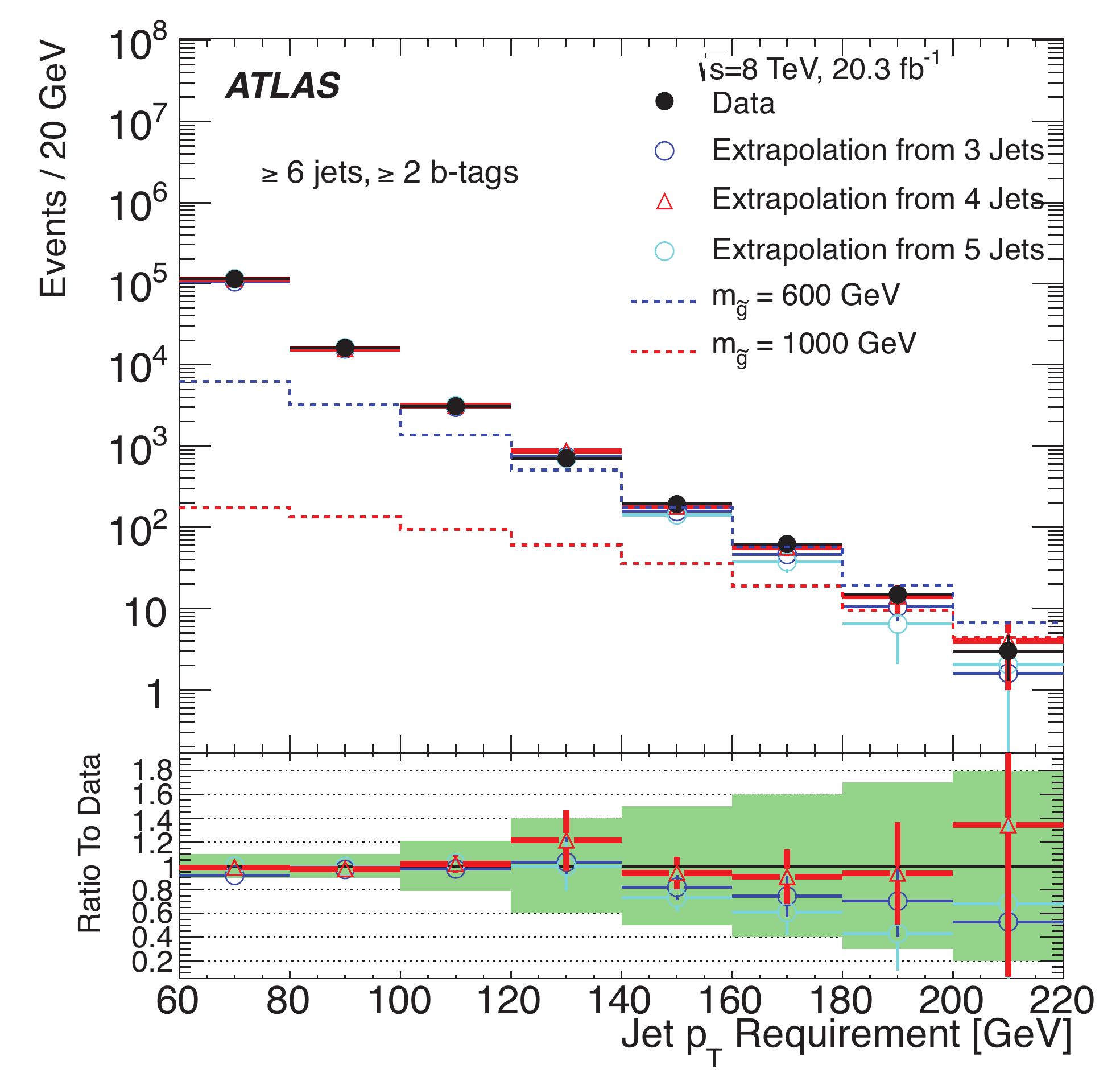}
    }\\
    \subfigure[$\ge7$ jets, $\ge2$ $b$-tagged jets] {
        \includegraphics[width=0.4\textwidth]{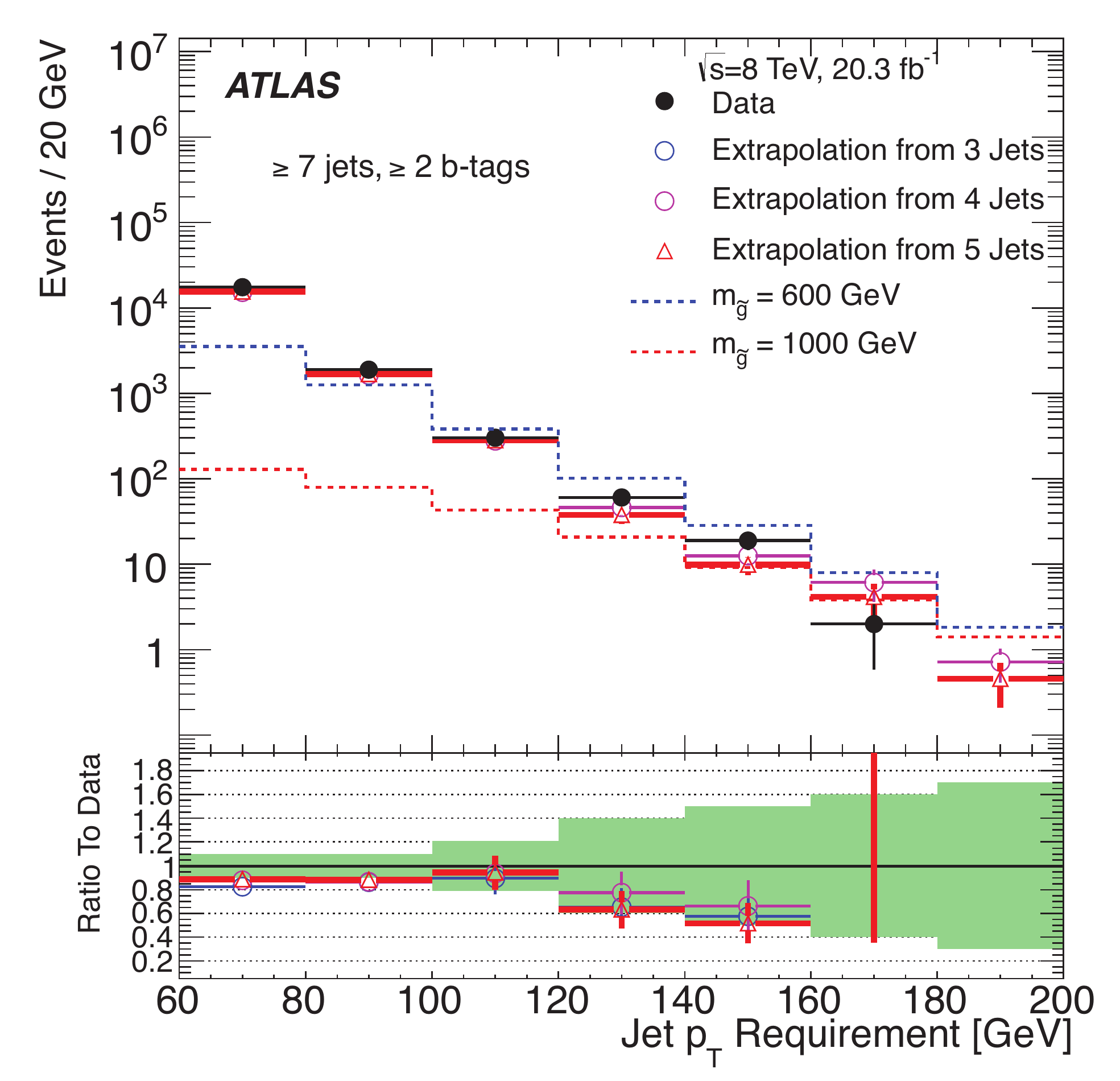}
    }
    \caption{Distributions shown here are as in \figref{projections0b} but with $\ge2$ $b$-tagged jets required.}
    
    \label{fig:projections2b}
\end{figure}

\begin{figure*}[!htb]
    \centering
    \subfigure[Exactly $5$ jets, $\ge0$ $b$-tagged jets] {
        \includegraphics[width=0.47\textwidth]{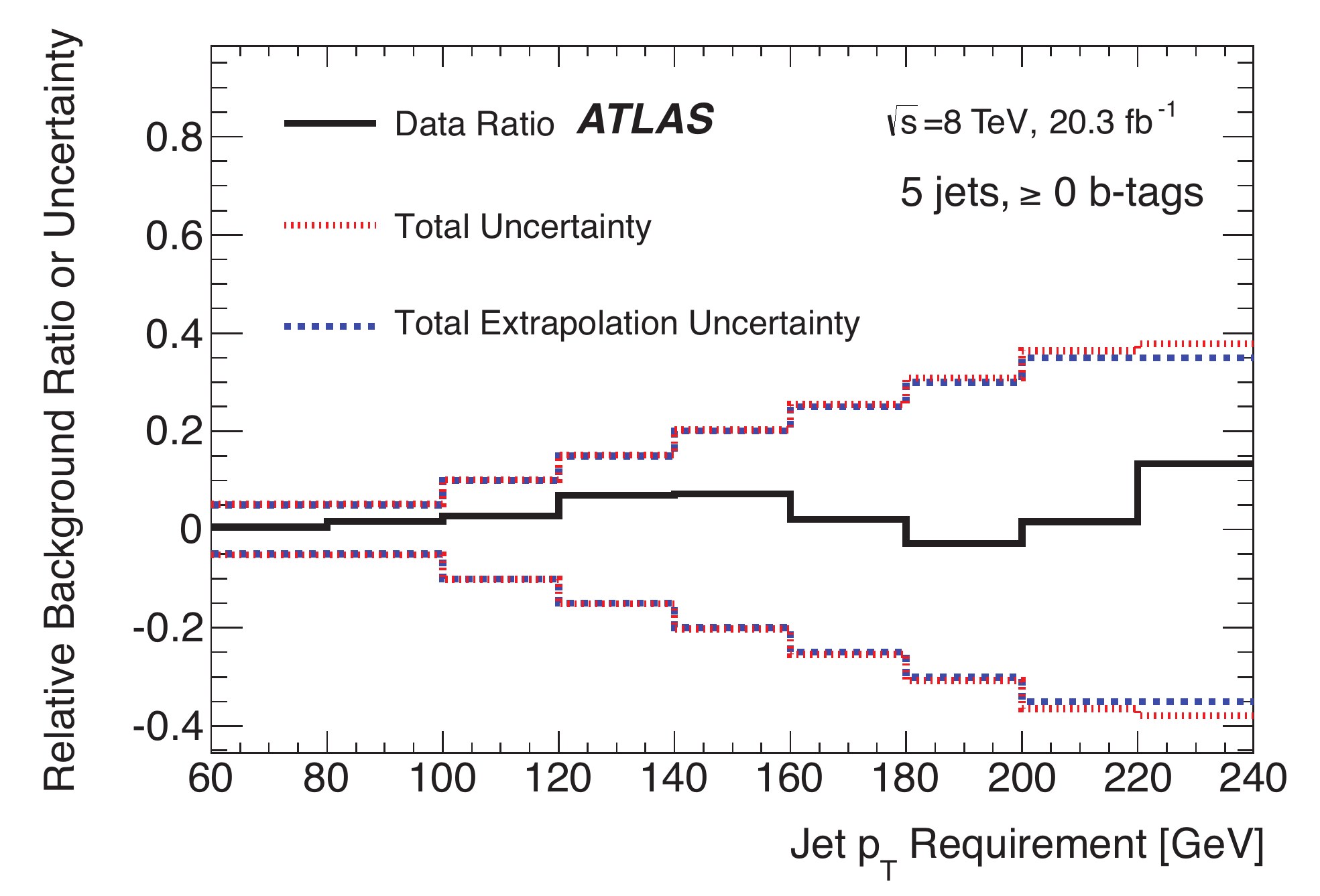}
    }\label{fig:5jetvalidationbias:0btag}
    \subfigure[$\ge6$ jets, $\ge0$ $b$-tagged jets] {
        \includegraphics[width=0.47\textwidth]{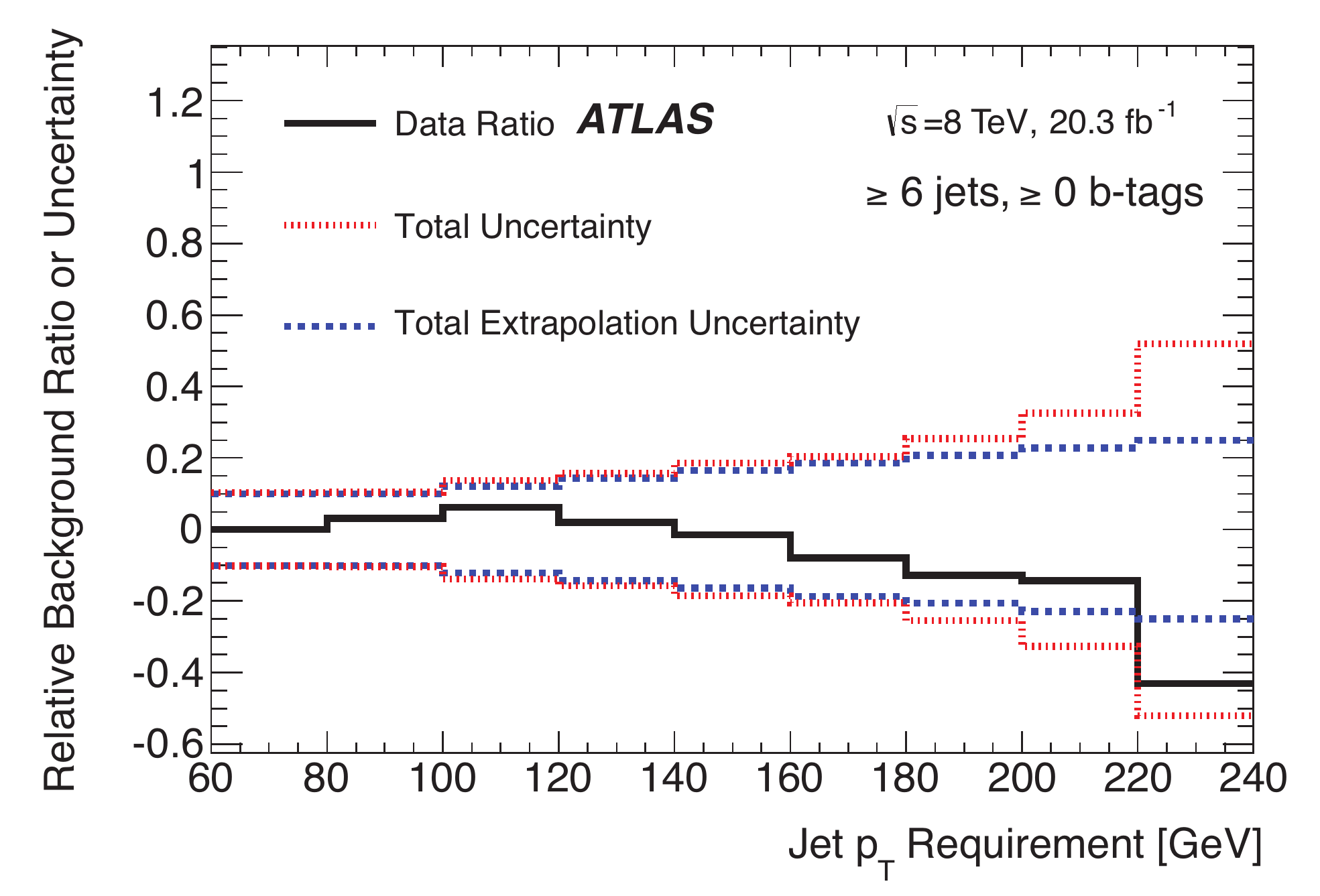}
    }
    \subfigure[$\ge7$ jets, $\ge0$ $b$-tagged jets] {
        \includegraphics[width=0.47\textwidth]{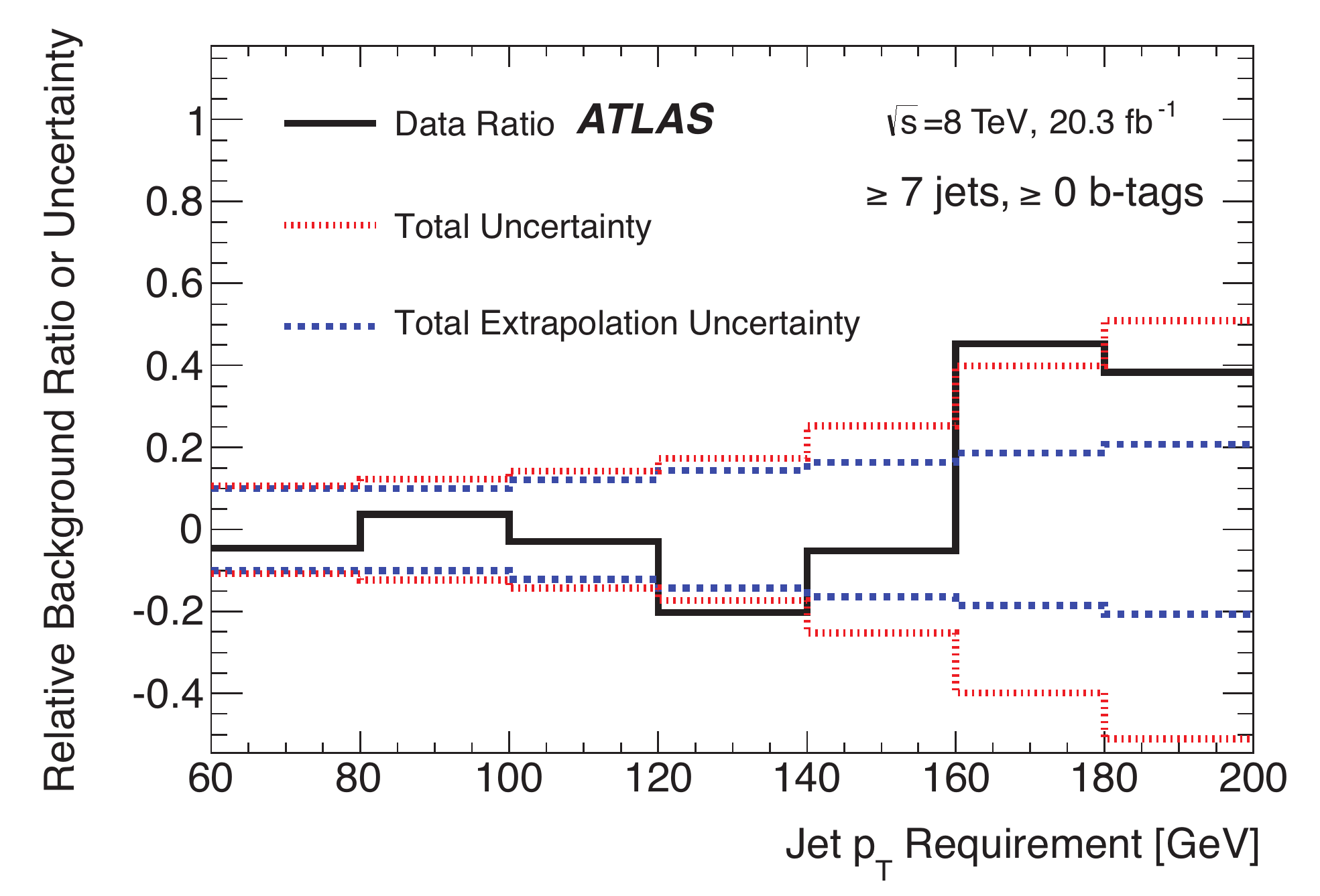}
    }
    \caption{Comparisons of the deviation between data and expectations in the control and signal regions without $b$-tagging requirements are shown, as a function of the jet \pT requirement. The solid black line shows the relative difference between the observed data and the predicted background. The coarsely dashed blue distribution shows the relative systematic uncertainty on the background estimation. The finely dashed red distribution shows the total uncertainty on the comparison between background and data, including the background systematic uncertainty and all sources of statistical uncertainty from the data and simulation.}
    \label{fig:projectionuncs}
\end{figure*}

\subsection{Validation and systematic uncertainties}
\label{sec:reolved:systematics}

Since the 3-, 4-, and 5-jet-multiplicity bins have minimal expected signal contamination they are used to validate the background model based on the MC simulation. The initial validation of the background prediction is performed by extrapolating the background from either the $m=3$ or $m=4$ jets control region into the $n=5$ jets control region and comparing with the data. This comparison is presented in \figref{5JetPythiaValidation}, which shows the number of events passing a given jet \pt requirement with a 5-jet requirement. This procedure is shown to be accurate in the extrapolations to the 5-jet bin in data, both with and without the requirement of $b$-tagging. The conclusion of this validation study is that \equref{ProjectionFormula} can be used with no correction factors, but a systematic uncertainty on the method is assigned to account for the discrepancies between data and the prediction in the control regions. This systematic uncertainty is assigned to cover, per \ptjet\ bin, the largest discrepancy that is observed between data and the prediction when extrapolating from either the 3-jet or 4-jet bins into the 5-jet control region, as well as from extrapolations to higher jet multiplicity as discussed below.

Alternative MC models of extra-jet production such as those given by \Sherpa, \Herwigpp, and additional parameter tunes in \Pythia were studied and either did not satisfy the criterion that the model be consistent through control and signal regions (e.g. the model must not describe the control regions with a matrix element calculation and the signal regions with a parton shower model giving unreliable projections) or disagreed significantly with the data in the validations presented here. The internal spread of predictions given by each of these background models in various extrapolations is considered when assigning systematic uncertainties. In all cases, this spread is consistent with the systematic uncertainties obtained using \Pythia in the manner described above.

In addition to the extrapolation factor described by \equref{ProjectionFormula}, it is possible to also study the extrapolation along the jet \pt degree of freedom. In this case, the $n$-jet event yield for a given high jet-\pt selection is predicted using extrapolation factors from lower jet-\pt selections determined from MC simulation. This method is tested exclusively in a low $n$-jet region for the high jet-\pt requirement and the spread is compared to the baseline systematic uncertainty, which is increased in case of disagreement larger than this baseline.

Additional control regions can be constructed from exclusive 6-jet regions with low jet-\pt requirements. Any region with an expected signal contribution less than 10\% for the $\mgluino=600$ GeV six-quark model is used as additional control region in the evaluation of the background systematic uncertainties. These regions are used to ensure that the jet-multiplicity extrapolation continues to accurately predict the event rate at higher jet multiplicities, as shown in \figref{validation:lowpt}, without looking directly at possible signal regions. This procedure allows the exclusive 6-jet, low jet-\pt\ region to be probed and shows that the jet-multiplicity extrapolations continue to provide accurate predictions at higher jet multiplicities.

To extend this validation, a requirement that the average jet pseudorapidity $\langle|\eta|\rangle > 1.0$ is applied to create a high-pseudorapidity control region to reduce the signal contribution to a level of less than approximately 10\% while retaining a reasonable number of events. Results of these extrapolations are shown for the exclusive 7-jet bin in \figref{validation:deta}. The largest deviations from the expected values are found to be a few percent larger than for the 5-jet extrapolations.

The uncertainty due to any mismodeling of contributions from backgrounds such as \ttbar, single top, and $W/Z$+jet processes is expected to be small and is covered by the procedure above since these contributions are included in the extrapolation. Therefore, any mismodeling of these sources results in increased systematic uncertainty on the entire background model in this procedure.

Distributions for data in the inclusive $\ge6$-jet and $\ge7$-jet signal regions are shown in \figrange{projections0b}{projections2b} compared with background predictions determined using extrapolations from three different jet-multiplicity bins. In each case, the distributions representing the extrapolations across two jet-multiplicity bins (i.e. $4\rightarrow6$ and $5\rightarrow7$) are used as the final background prediction whereas the other extrapolations are simply considered as additional validation. Contributions from higher jet-multiplicity regions are summed to construct an inclusive sample. The systematic uncertainty is constructed from the maximum deviation given by the various validations and for most signal regions is dominated by the baseline uncertainty obtained from the $n\le5$ jet regions. Results using the three $b$-tagging selections ($\ge0$, $\ge1$, $\ge2$ $b$-tagged jets) are shown in \figrange{projections0b}{projections2b}. The background systematic uncertainties determined from the control regions in the data are shown as the green shaded region in the ratio plots of these figures. This procedure results in a background systematic uncertainty in the $\ptjet\ge120$~GeV, $\ge7$-jet region of 14\%, 15\%, and 40\% for $\ge0$, $\ge1$, $\ge2$ $b$-tagged jets, respectively.

The bins in these distributions that were not assigned as control regions represent possible signal regions, which may be chosen as a signal region for a particular model under the optimization procedure described in \secref{results:resolved}. The level of disagreement between the expectation and data is shown in \figref{projectionuncs} for the $\ge0$ $b$-tagged jets control and signal regions. In the $b$-tagged signal regions similar agreement is observed between data and the predicted background, within the assigned uncertainties. In practice, it is seen that for most signals, the $\ge7$-jet bin is preferred by the optimization procedure as a signal region. The data in each distribution show good agreement with background predictions within uncertainties.

Systematic uncertainties on the jet-counting background estimation using the extrapolation method are determined directly from the data as part of the background validation and, by design, account for all uncertainties on the technique and on the reference model used in the projection. In contrast, systematic uncertainties on the signal predictions are determined from several sources of modeling uncertainties. The largest systematic uncertainties are those on the background yield, the jet energy scale uncertainties on the signal yield (10--20\% for most signal regions), and the uncertainty in $b$-tagging efficiencies for many signal regions that require the presence of $b$-tagged jets (between 15--20\% for signal regions requiring at least two $b$-tags). 

An additional systematic uncertainty is included in these estimates in order to cover possible contamination of signal in the control regions for the extrapolation. The analysis is repeated with signal injected into the control regions and the backgrounds are re-computed. The resulting bias depends on the signal model and is found to be less than 5\% in all cases. 

Given the good agreement between the data and the predictions from the jet-counting background estimation, there is no evidence of new physics.

\section{Total-jet-mass analysis}
\label{sec:Merged}

\begin{figure}[!htb]
  \centering
  
  \subfigure[\MJ for fixed $\mninoone = 175$~GeV]{
   \includegraphics[width=0.47\textwidth]{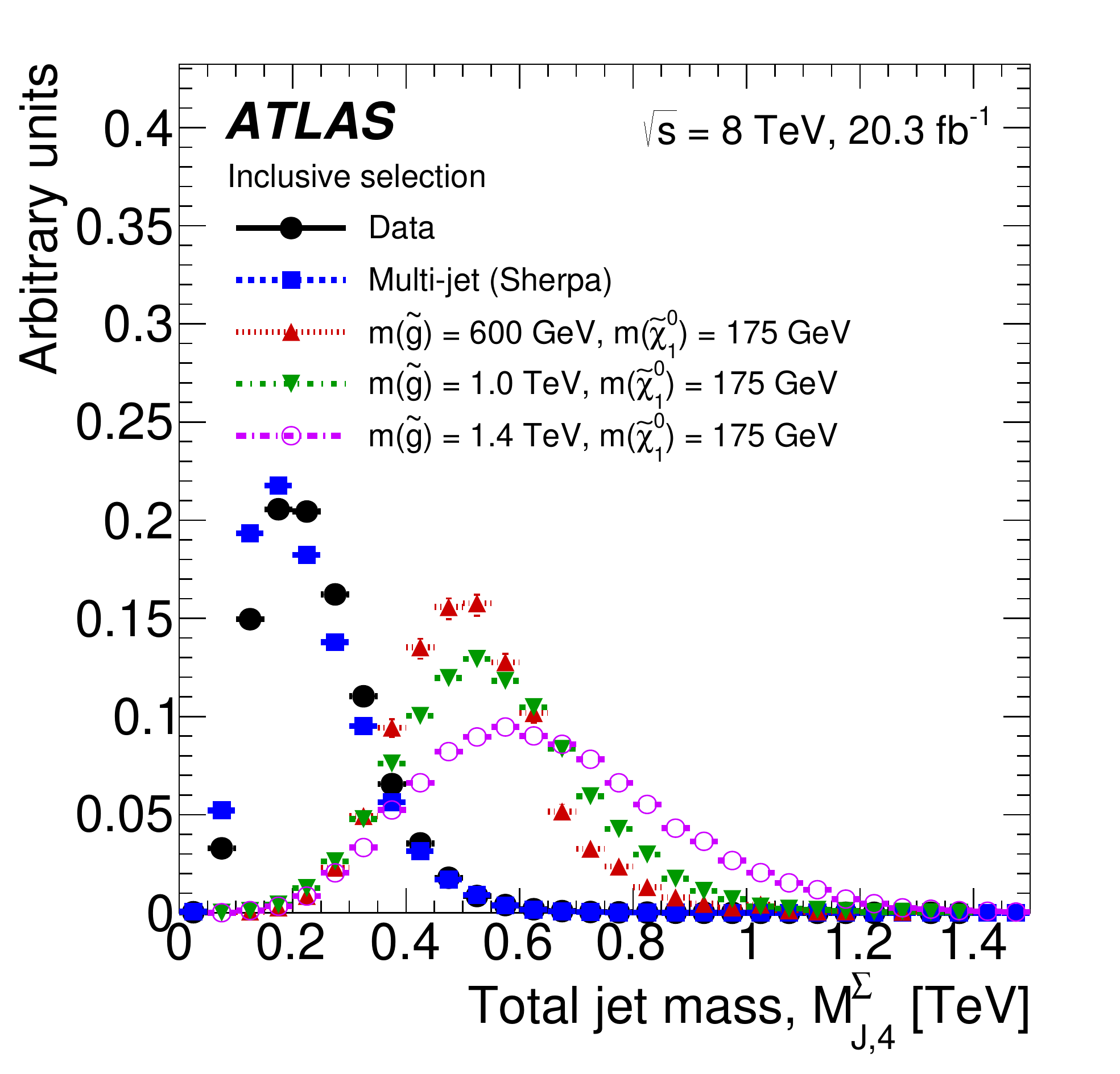}
    \label{fig:substructure:SvsB:MJ}}
  \subfigure[\Deta for fixed $\mninoone = 175$~GeV]{
   \includegraphics[width=0.47\textwidth]{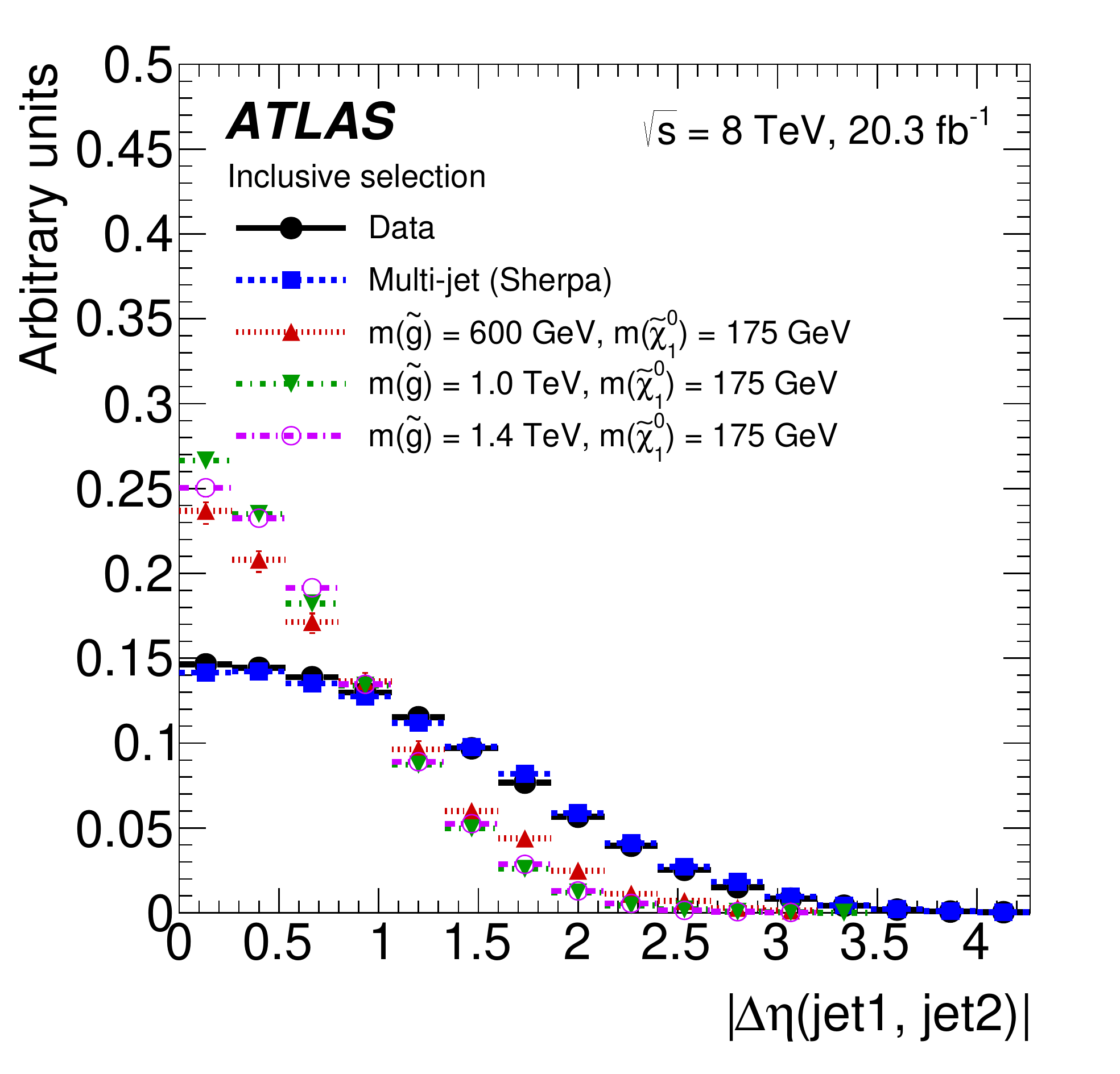}
    \label{fig:substructure:SvsB:DEta}}
    
  \caption{Comparison between signal and background for \subref{fig:substructure:SvsB:MJ} the scalar sum of the masses of the four leading \largeR jets \MJ and \subref{fig:substructure:SvsB:DEta} the difference in pseudorapidity between the two leading \largeR jets \Deta. Several typical signal points are shown, as well as the distributions obtained from the data. All distributions are normalized to the same area. The selection requires four or more jets, similar to the 4j regions but inclusive in \Deta.
           }
           
  \label{fig:substructure:SvsB}
\end{figure}

\subsection{Method and techniques}
\label{sec:merged:backgrounds}

The total-jet-mass analysis uses a topological observable \MJ as the primary distinguishing characteristic between signal and background. The observable \MJ~\cite{Hook2012, Hedri:2013pvl, Cohen:2014epa} is defined as the scalar sum of the masses of the four leading \largeR\ jets reconstructed with a radius parameter $R=1.0$, $\pT>100$~GeV and $|\eta|<\fatjetetacut$,

\begin{equation}
\MJ = \sum^4_{{\pT}>100\,\textrm{GeV} \atop |\eta|\leq\fatjetetacut}\mjet .
\label{eq:defMJ}
\end{equation}

\noindent This observable was used for the first time in the \sqseight search by the \ATLAS\ Collaboration for events with many jets and missing transverse momentum~\cite{Aad:2013wta} and provides significant sensitivity for very high-mass gluinos. Four-jet (or more) events are used as four \largeR jets cover a significant portion of the central region of the calorimeter, and are very likely to capture most signal quarks within their area. This analysis focuses primarily on the ten-quark models mentioned in \secref{introduction}.

Simulation studies show that \MJ provides greater sensitivity than variables such as \HT, the scalar sum of jet \pt: the masses contain angular information about the events by definition, whereas a variable like \HT\ simply describes the energy (or transverse momentum) in the event. A large \MJ implies not only high energy, but also rich angular structure. Previous studies at the Monte Carlo event generator level have demonstrated the power of the \MJ variable in the high-multiplicity events that this analysis targets~\cite{Hook2012,Hedri:2013pvl}.

\Figref{substructure:SvsB} presents examples of the discrimination that the \MJ observable provides between the background (represented here by \Sherpa multi-jet MC simulation) and several signal samples, as well as the comparison of the data to the \Sherpa multi-jet background. Three signal samples, each with $\mninoone=175$~GeV and several gluino masses $\mgluino$ in the range $0.6-1.4$~TeV are shown. In each case, the discrimination in the very high \MJ region is similar and is dictated primarily by the gluino mass, but is also sensitive to the mass splitting, $\mgluino-\mninoone$. Larger \mgluino results in larger \MJ, as expected. However, for the same \mgluino, \MJ is largest for $\mninoone\approx\mgluino/2$. This is due to the partitioning of the energy in the final state. For very large \mninoone, with $\mninoone\lesssim\mgluino$, the two quarks from the decay of the \gluino\ are very soft and the partons from the decay of the \ninoone\ are relatively isotropic, slightly reducing the efficacy of the approach. For very low \mninoone, $\mninoone \ll \mgluino$, the opposite occurs: the two quarks from the gluino decay have very high \pT and the neutralino is Lorentz-boosted, often to the point that the decay products merge completely, no longer overlapping with quarks from other parts of the event, and the mass of the jet is substantially reduced.\footnote{While the complete merging of the decay products of a \ninoone into a single jet may suggest that the most effective variable at low \mninoone might be the jet mass itself, typically only the lightest \ninoone have enough \pt\ to be strongly collimated. Such jets thereby have very low jet masses. These low jet masses similar to what is expected from QCD radiation, making discrimination very difficult, and so the nominal total-jet-mass technique is maintained even in these regions.} In both cases, although the sensitivity of \MJ is reduced, the overall approach still maintains good sensitivity.

Another discriminating variable that is independent of \MJ is necessary in order to define suitable control regions for the analysis. As in the jet-counting analysis, the signal is characterized by a considerably higher rate of central jet events as compared to the primary multi-jet background. This is expected due to the difference in the production processes that is predominantly $s$-channel for the signal, while the background can also be produced through $u$- and $t$-channel processes. \Figref{substructure:SvsB} additionally shows the distribution of the pseudorapidity difference between the two leading \largeR\ jets, \Deta. The discrimination between the signal samples and the background is not nearly as significant for \Deta as for \MJ. However, the lack of significant correlation (Pearson linear correlation coefficient of approximately 1\%) between the two observables makes \Deta effective as a means to define additional control regions in the analysis. It is also observed that the shape of the distribution is relatively independent of the \gluino\ and \ninoone\ masses and mass splittings.

The ability of several other observables to discriminate between signal and background was also tested. In particular, the possibility of using more detailed information about the substructure of jets (e.g. the subjet multiplicity or observables such as $N$-subjettiness, \tauThrTwo~\cite{Thaler:2010tr, Thaler:2011gf}) was investigated. Although some additional discrimination is possible using more observables, these significantly complicated the background estimation techniques and only marginally increase the sensitivity of the analysis.

The use of \MJ in this analysis provides significant sensitivity as well as the opportunity to complement the jet-counting analysis described in \secref{Resolved} with a fully data-driven background estimation that does not require any input from MC simulation. A \textit{template method} is adopted in which an expected \MJ distribution is constructed using individual jet mass templates. Single-jet mass templates are extracted jet-by-jet from a signal-depleted 3-jet control region (3jCR), or \textit{training sample}. These jet mass templates are binned in jet \pT and $\eta$, which effectively provides a \textit{probability density function} that describes the relative probability for a jet with a given \pT\ and $\eta$ to have a certain mass. This template is randomly sampled 2500 times for a single jet \pT\ and $\eta$, and a precise predicted distribution of possible masses for the given jet is formed.\footnote{2500 times was found to be the best balance between the precision of the result and computational time.} For an event with multiple jets, the jet mass templates are applied to each jet and the resulting predicted mass distributions are combined to predict the total-jet-mass \MJ for that ensemble of jets.

Jet mass templates are applied to jets in events in orthogonal regions, typically with at least four \largeR\ jets -- the control (4jCR), validation (4jVR), and signal regions (4jSR) -- but also in the 3jCR to test the method. Samples used in this way are referred to as the \textit{kinematic samples}. The only information used is the jet \pT\ and $\eta$, which are provided as inputs to the templates. The result is referred to as a \textit{dressed sample}, which provides an SM prediction of the individual jet mass distributions for the jets in the kinematic sample. An SM prediction for the total-jet-mass can then be formed by combining the individual dressed jet mass distributions. The normalization of the \MJ prediction -- the dressed sample -- is preserved such that the total expected yield is equal to the number of events in the kinematic sample. The procedure can be summarized as~\cite{Cohen:2014epa}:
%
\begin{enumerate}
  \item Define a control region to obtain the training sample from which jet mass templates are to be constructed;
  \item Derive a jet mass template binned in jet $\eta$ and $\pt$ using a smoothed Gaussian kernel technique;
  \item Define a \textit{kinematic} sample as either another control region or the signal region;
  \item Convolve the jet mass template with the kinematic sample using only the jet \pt\ and $\eta$; 
  \item Obtain a sample of \textit{dressed} events which provides the data-driven background estimate of \MJ.
\end{enumerate}

The key assumption in this approach is that the jet kinematics factorize and are independent of the other jets in the event. Deviations from this approximation may occur due to effects that are not included in the derivation of the jet mass templates. In particular, the composition of quarks and gluons can vary across different samples~\cite{Gallicchio:2011xc}, and quark and gluon jets have been observed to have different radial energy distributions~\cite{Aad:2014gea}. Other experimental affects, arising from close-by or overlapping jets, can also have an effect. For this reason, extensive tests are performed in the 4jCR and 4jVR, as defined in \secref{merged:SRCR}, to estimate the size of the correction factors needed to account for any sample dependence, and to assess systematic uncertainties. The entire procedure is tested first in \Sherpa multi-jet MC simulation, which shows minimal differences between the template prediction and observed mass spectrum.

  \begin{table}[!htb]
    \begin{center}
	  \begin{tabular}{r|c|c|c|c|c}
	    \hline \hline
Region  &   \Njet       & $\Deta$    & \ptthr & \ptfour & \MJ \\ 
Name  &         &     & [GeV] & [GeV] & [GeV]\\ \hline   
 		 3jCR  & $\Njet=3$     & --         &              & --            & -- \\ \hline
		 \multirow{2}{*}{4jCR}  & \multirow{2}{*}{$\Njet\geq4$}  & \multirow{2}{*}{$>1.40$}    & $>100$       & \multirow{2}{*}{$>100$}        & -- \\
		 	   &               &            & $>250$       &               & -- \\ \hline
		 \multirow{2}{*}{4jVR}  & \multirow{2}{*}{$\Njet\geq4$}  & \multirow{2}{*}{1.0--1.40} & $>100$       & \multirow{2}{*}{$>100$}        & -- \\
		 	   &               &            & $>250$       &               & -- \\ \hline
		 SR1   &               &            & $>250$       &               & $>625$ \\
		 SR100 & $\Njet\geq4$  & $<0.7$     & $>100$       & $>100$        & $>350$ (binned) \\
		 SR250 &               &            & $>250$       &               & $>350$ (binned) \\
		\hline \hline
	  \end{tabular}
    \end{center}
  \caption{Control (CR), validation (VR), and signal regions (SR) used for the analysis. \ptthr and \ptfour represent the transverse momentum of the third and fourth jet in \pT, respectively.}
  \label{tab:signal-control:CRselections}
  \end{table}%

\begin{figure}[htb]
  \centering
  \subfigure[\MJ in 4jVR using $\ptthr>100$~GeV]{
    \includegraphics[width=0.4\textwidth]{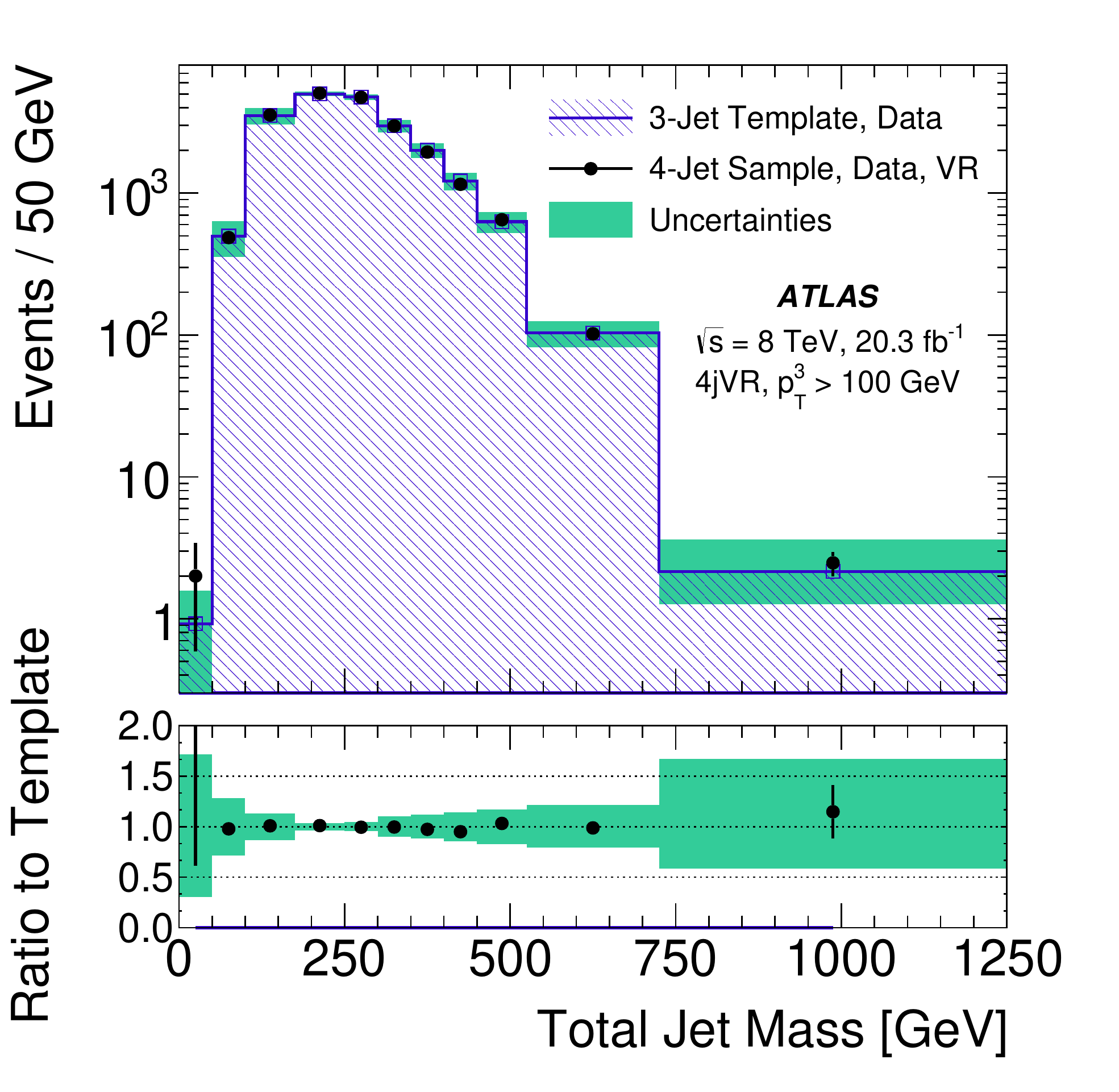}
    \label{fig:template:performance:4jvrTotal}
  }
  \subfigure[\MJ in 4jVR using $\ptthr>250$~GeV]{
    \includegraphics[width=0.4\textwidth]{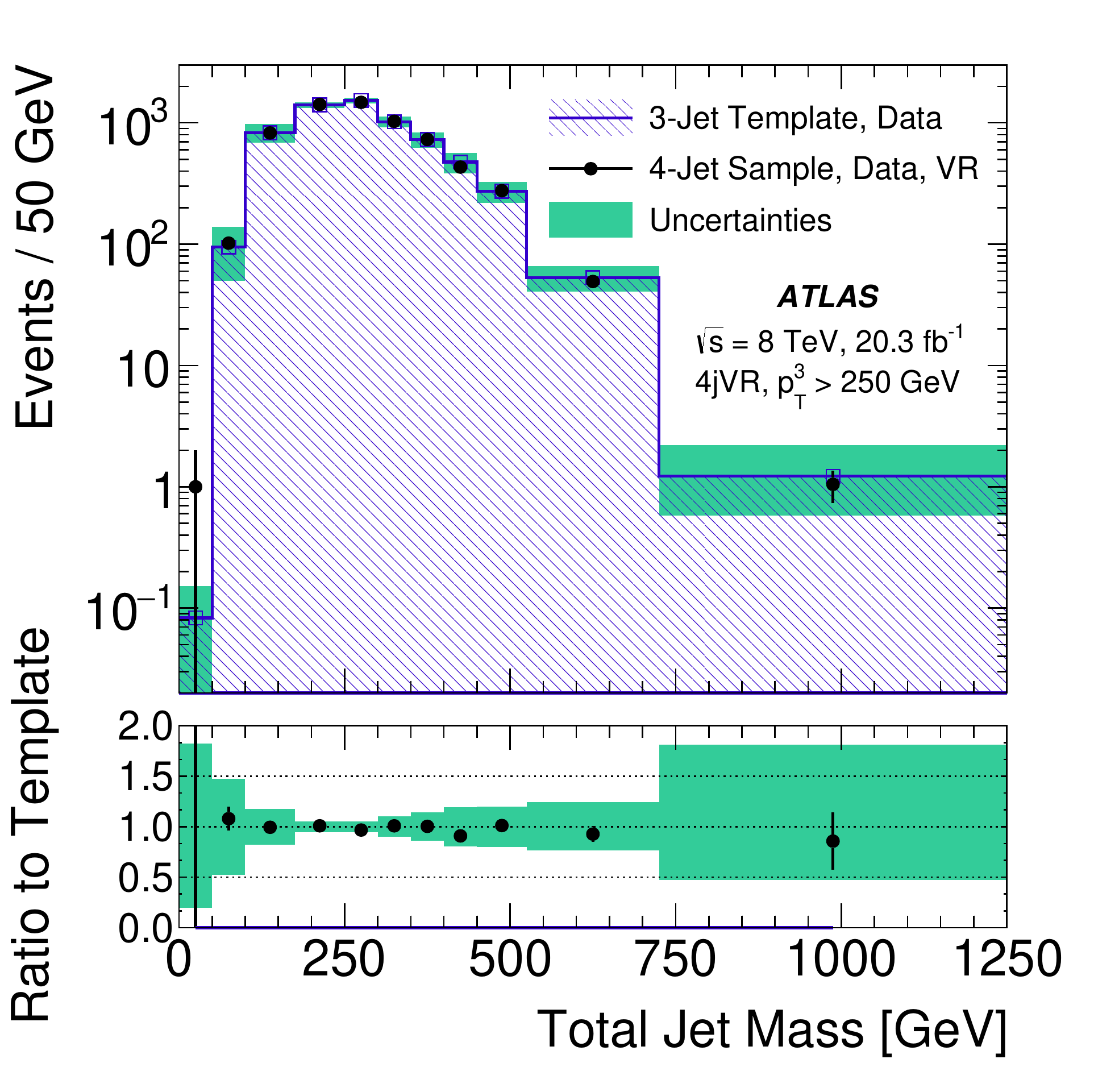}
    \label{fig:template:performance:4j250vrTotal}
  }
  \caption{\subref{fig:template:performance:4jvrTotal} Total-jet-mass in the 4jVR with $\ptthr>100$~GeV. The reweighted template is shown in the hatched blue histogram. \subref{fig:template:performance:4j250vrTotal} Total-jet-mass in the 4jVR with $\ptthr>250$~GeV. The 4jVR \MJ spectra are shown in the open black squares.  The total systematic uncertainty due to the smoothing procedure, finite statistics in control regions, and the difference between template prediction and the data observed in the 4jCR is shown in green.}
  \label{fig:template:performance:4jTotal}
\end{figure}

%
\begin{figure}[htb]
  \centering
  \subfigure[\MJ in SR100 using $\ptthr>100$~GeV]{
    \includegraphics[width=0.4\textwidth]{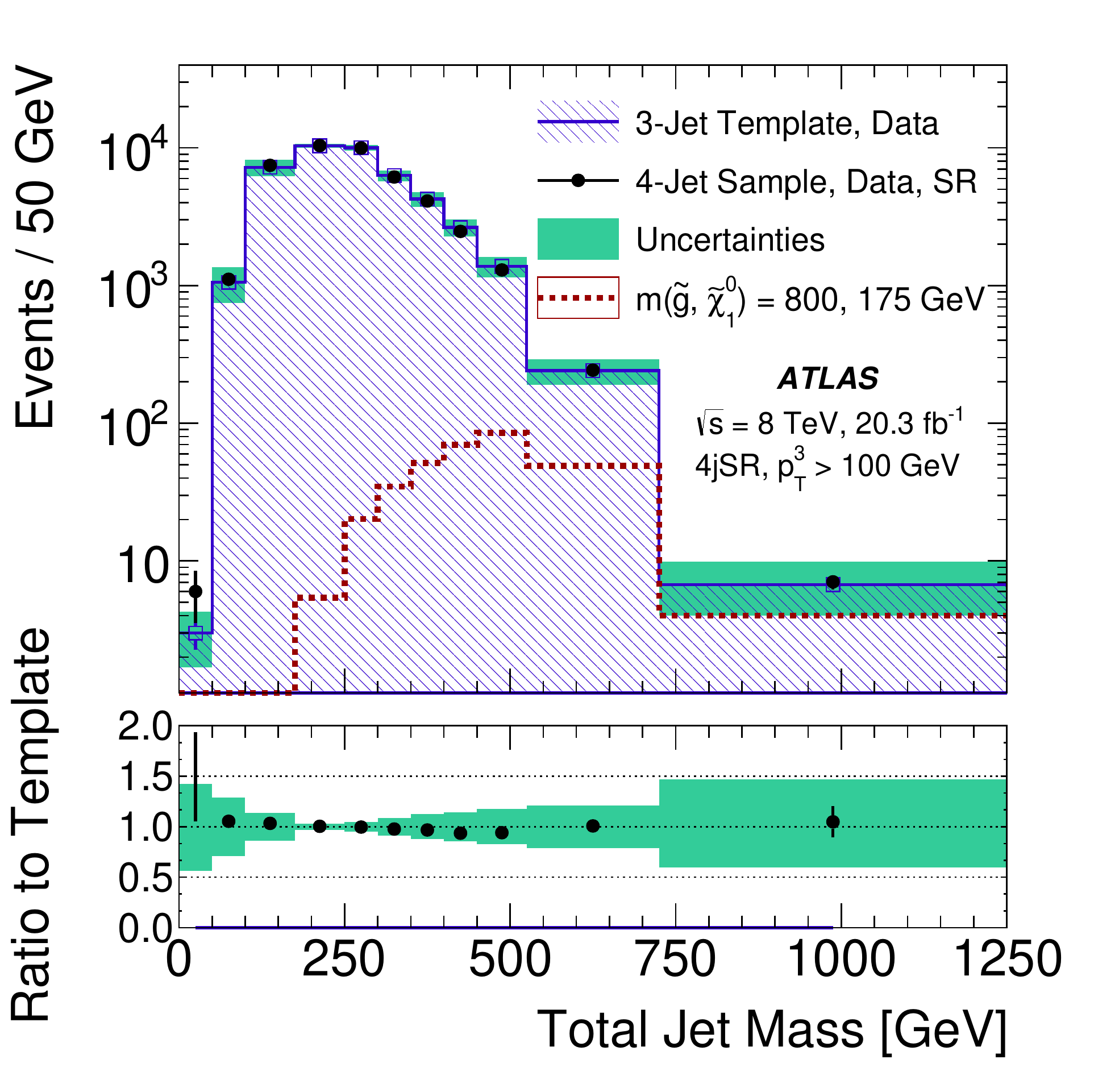}
    \label{fig:template:performance:4jsrTotal:SR100}
  }
  \subfigure[\MJ in SR250 using $\ptthr>250$~GeV]{
    \includegraphics[width=0.4\textwidth]{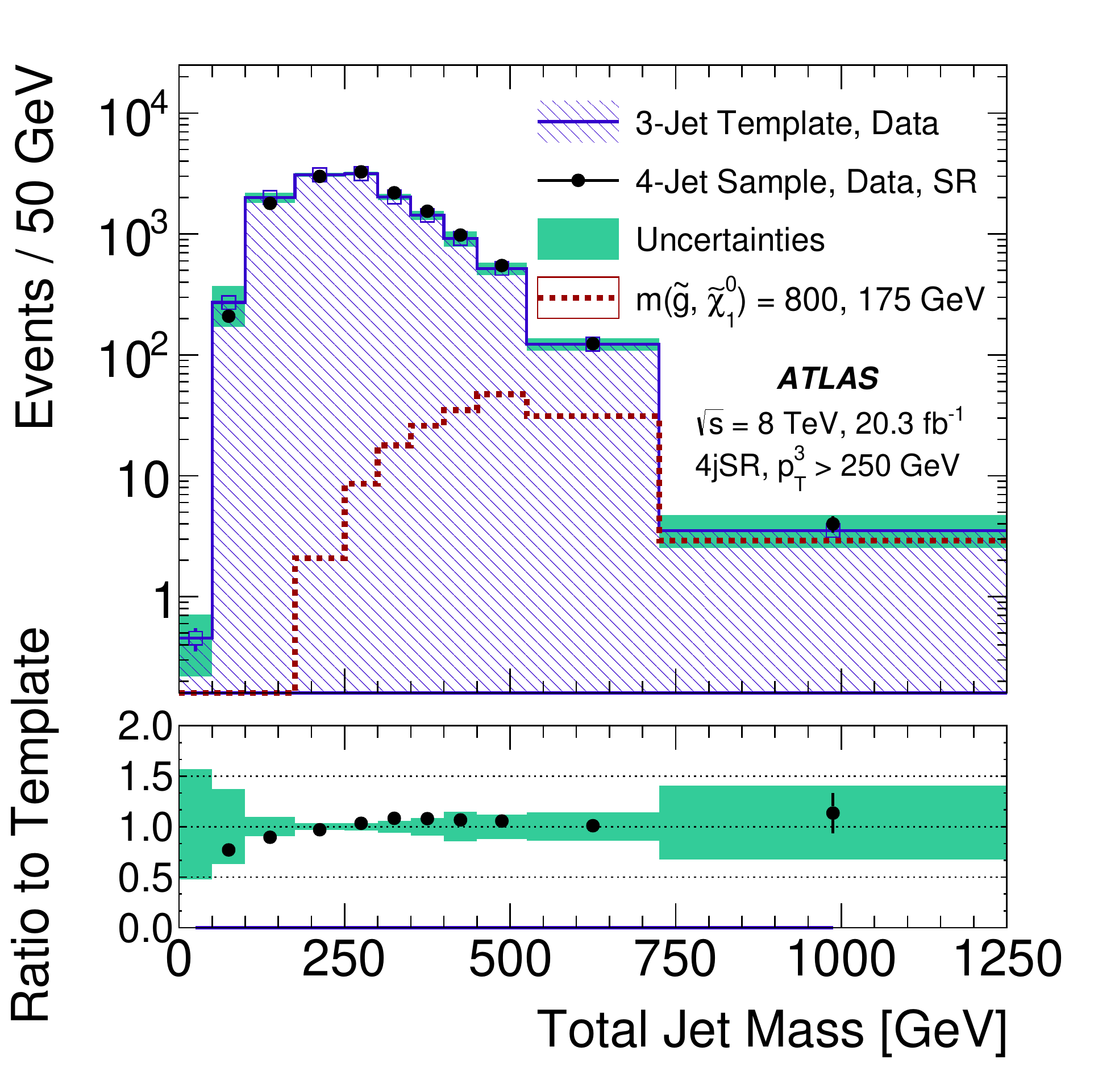}
    \label{fig:template:performance:4jsrTotal:SR250}
  }
  \caption{Total-jet-mass in the 4jSR \subref{fig:template:performance:4jsrTotal:SR100} using $\ptthr>100$~GeV (SR100) and \subref{fig:template:performance:4jsrTotal:SR250} using $\ptthr>250$~GeV (SR250). For the SR100 selection, the reweighted template (built in the 3jCR, and reweighted jet-by-jet in the 4jCR) is shown in the hatched blue histogram.   The total systematic uncertainty due to the smoothing procedure, finite statistics in control regions, and the difference between template prediction and the data observed in the 4jCR is shown in green.}
  \label{fig:template:performance:4jsrTotal}
\end{figure}

\subsection{Signal and control region definitions}
\label{sec:merged:SRCR}

The \MJ and \Deta observables form the basis for the signal region definition for the analysis, where \Deta is used to define control regions for testing the background estimation in data. A requirement of $\Deta < 0.7$ is found to have the best signal sensitivity over the entire plane of (\mgluino, \mninoone). In this optimization, the background contribution is modeled by multi-jet events simulated with \Sherpa.

An optimization study indicated that when using a single \MJ selection, $\MJ > 625$~GeV provides the best sensitivity to many signal hypotheses, and gives the best expected sensitivity at high \mgluino. A single-bin signal region (SR1) is therefore defined with $\MJ>625$~GeV and a 250 GeV \pt\ threshold applied to the third leading in \pT \largeR\ jet. This region has an acceptance of $0.26\%$ for the $\mgluino=600$~GeV, $\mninoone=50$~GeV signal point. This acceptance grows rapidly with gluino mass to $11\%$ for the point $\mgluino=1000$~GeV, $\mninoone=600$~GeV, and is only weakly dependent on the neutralino mass.

A second set of signal regions is used to further improve the power of the analysis by making use of the shape of the \MJ distribution. Two selections on the third leading jet in \pt (\ptthr) are used, $\ptthr>100$~GeV (SR100) and $\ptthr>250$~GeV (SR250). This provides better sensitivity to the full range of gluino masses considered, compared to SR1. The lower \pt region, SR100, has better sensitivity for lower gluino masses, whereas SR250 has improved sensitivity for higher masses. All other selections are unchanged. In this case, a lower threshold of $\MJ > 350$~GeV is used and the observed data are compared to the template predictions in bins of \MJ. The improvements in the sensitivity obtained by adding these additional signal regions and using the shape of the \MJ spectrum are described below. The full set of selection criteria is listed in \tabref{signal-control:CRselections}.

The jet multiplicity and \Deta are used to define the control regions. The 3jCR, with exactly three jets, is used to train the background templates previously discussed. In the remaining control and validation regions, each requiring $\geq$ 4 jets, the $\Deta$ selection suppresses the signal contribution and is used to define the 4jCR and 4jVR. In the $\geq$ 4-jet regions, the \Deta selection value for the control regions is chosen to be larger than an inversion of the signal region selection, resulting in the selections presented in \tabref{signal-control:CRselections}. These control region definitions permit studies of the full \MJ spectrum as well as comparisons of data and SM predictions without significant signal contamination.

\subsection{Validation and systematic uncertainties}
\label{sec:merged:systematics}

Many tests are performed using the 3jCR as both the training sample and the kinematic sample in order to determine the robustness of the method. The selection requires that there be exactly three \largeR\ jets in the event, as described in \tabref{signal-control:CRselections}. The dependence of the template on the jet in question (leading, subleading, etc) is tested, as well as the dependence of the template on the jet kinematics. It is determined that it is optimal to define separate templates for each of the three jet categories (leading, subleading, and third jet) and to bin the templates according to the jet \pt and $\eta$.\footnote{It is observed that the difference between the leading and subleading jet templates is minimal, but that the third jet exhibits qualitatively different masses as function of the jet \pT.} In the 4-jet regions, the fourth jet uses the template derived for the third jet in the 3jCR: tests in the 4jCR and 4jVR indicate very good agreement between this template and the observed spectrum. As a first test, the \MJ template constructed from the 3-jet kinematic sample is compared to the actual \MJ distribution in 3-jet events, and very good agreement is observed. 
 
There are two intrinsic sources of systematic uncertainty associated with the template procedure: the uncertainty due to finite statistics in the 3jCR training sample (the variance), and the uncertainty due to the smoothing procedure in the template derivation (the bias). The former is estimated by generating an ensemble of \MJ templates and taking the $\pm 1\sigma$ deviations (defined as the $\pm 34\%$ quantile) with respect to the median of those variations as the uncertainty, bin by bin. The systematic uncertainty due to the smoothing procedure is determined using the fact that a Gaussian kernel smoothing is applied to the template. The full difference between the nominal template and a template constructed using a leading-order correction for the bias, derived analytically in Ref.~\cite{Cohen:2014epa}, is taken as the systematic uncertainty. The systematic uncertainty due to finite control region statistics is chosen to be larger (by setting the size of the kernel smoothing) than that due to the smoothing procedure since the former is more accurately estimated.

A small level of disagreement (between 5 to 15\%) is observed when comparing the observed mass to the predicted mass in the 4jCR: a reweighting derived in the 4jCR (as a function of each individual jet mass) is then applied to the individual jet masses prior to the construction of the \MJ for each event. After the reweighting the agreement is substantially improved at high total-jet-mass. \Figref{template:performance:4jTotal} presents the total-jet-mass \MJ in the 4jVR using $\ptthr>100$~GeV. The reweighted template agrees very well with the observed \MJ distribution in the 4jVR--- a sample completely independent from where the reweighting was derived--- validating both the template method and the reweighting.
The full magnitude of the reweighting on the total-jet-mass distribution is taken as a systematic uncertainty of the method. The total systematic on the background prediction therefore includes both the intrinsic systematic uncertainty given by the variance and the bias, as well as the difference observed in the 4jCR.
The \MJ distribution is also shown for the 4jVR for the case in which $\ptthr>250$~GeV. No reweighting is required when using the significantly higher \ptthr\ selection since the observed effects due to topological differences in the training sample compared to the kinematic sample are suppressed. In order to account for any remaining disagreement, the difference between the data and template prediction in the 4jCR is applied as a further systematic. The total uncertainty therefore includes again both the instrinsic background estimation uncertainties and the disagreement observed in the 4jCR.

One possible concern for the template technique is that it assumes that the same mechanism is responsible for generating the individual jet masses in both the control and signal regions. In order to test the extent to which a different composition of processes may affect the derived templates, the assumption that multi-jet events are the only background in the 3jCR and 4j regions is modified by injecting separately a sample of \Sherpa \ttbar MC simulation events (assuming SM cross-sections) into the full procedure. The resulting background estimates are fully consistent with the prediction without the injection -- indicating that the technique is not sensitive to contamination from top quark production -- and thus no additional systematic uncertainty is assessed for the potential presence of specific background processes. A similar procedure is performed for signal processes (assuming standard \gluino production cross-sections) and again no impact of signal contamination on the constructed background templates is observed.

\Figref{template:performance:4jsrTotal} shows the total-jet-mass in the 4jSR compared to the template prediction. For both SR100 and SR250, the total systematic error on the template method is also shown in the ratio plot in the lower panel of each distribution. The template predictions are clearly consistent with the observed data. Thus there is no indication of new physics in these results.

Systematic uncertainties associated with the scale and resolution of \largeR\ jet mass and energy~\cite{Aad:2013gja} are significantly reduced by the use of a data-driven background estimate: residual effects may remain due to differences between the 3jCR and the 4j regions, and these are reflected in the systematic uncertainties assessed by the difference between the template prediction and observed spectrum in the 4jCR. The uncertainties due to the background estimation method are dominated by propagation of the statistical uncertainty from the 3jCR: these are typically 5--10\%, except in the highest \MJ bins of SR100 and SR250, where they can extend to 20-40\%. In addition, the observed difference systematic uncertainty from the 4jCR varies from 5\% to 15\%. Signal reconstruction -- both in terms of selection efficiency and the \MJ spectrum predicted for a given $\mgluino,\mninoone$ combination -- is sensitive to the kinematic uncertainties associated with the final state jets in the analysis. The impacts of these systematic uncertainties are directly assessed by varying the kinematics within the uncertainties and reported in \secref{results}. Jet mass scale uncertainties have the largest effect, which for SR1 range from 30\% for very low \mgluino to 15\% for very high \mgluino. In the cases of SR100 and SR250, the impact of the jet mass scale uncertainty also dominates, and varies across the \MJ spectrum from 10--20\% at lower \MJ up to 50\% for the very highest \MJ bin in the spectrum for low \mgluino. The luminosity uncertainty of $3\%$ also affects the signal only.

\section{Results and interpretations}
\label{sec:results}

As no significant excess is observed in data in either analysis, a procedure to set limits on the models of interest is performed. A profile likelihood ratio combining Poisson probabilities for signal and background is computed to determine the confidence level (CL) for consistency of the data with the signal-plus-background hypothesis (\CLsb). A similar calculation is performed for the background-only hypothesis (\CLb). From the ratio of these two quantities, the confidence level for the presence of signal (\CLs) is determined~\cite{HistFitter}. Systematic uncertainties are treated via nuisance parameters assuming Gaussian distributions. In all cases, the nominal signal cross-section and uncertainty are taken from an envelope of cross-section predictions using different PDF sets and factorization and renormalization scales, as described in Ref.~\cite{Kramer:2012bx}. As discussed in \secref{data-mc}, the region with $(\mgluino-\mninoone)<100$~GeV is not considered in this analysis in order to ensure that the results are insensitive to the effects of ISR, since the uncertainties cannot be assessed for the UDD decays considered here. 

The total-jet-mass analysis is designed to be agnostic to the flavor composition of the signal process and to remove any reliance on MC simulations of these complex hadronic final states. The jet-counting analysis provides the opportunity to enhance sensitivity to specific heavy-flavor compositions in the final state and to explore various assumptions on the branching ratios of the benchmark signal processes studied in this paper. The results obtained from the total-jet-mass analysis in the inclusive final state are presented first, and then the specific sensitivity provided by the jet-counting analysis to the full branching ratio space is presented.

\begin{figure}[!htbp]
  \centering  
    \includegraphics[width=0.99\columnwidth]{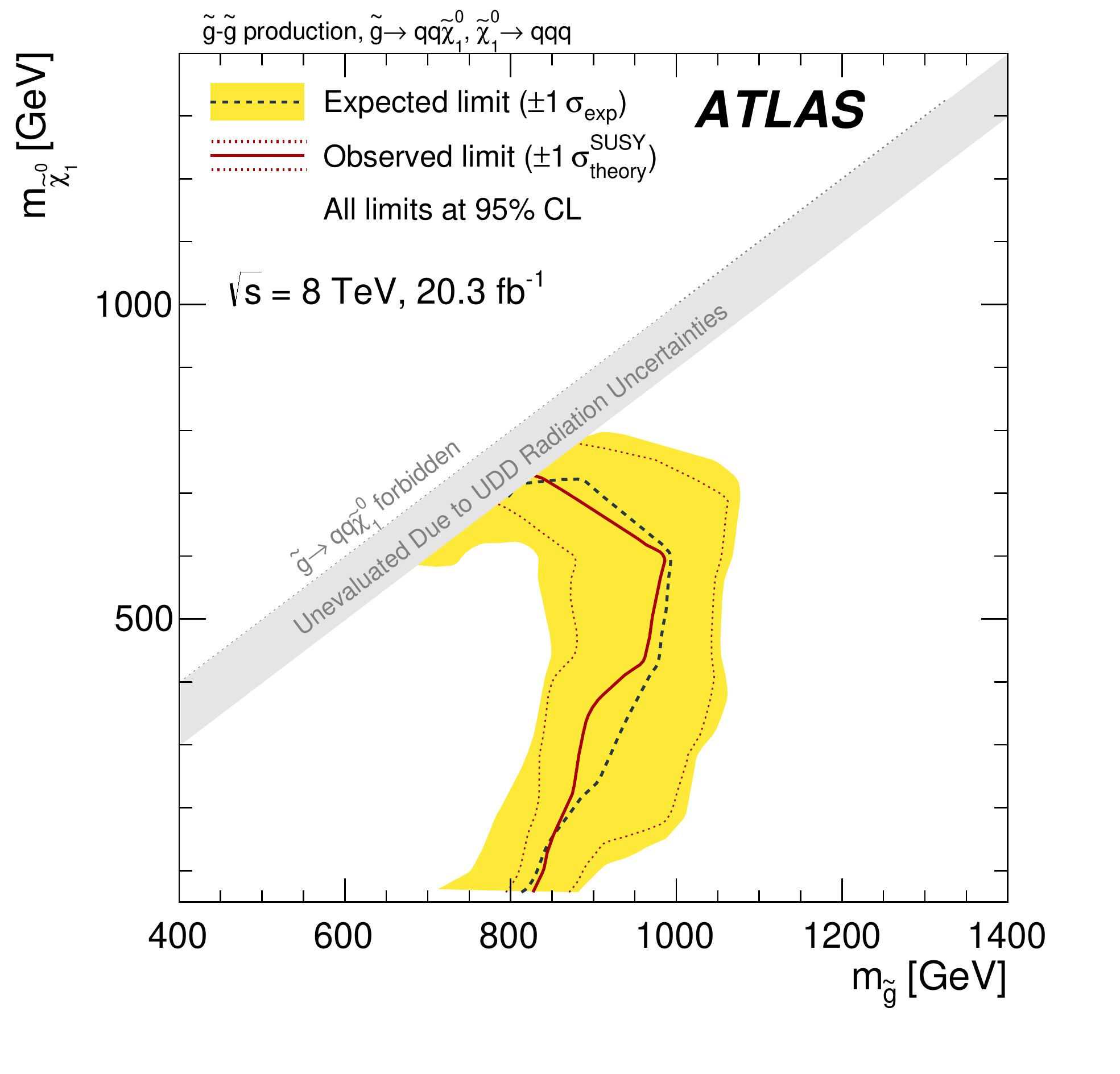}
  \caption{Expected and observed exclusion limits in the (\mgluino, \mninoone) plane for the ten-quark model given by the total-jet-mass analysis. Limits are obtained by using the signal region with the best expected sensitivity at each point. The dashed black lines show the expected limits at 95\% CL, with the light (yellow) bands indicating the 1$\sigma$ excursions due to experimental and background-only theory uncertainties. Observed limits are indicated by medium dark (maroon) curves, where the solid contour represents the nominal limit, and the dotted lines are obtained by varying the signal cross-section by the renormalization and factorization scale and PDF uncertainties.}
  \label{fig:results:merged}
\end{figure}

\begin{table*}[htb]
\begin{center}
\footnotesize
\begin{tabular}{r|c|c|c|c|c} 
\multicolumn{6}{c}{Summary yield table for SR1}\\
\hline \hline

\multirow{2}{*}{\textbf{\MJ Bin}} & \multirow{2}{*}{\textbf{Expected SM}} & \multirow{2}{*}{\textbf{Obs.}} & $\mgluino=600$~GeV & $\mgluino=1$~TeV & $\mgluino=1.4$~TeV\\
 & & & $\mninoone = 50$~GeV & $\mninoone = 600$~GeV & $\mninoone = 900$~GeV \\ 
\hline
    $>$ 625 GeV & 160$\pm$9.7 $^{+40}_{-34} $ & 176 & 70$\pm$4.2 $\pm$25$\pm$30 (0.26\%) & 55$\pm$0.51 $\pm$8.6 $\pm$14 (11\%) & 6.3$\pm$0.07 $\pm$0.46$\pm$2.5 (35\%)\\
\hline \hline

\end{tabular} 

\caption{Table showing the predicted in the SM and observed number of events in SR1 as well as three representative signal scenarios. Acceptances (including efficiency) of the various signals are listed in parentheses. The background uncertainties are displayed as statistical + systematic; the signal uncertainties are displayed as statistical + systematic + theoretical. 
 \label{tab:results:yields:sr1}}
\end{center}
\end{table*}

\begin{table*}[htb]
\begin{center}
\footnotesize
\begin{tabular}{r|c|c|c|c|c} 
\multicolumn{6}{c}{Summary yield table for SR100}\\
\hline \hline

\multirow{2}{*}{\textbf{\MJ Bin}} & \multirow{2}{*}{\textbf{Expected SM}} & \multirow{2}{*}{\textbf{Obs.}} & $\mgluino=600$~GeV & $\mgluino=1$~TeV & $\mgluino=1.4$~TeV\\
 & & & $\mninoone = 50$~GeV & $\mninoone = 600$~GeV & $\mninoone = 900$~GeV \\ 
\hline
350 - 400 GeV & 4300$\pm$78 $^{+510}_{-500} $ & 5034 & 200$\pm$7.2$\pm$22$\pm$35 & 5.8$\pm$0.17$\pm$1.3$\pm$1.5 & 0.19$\pm$0.01$\pm$0.04$\pm$0.07\\

400 - 450 GeV & 2600$\pm$49 $^{+380}_{-380} $ & 2474 & 200$\pm$7.1$\pm$9.5$\pm$35 & 9.7$\pm$0.21$\pm$2.2$\pm$2.5 & 0.31$\pm$0.02$\pm$0.07$\pm$0.12\\

450 - 525 GeV & 2100$\pm$42 $^{+360}_{-360} $ & 1844 & 280$\pm$8.4$\pm$13$\pm$49 & 26$\pm$0.35$\pm$4.3$\pm$6.7 & 0.88$\pm$0.03$\pm$0.14$\pm$.34\\

525 - 725 GeV & 960$\pm$25 $^{+200}_{-200} $ & 1070 & 280$\pm$8.4$\pm$57$\pm$49 & 77$\pm$0.60$\pm$3.2 & 3.6$\pm$0.05$\pm$0.36$\pm$1.4\\

    $>$ 725 GeV & 71$\pm$7.0 $^{+32}_{-27} $ & 79 & 35.$\pm$2.9$\pm$18$\pm$6.0 & 35$\pm$0.40$\pm$9.9$\pm$9.0  & 4.8$\pm$0.06$\pm$0.61$\pm$1.9\\
    
\hline \hline
\end{tabular} 
\caption{Table showing the predicted in the SM and observed number of events in SR100 as well as three representative signal scenarios. The background uncertainties are displayed as statistical + systematic; the signal uncertainties are displayed as statistical + systematic + theoretical.  
\label{tab:results:yields:sr100}}
\end{center}
\end{table*}

\begin{table*}[htb]
\begin{center}
\footnotesize
\begin{tabular}{r|c|c|c|c|c} 
\multicolumn{6}{c}{Summary yield table for SR250}\\
\hline \hline

\multirow{2}{*}{\textbf{\MJ Bin}} & \multirow{2}{*}{\textbf{Expected SM}} & \multirow{2}{*}{\textbf{Obs.}} & $\mgluino=600$~GeV & $\mgluino=1$~TeV & $\mgluino=1.4$~TeV\\
 & & & $\mninoone = 50$~GeV & $\mninoone = 600$~GeV & $\mninoone = 900$~GeV \\ 
\hline
350 - 400 GeV & 1400$\pm$35 $^{+120}_{-134} $ & 1543 & 83$\pm$4.6 $\pm$15$\pm$14 & 3.3$\pm$0.12 $\pm$0.78$\pm$0.85 & 0.17$\pm$0.01 $\pm$0.03$\pm$0.07\\

400 - 450 GeV & 920$\pm$33 $^{+140}_{-140} $ & 980 & 92$\pm$4.8 $\pm$11$\pm$16 & 5.6$\pm$0.16 $\pm$1.5$\pm$1.5 & 0.27$\pm$0.01 $\pm$0.07$\pm$0.11\\

450 - 525 GeV & 780$\pm$33 $^{+94}_{-94} $ & 823 & 140$\pm$5.8 $\pm$15$\pm$23 & 17$\pm$0.28 $\pm$3.3$\pm$4.4 & 0.79$\pm$0.02 $\pm$0.13$\pm$0.31\\

525 - 725 GeV & 490$\pm$24 $^{+67}_{-67} $ & 495 & 160$\pm$6.2 $\pm$30.$\pm$27 & 56$\pm$0.51 $\pm$4.1$\pm$15 & 3.3$\pm$0.05 $\pm$0.34$\pm$1.3\\

    $>$ 725 GeV & 37$\pm$5.5 $^{+16}_{-12} $ & 42 & 22$\pm$2.3 $\pm$9.1$\pm$3.9 & 27$\pm$0.36 $\pm$7.4$\pm$7.0 & 4.4$\pm$0.06 $\pm$0.56$\pm$1.7\\
    
\hline \hline
\end{tabular} 
\caption{Table showing the predicted in the SM and observed number of events in SR250 as well as three representative signal scenarios. The background uncertainties are displayed as statistical + systematic; the signal uncertainties are displayed as statistical + systematic + theoretical.  
\label{tab:results:yields:sr250}}
\end{center}
\end{table*}

\subsection{Total-jet-mass analysis}
\label{sec:results:merged}

The observed and expected event yields are presented in \tabref{results:yields:sr1}, \ref{tab:results:yields:sr100}, and \ref{tab:results:yields:sr250} for the three signal regions SR1, SR100 and SR250 respectively. The single-bin signal region selection (SR1) is reported in addition to the binned \MJ results in SR100 and SR250 in order to provide yields that can be easily reinterpreted for other signal hypotheses. In the case of the binned \MJ signal regions, a binned fit (where the number and size of the bins were optimized) is performed that takes into account the predictions for each \MJ range. This approach provides greater sensitivity to small deviations from the template predictions. The correlation of the uncertainties in the bins of the \MJ spectrum are accounted for by evaluating the full correlation matrix. The result leads the analysis to treat the different bins as fully uncorrelated for the variance, which is the largest component of the background uncertainties. All other uncertainties treat the bins of the \MJ spectrum as fully correlated.

\Figref{results:merged} shows both the expected and observed 95\% CL limits in the (\mgluino, \mninoone) mass plane when the signal region that provides the best expected exclusion is used for each mass combination. The dashed black line shows the expected exclusion limits, and the yellow band represents the experimental uncertainties on this limit. The solid line shows the observed limit, with the finely dashed lines indicating the $\pm1 \sigma$ variations due to theoretical uncertainties on the signal production cross-section given by renormalization and factorization scale and PDF uncertainties. All mass limits are reported conservatively assuming the $-1\sigma^{\text{SUSY}}_\text{theory}$ signal production cross-section. At low \mninoone, the region with gluino mass $\mgluino\lesssim750$~GeV is excluded. Excluded $\mgluino$ masses rise with increasing $\mninoone$, up to a maximum exclusion of approximately $\mgluino\lesssim 870$~GeV at $\mninoone=600$~GeV. No models with $\mninoone > 650$~GeV are excluded.

\begin{figure}[!tbph]
\centering
\subfigure[][(BR($t$), BR($b$), BR($c$))=(0\%, 0\%, 0\%)] {
    \includegraphics[width=0.4\textwidth]{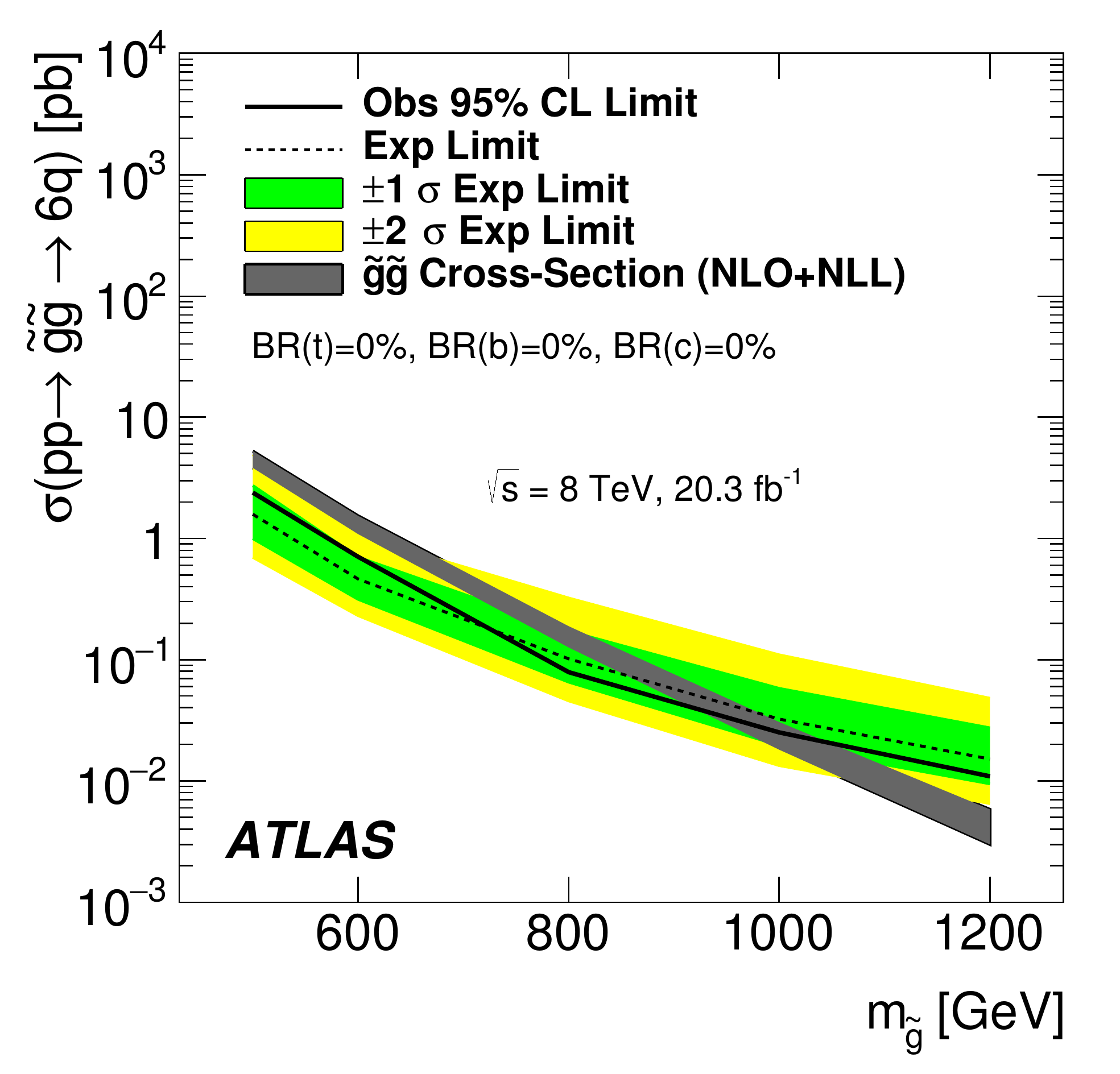}
    \label{fig:1D6qResults:light}
}\\
\subfigure[][(BR($t$), BR($b$), BR($c$))=(0\%, 100\%, 0\%)] {
    \includegraphics[width=0.4\textwidth]{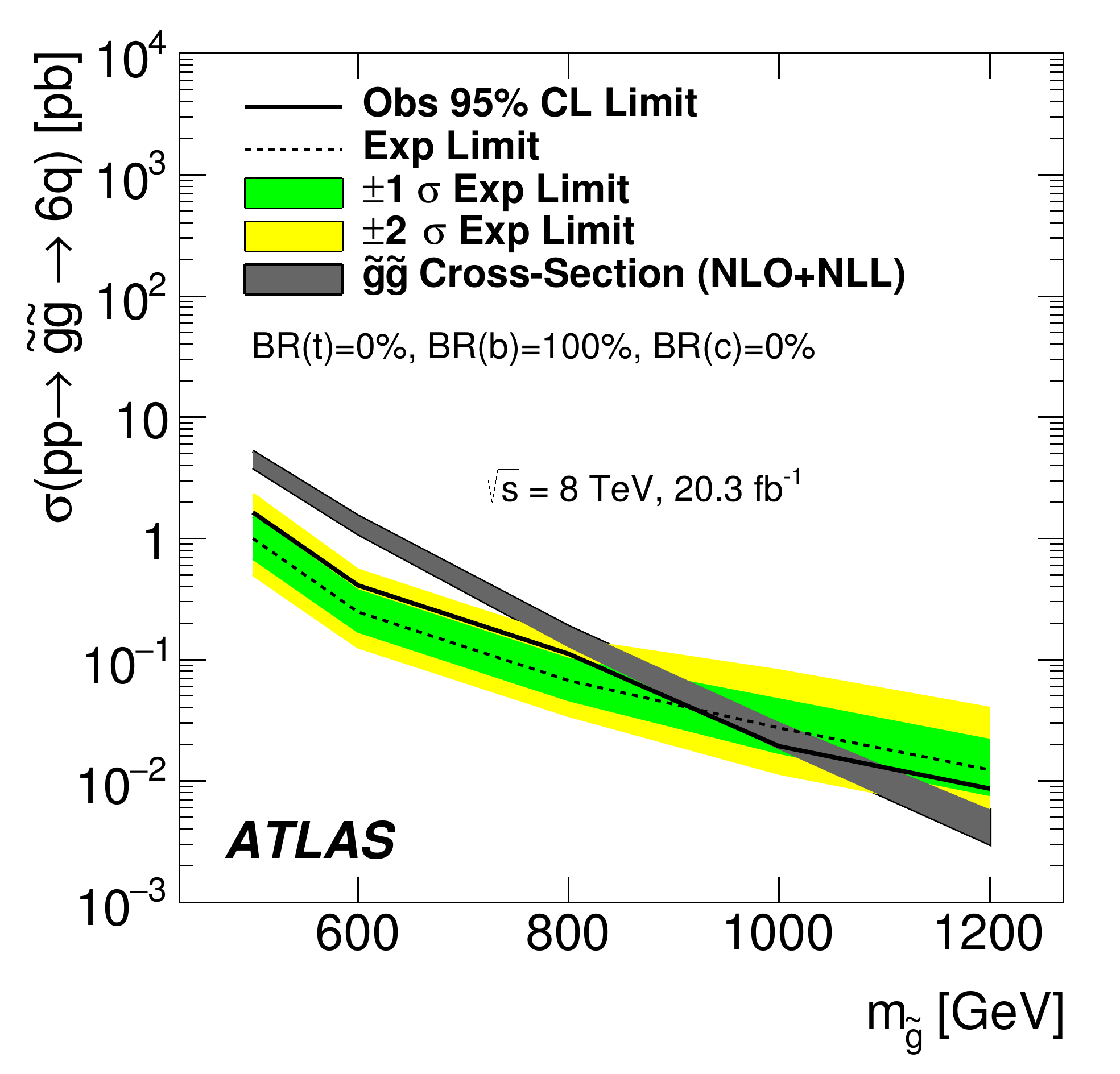}
    \label{fig:1D6qResults:b}
}
\caption{Expected and observed cross-section limits for the six-quark gluino models for \subref{fig:1D6qResults:light} the case where no gluinos decay into heavy-flavor quarks, and \subref{fig:1D6qResults:b} the case where every gluino decays into a $b$-quark in the final state. \label{fig:1D6qResults1}}
\end{figure}

\begin{figure}[!tbph]
\centering
\subfigure[][(BR($t$), BR($b$), BR($c$))=(100\%,0\%,0\%)] {
    \includegraphics[width=0.4\textwidth]{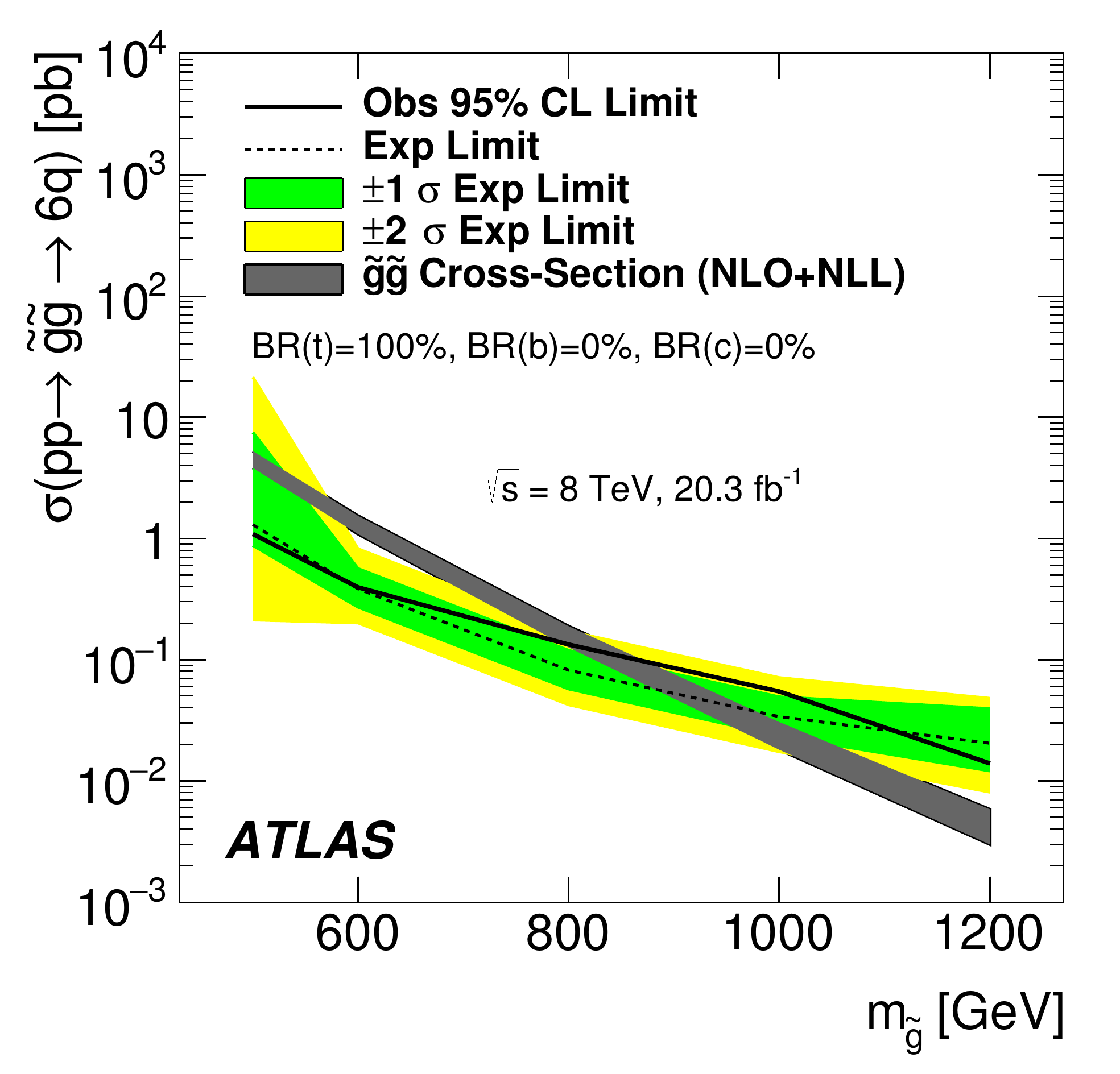}
    \label{fig:1D6qResults:top}
}\\
\subfigure[][(BR($t$), BR($b$), BR($c$))=(100\%,100\%,0\%)] {
    \includegraphics[width=0.4\textwidth]{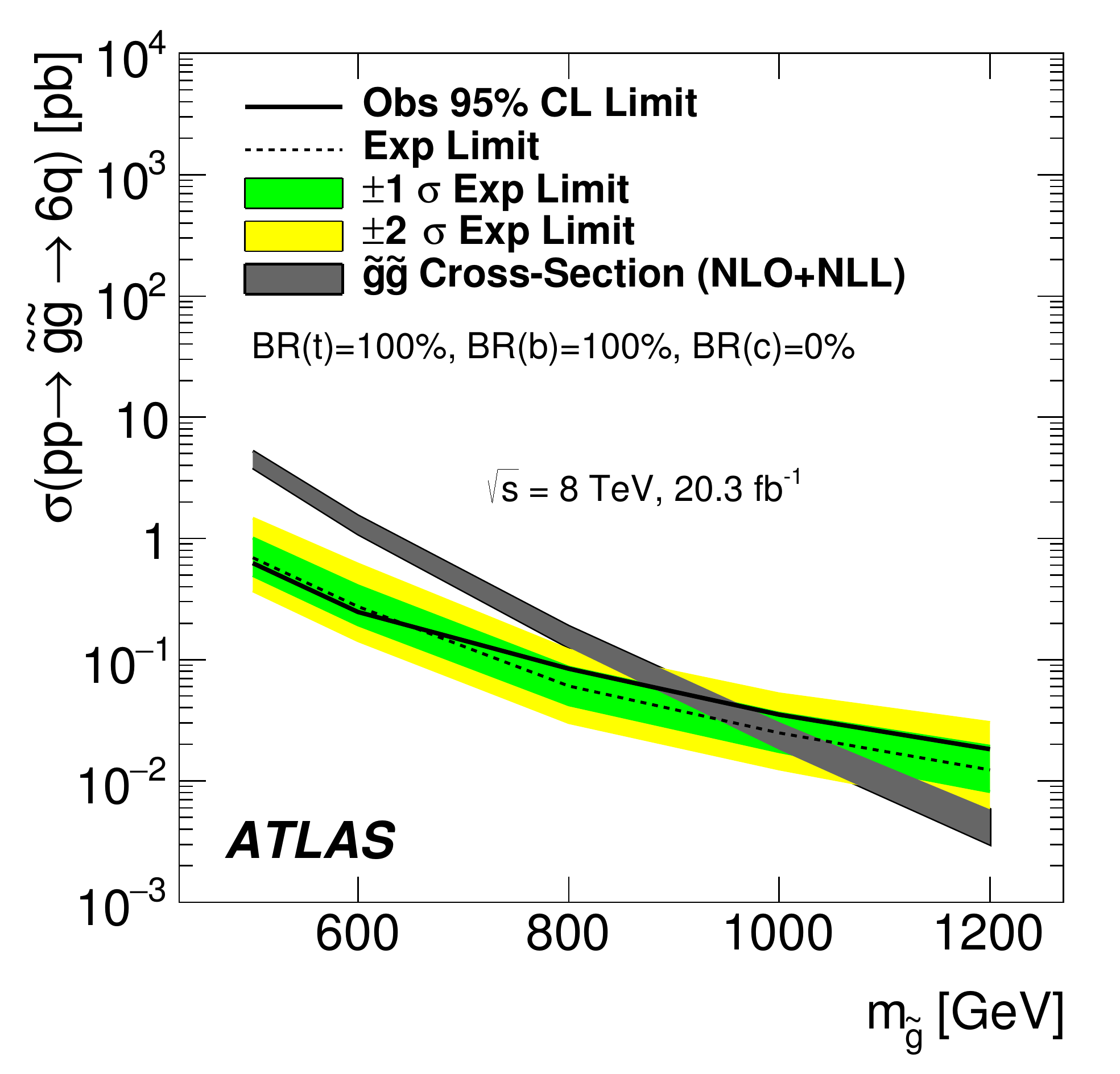}
    \label{fig:1D6qResults:btop}
}
\caption{Expected and observed cross-section limits for the six-quark gluino models for \subref{fig:1D6qResults:top} the case where each gluino is required to decay into a top quarks, and \subref{fig:1D6qResults:btop} the case where every gluino decays into a $b$-quark and a top quark. \label{fig:1D6qResults2}}
\end{figure}

\begin{figure}[!htb]
\centering
\subfigure[][BR($c$)=0\% - Expected] {
    \includegraphics[width=0.4\textwidth]{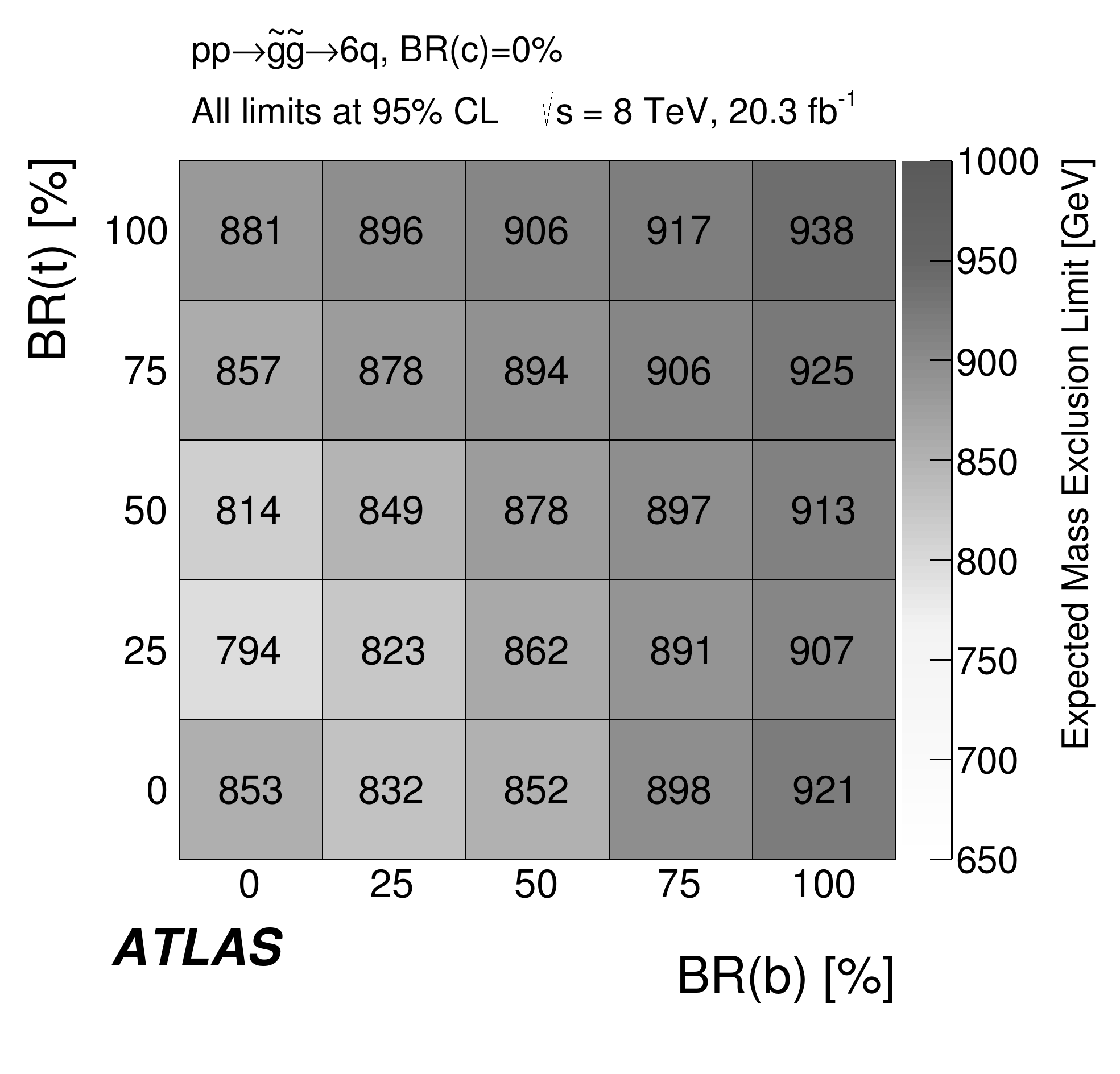}
}\\
\subfigure[][BR($c$)=0\% - Observed] {
    \includegraphics[width=0.4\textwidth]{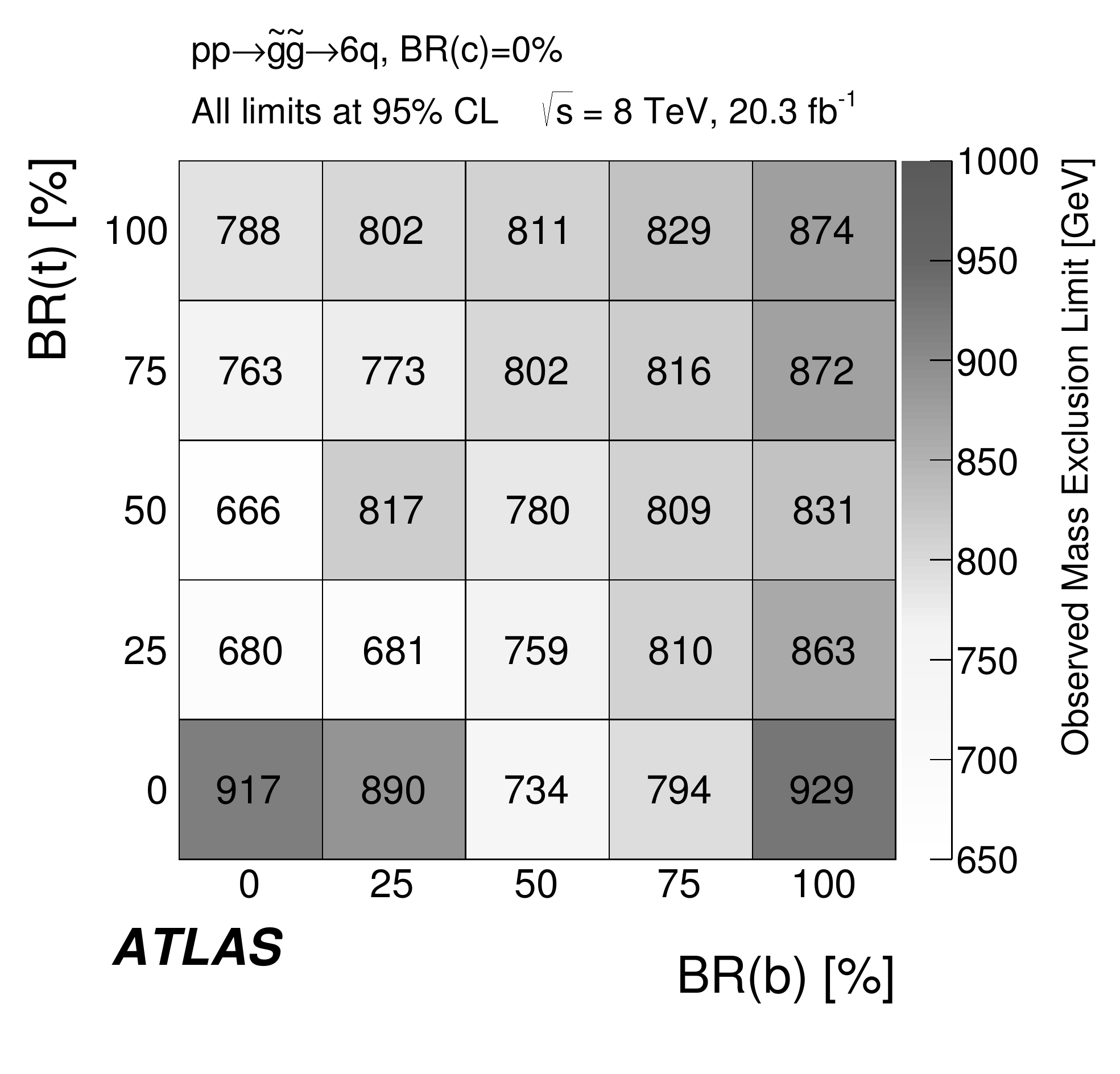}
}
\caption{Expected and observed mass exclusions at the 95\% CL in the BR($t$) vs. BR($b$) space for BR($c$)=0\%. Each point in this space is individually optimized and fit. Masses below these values are excluded in the six-quark model. Bin centers correspond to evaluated models.}
\label{fig:massSummary6q_0}
\end{figure}

\begin{figure}[!htb]
\centering
\subfigure[][BR($c$)=50\% - Expected] {
    \includegraphics[width=0.4\textwidth]{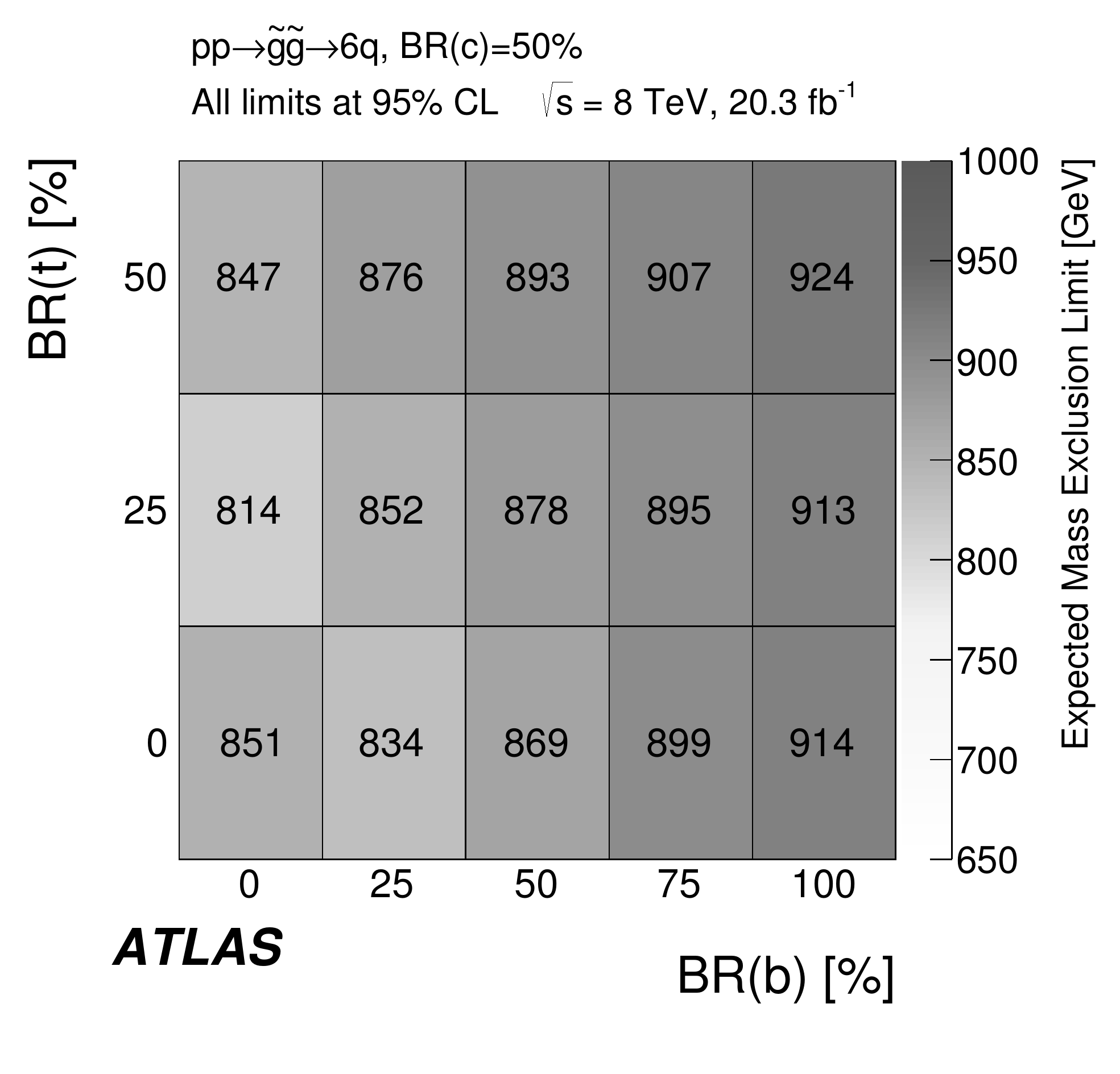}
}\\
\subfigure[][BR($c$)=50\% - Observed] {
    \includegraphics[width=0.4\textwidth]{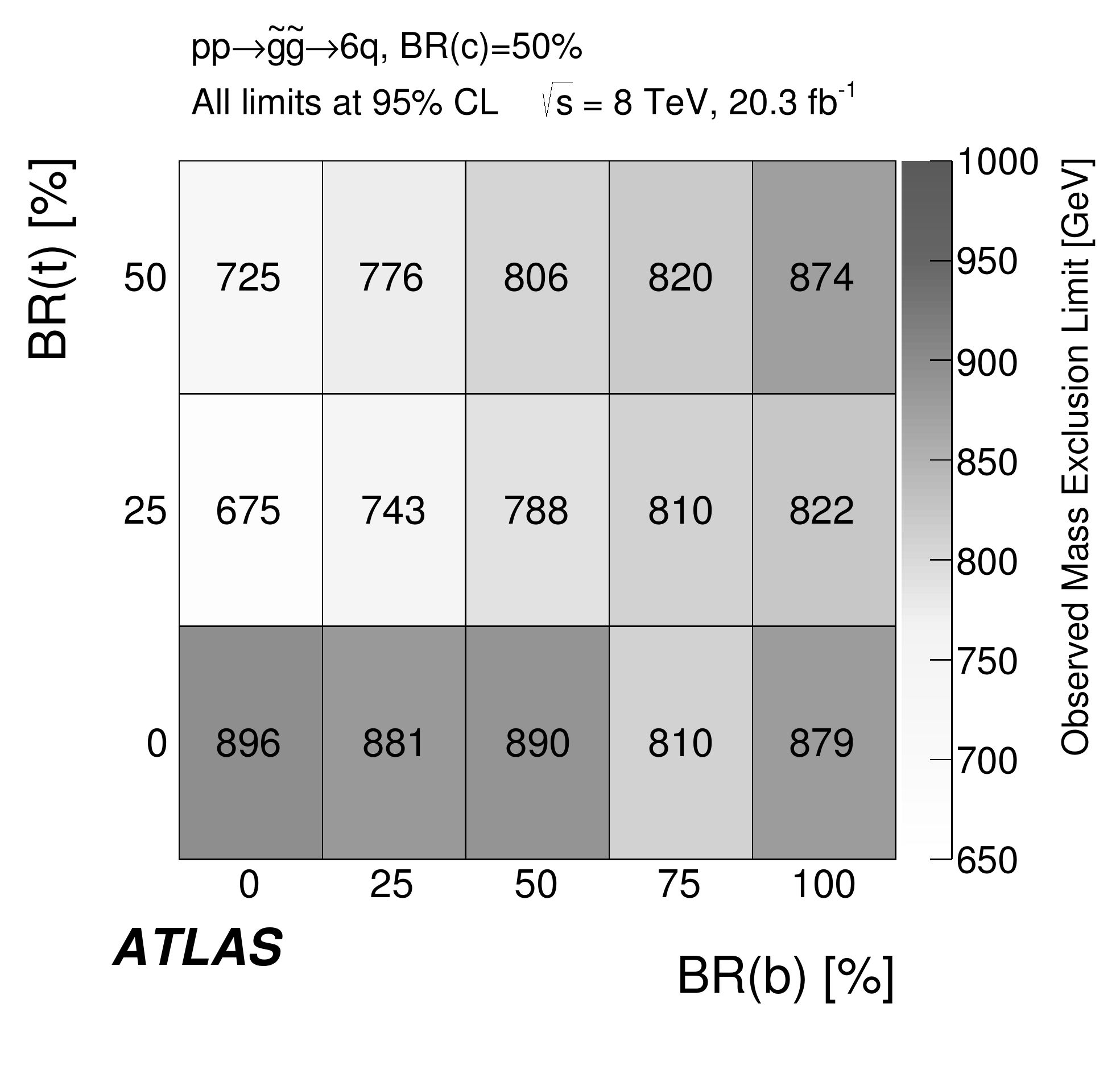}
}
\caption{Expected and observed mass exclusions at the 95\% CL in the BR($t$) vs. BR($b$) space for BR($c$)=50\%. Each point in this space is individually optimized and fit. Masses below these values are excluded in the six-quark model. Bin centers correspond to evaluated models.}
\label{fig:massSummary6q_50}
\end{figure}

\begin{figure}[!htb]
\centering
\subfigure[][Without $b$-tagging optimization] {
    \includegraphics[width=0.45\textwidth]{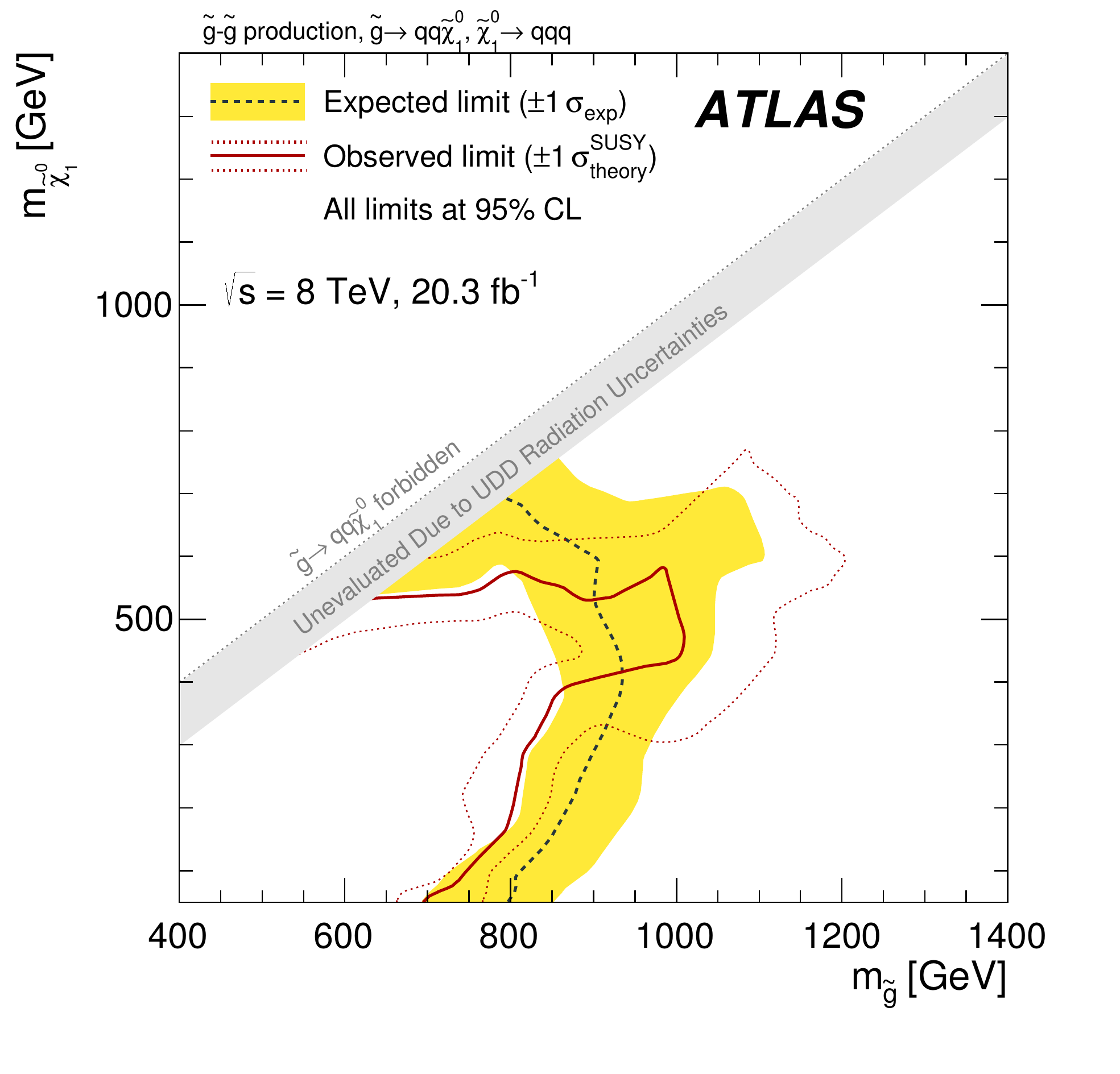}
    \label{fig:massExclusion10q:light}
}\\
\subfigure[][With $b$-tagging optimization] {
    \includegraphics[width=0.45\textwidth]{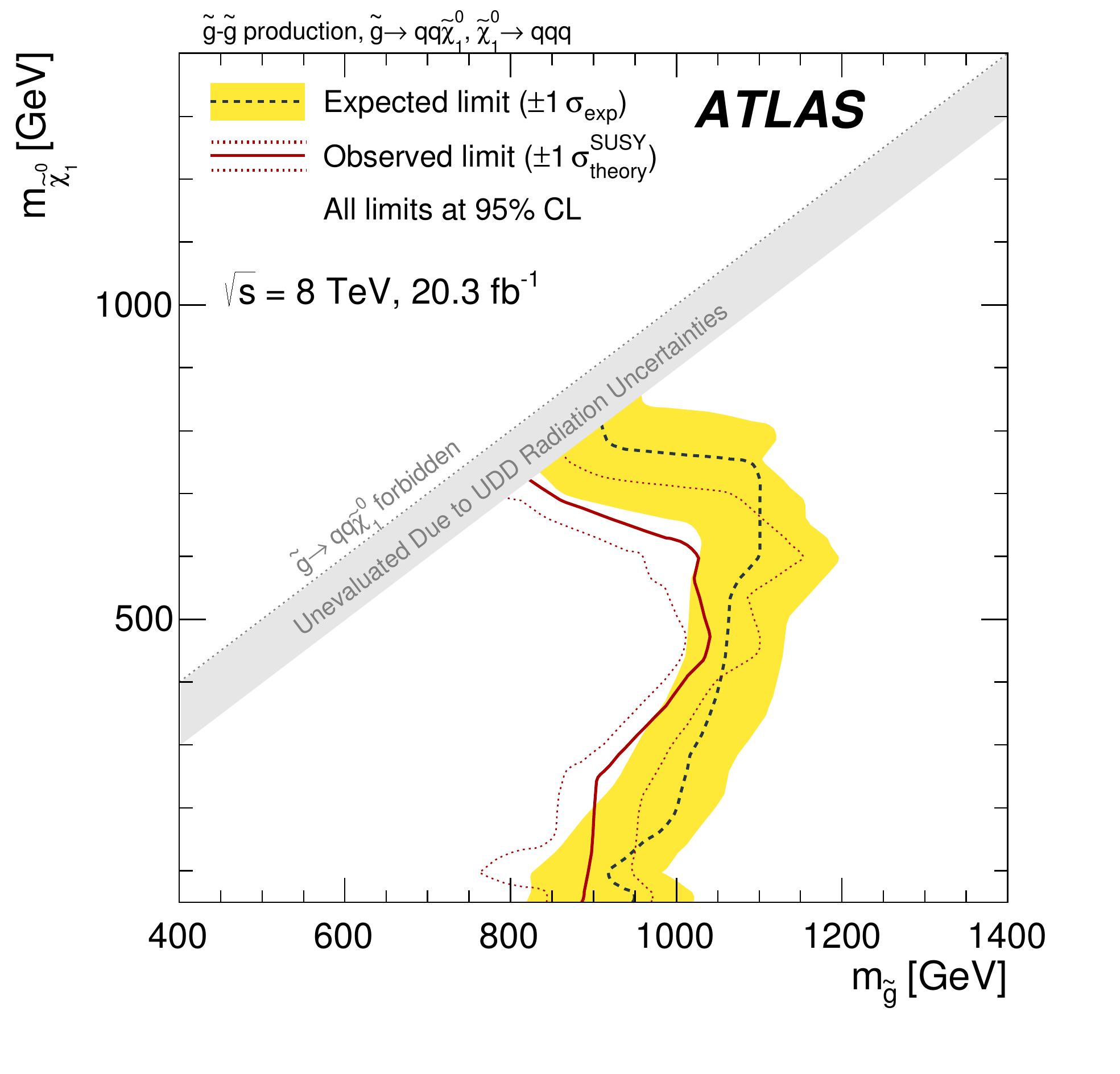}
    \label{fig:massExclusion10q:b}
}
\caption{Expected and observed exclusion limits in the (\mgluino, \mninoone) plane for the ten-quark model given by the jet-counting analysis. \subref{fig:massExclusion10q:light} shows the results when the branching ratios for the RPV decay are considered inclusively, without any $b$-tagging requirements applied. This figure is analogous to \figref{results:merged}. \subref{fig:massExclusion10q:b} shows the exclusion results when $b$-tagging requirements are allowed to enter into the optimization procedure, improving limits significantly. \label{fig:massExclusion10q}}
\end{figure}

\begin{table}[htp]
\scalebox{0.90}{
\centering 
\begin{tabular}{r|c|c|c|c|c|c}
\hline \hline
 Sample    & Jet \pt\     & \# of & \# of                  & Signal             & Back-     & Data  \\ \relax
 $m_{\tilde{g}}$              &   req.        &  jets &    $b$-tags          & (Acceptance)       & ground    &       \\ \relax
  [GeV]                     &   [GeV]      &       &                    &                    &            &       \\ 
\hline \hline
\multicolumn{7}{c}{(BR($t$), BR($b$), BR($c$))=(0\%, 0\%, 0\%)} \\
\hline
500   & 120 & 7 & 0 & 600$\pm$230  (0.7\%) & 370$\pm$60 & 444 \\  
600   & 120 & 7 & 0 & 410$\pm$100  (1.5\%)  & 370$\pm$60 & 444 \\  
800   & 180 & 7 & 0 & 13$\pm$4     (0.4\%)  & 6.1$\pm$2.2 & 4 \\  
1000  & 180 & 7 & 0 & 6.8$\pm$2.3  (1.4\%)  & 6.1$\pm$2.2 & 4 \\  
1200  & 180 & 7 & 0 & 2.7$\pm$0.5  (3.0\%)    & 6.1$\pm$2.2 & 4 \\  
\hline
\multicolumn{7}{c}{(BR($t$), BR($b$), BR($c$))=(0\%, 100\%, 0\%)} \\
\hline
500    & 80 & 7 & 2 & 1900$\pm$400    (2.1\%)  & 1670$\pm$190 & 1560 \\  
600    & 120 & 7 & 1 & 300$\pm$60     (1.1\%)  & 138$\pm$26 & 178 \\  
800    & 120 & 7 & 1 & 131$\pm$25     (4.1\%)  & 138$\pm$26 & 178 \\  
1000   & 180 & 7 & 1 & 4.4$\pm$1.0    (0.9\%)  & 2.3$\pm$1.0 & 1 \\  
1200   & 180 & 7 & 1 & 1.86$\pm$0.31  (2.1\%)  & 2.3$\pm$1.0 & 1 \\  
\hline
\multicolumn{7}{c}{(BR($t$), BR($b$), BR($c$))=(100\%, 0\%, 0\%)} \\
\hline
500   & 80 & 7 & 1 & 4600$\pm$800  (5.0\%)  & 5900$\pm$700 & 5800 \\  
600   & 100 & 7 & 1 & 940$\pm$190  (3.5\%)  & 940$\pm$140 & 936 \\  
800   & 120 & 7 & 1 & 108$\pm$18   (3.4\%)  & 138$\pm$26 & 178 \\  
1000  & 120 & 7 & 1 & 42$\pm$6     (8.5\%)  & 138$\pm$26 & 178 \\  
1200  & 180 & 7 & 1 & 1.3$\pm$0.4  (1.5\%)  & 2.3$\pm$1.0 & 1 \\ 
\hline
\multicolumn{7}{c}{(BR($t$), BR($b$), BR($c$))=(100\%, 100\%, 0\%)} \\
\hline
500   & 80 & 7 & 2 & 3600$\pm$600  (3.9\%)   & 1670$\pm$190 & 1560 \\  
600   & 80 & 7 & 2 & 2300$\pm$400  (8.6\%)   & 1670$\pm$190 & 1560 \\  
800   & 120 & 7 & 2 & 94$\pm$15    (3.0\%)   & 38$\pm$17 & 56 \\  
1000  & 120 & 7 & 2 & 37$\pm$6     (7.5\%)   & 38$\pm$17 & 56 \\  
1200  & 140 & 7 & 2 & 5.5$\pm$1.0  (6.2\%)   & 10$\pm$5 & 18 \\
\hline \hline
\end{tabular}
}
\caption{Requirements as optimized for the six-quark model under a variety of gluino mass hypotheses when the RPV vertex has various branching ratio combinations corresponding to respective RPV terms given by $\lampp_{ijk}$ being nonzero. The optimized signal region selection requirements are shown along with the resulting background and signal expectations and the number of observed data events. The nominal signal acceptance (including efficiency) is also shown for each result. Quoted errors represent both the statistical and systematic uncertainties added in quadrature.}\label{tab:Body6qTable1}
\end{table}

\begin{table*}[htb]
\centering 
\begin{tabular}{r|c|c|c|c|c|c}
\hline \hline
Sample &  Jet \pt\ req.& \# jets & \# $b$-tagged jets & Signal & Background & Data  \\ 
($m_{\tilde{g}},m_{\tilde{\chi}^{0}_{1}}$)       &   [GeV]         &         &                    & (Acceptance) & &   \\ 
\hline \hline
(400 GeV, 50 GeV)   & 80 & 7 & 2 & 1900$\pm$400      (0.5\%)  & 1670$\pm$190 & 1558 \\  
(400 GeV, 300 GeV)   & 80 & 7 & 2 & 2500$\pm$600     (0.7\%)  & 1670$\pm$290 & 1558 \\  
(600 GeV, 50 GeV)   & 120 & 7 & 1 & 180$\pm$40       (0.7\%)  & 138$\pm$26 & 178 \\  
(600 GeV, 300 GeV)   & 80 & 7 & 2 & 2200$\pm$350     (8.3\%)  & 1670$\pm$200 & 1558 \\  
(800 GeV, 50 GeV)   & 120 & 7 & 1 & 95$\pm$16        (3.0\%)  & 138$\pm$26 & 178 \\  
(800 GeV, 300 GeV)   & 120 & 7 & 1 & 172$\pm$28      (5.4\%)  & 138$\pm$26 & 178 \\  
(800 GeV, 600 GeV)   & 120 & 7 & 1 & 150$\pm$23      (4.7\%)  & 138$\pm$26 & 178 \\  
(1000 GeV, 50 GeV)   & 220 & 6 & 1 & 7.0$\pm$1.3     (1.4\%)  & 3.8$\pm$3.0 & 5 \\  
(1000 GeV, 300 GeV)   & 120 & 7 & 1 & 67$\pm$8       (14\%)  & 138$\pm$26 & 178 \\  
(1000 GeV, 600 GeV)   & 120 & 7 & 1 & 101$\pm$13     (20\%)  & 138$\pm$26 & 178 \\  
(1000 GeV, 900 GeV)   & 120 & 7 & 1 & 33$\pm$4       (6.7\%)  & 138$\pm$26 & 178 \\  
(1200 GeV, 50 GeV)   & 220 & 6 & 1 & 3.8$\pm$0.7     (4.3\%)  & 3.8$\pm$3.0 & 5 \\  
(1200 GeV, 300 GeV)   & 180 & 7 & 1 & 2.01$\pm$0.32  (2.3\%)  & 2.3$\pm$1.0 & 1 \\  
(1200 GeV, 600 GeV)   & 140 & 7 & 1 & 18.9$\pm$2.3   (21\%)  & 41$\pm$12 & 45 \\  
(1200 GeV, 900 GeV)   & 140 & 7 & 1 & 12.6$\pm$1.5   (14\%)  & 41$\pm$12 & 45 \\  
\hline \hline
\end{tabular}
\caption{Requirements as optimized for the ten-quark model under a variety of mass hypotheses when all $\lampp$ couplings are nonzero and equal and $b$-tagging requirements are considered as part of the optimization procedure. The optimized signal region selection requirements are shown along with the resulting background and signal expectations and the number of observed data events. The nominal signal acceptance (including efficiency) is also shown for each result. Quoted errors represent both the statistical and systematic uncertainties added in quadrature.}\label{tab:Body10qTable1_AllLambda_BTags}
\end{table*}

\subsection{Jet-counting analysis}
\label{sec:results:resolved}

In order to set limits on individual branching ratios, it is necessary to refer to the structure of the couplings that are allowed. From \equref{rpvandudd:wrpv}, it is clear that each RPV decay produces exactly two down-type quarks of different flavor and one up-type quark. Since the cross-section for gluino production is not dependent upon the $\lamppijk$ parameters, it is not possible to directly probe or set limits upon any individual $\lamppijk$ parameter. Instead, results are categorized based upon the probability for an RPV decay to produce a $t$-quark, a $b$-quark, or a $c$-quark. These branching ratios are denoted by BR($t$), BR($b$), and BR($c$), respectively. These branching ratios are partially constrained. The branching ratios for decays including $u$-, $c$-, and $t$-quarks (all given by the flavor index $i$ in the \lamppijk couplings) must sum to one and must therefore satisfy BR($t$) + BR($c$) $\le 1$. The branching ratios to decays including each down-type quark (as given by the flavor indices $j$ and $k$ in the \lamppijk couplings) are independent of the up-type branching ratios. At most, one $b$-quark can be produced in such an RPV decay. Simultaneous nonzero \lamppijk values can result in nontrivial branching ratio combinations.

Results using the jet-counting analysis are determined for different hypotheses on the branching ratios of RPV decays to $t$, $b$, $c$, and light-flavor quarks. The selection requirements for the signal regions are optimized separately for each of these hypotheses. When running the optimization, the full limit-setting procedure is performed under the assumption that the expected number of background events is observed in the data, taking all statistical and systematic uncertainties into account. The results of this optimization are provided in \tabref{Body6qTable1}. The first portion of \tabref{Body6qTable1} shows the optimization results and the comparison of the data with background predictions for the six-quark signal models under the assumption that (BR($t$), BR($b$), BR($c$))=(0\%, 0\%, 0\%). In this simple model, it is equivalent to say that only the term given by $\lampp_{112}$ is nonzero. Explicitly, this flavor hypothesis forces the RPV decays to result only in light quarks. Below this, the table shows the same comparisons under the assumption that (BR($t$), BR($b$), BR($c$))=(0\%, 100\%, 0\%) corresponding to only RPV terms given by $\lampp_{113}$ and $\lampp_{123}$. The second half of \tabref{Body6qTable1} is analogous to the first, only with BR($t$)=100\%. The signal acceptance is largely affected by BR($t$) and BR($b$) due to the presence of signal regions with $b$-tagged jets. Because this search requires many high-\pt\ jets, increased BR($t$) results in a lower acceptance from larger energy sharing in a higher multiplicity of final state objects. For this reason, the corners of the BR($t$) vs. BR($b$) space are shown here. Since the sensitivity to increased BR($c$) comes from $b$-tagging configurations that are designed to efficiently select $b$-jets, the effect on the signal acceptance is dominated by BR($b$). For this reason, the focus of this discussion is on the BR($b$) degree of freedom. However, several results with various values of BR($c$) are presented below.

The results of performing the limit-setting procedure on the data in the signal regions are shown in \figref{1D6qResults1} and \ref{fig:1D6qResults2} for various flavor branching ratio hypotheses as a function of gluino mass for the six-quark model. These results show both the expected and observed cross-section limits in comparison to the predicted cross-section from the theory. Under the assumption that all RPV decays are to light-flavor quarks (BR($b$)=BR($t$)=BR($c$)=0\%), gluino masses of $\mgluino <  $ \ExpLimitLF\ (expected) and $\mgluino <  $ \ObsLimitLF\ (observed) are excluded at the 95\% CL. Alternatively for the scenario where BR($b$)=100\% while the other heavy-flavor branching ratios are zero, exclusions of $\mgluino < $ \ExpLimitB\ (expected) and $\mgluino < $ \ObsLimitB\ (observed) are found. Similarly, for the case where BR($b$)=BR($t$)=100\%, exclusions of $\mgluino < $ \ExpLimitBT\ (expected) and $\mgluino < $ \ObsLimitBT\ (observed) are found. More generally, excluded masses as a function of the branching ratios of the decays are presented in \figref{massSummary6q_0} and \ref{fig:massSummary6q_50} where each bin shows the maximum gluino mass that is excluded for the given decay mode.

The event selection is optimized separately for the ten-quark model. \Tabref{Body10qTable1_AllLambda_BTags} shows the results for the ten-quark model with all UDD couplings allowed, as in the total-jet-mass analysis, when the number of $b$-tagged jets is also used as a variable in the optimization procedure. For the flavor-agnostic model where all couplings are equal, \figref{massExclusion10q:light} shows both the expected and observed limits in the (\mgluino, \mninoone) mass plane when the signal region that provides the best expected exclusion is used for each mass point, not including signal regions containing $b$-tagged jets. The shapes of the contours are given by discontinuous changes in the optimized signal regions and fluctuations well within the given uncertainties. At $\mninoone\sim100$~GeV, models with $\mgluino\lesssim700$~GeV are excluded. \Figref{massExclusion10q:b} shows the exclusion when signal regions with $b$-tagged jets are considered as part of the optimization and increase the sensitivity up to $\mgluino\lesssim1$~TeV for moderate $\mgluino-\mninoone$ mass splittings. In the ten-quark model, there is a significant probability that the cascade decays of the gluinos produce at least one $b$- or $t$-quark, and so the requirement of a $b$-tagged jet improves the sensitivity of the analysis.

\subsection{Comparisons}
\label{sec:results:combined}

\begin{table*}[!htb]
\begin{center}
\footnotesize
\begin{tabular}{r|c|c|c|c|c|c|c} 
\multicolumn{8}{c}{}\\
\hline \hline

\multirow{2}{*}{Signal Region} & \multirow{2}{*}{Expected} & \multirow{2}{*}{Obs.} & \multirow{2}{*}{$p_0$} & $N_\text{non-SM}$ & $N_\text{non-SM}$ & $\sigma_\text{vis}$  [fb] & $\sigma_\text{vis}$ [fb]\\
 & & & & Exp. & Obv. & Exp. & Obv. \\ 
\hline
    SR1 (\MJ) & 160$^{+40}_{-34} $ & $176$ & $0.39$ & $49$ & $64$ & $2.4$ & $3.2$\\
    ($n_{\rm jet}$, \ptjet, $n_{b\rm -tags}$) = (7, 120 GeV, 0) & $370\pm60  $  & $444 $  & $0.12$   & $44$ & $38$ & $2.2$ & $1.9$ \\
    ($n_{\rm jet}$, \ptjet, $n_{b\rm -tags}$) = (7, 180 GeV, 0) & $6.1\pm2.2 $  & $4   $  & $\ge0.5$ & $19$ & $10$ & $0.9$ & $0.5$ \\
    ($n_{\rm jet}$, \ptjet, $n_{b\rm -tags}$) = (7, 120 GeV, 1) & $138\pm26  $  & $178 $  & $0.09$   & $56$ & $42$ & $2.8$ & $2.1$ \\
    ($n_{\rm jet}$, \ptjet, $n_{b\rm -tags}$) = (7, 180 GeV, 1) & $2.3\pm1.0 $  & $1   $  & $\ge0.5$ & $4$  & $4$  & $0.2$ & $0.2$ \\
    ($n_{\rm jet}$, \ptjet, $n_{b\rm -tags}$) = (7, \ 80 GeV, 2)& $1670\pm190$  & $1560$  & $\ge0.5$ & $38$ & $38$ & $1.9$ & $1.9$ \\
    ($n_{\rm jet}$, \ptjet, $n_{b\rm -tags}$) = (7, 120 GeV, 2) & $38\pm17   $  & $56  $  & $0.17$   & $36$ & $52$ & $1.8$ & $2.6$ \\
\hline \hline

\end{tabular} 

\caption{Table showing upper limits on the number of events and visible cross sections in various signal regions. Columns two and three show the expected and observed numbers of events. The uncertainties on the expected yields represent systematic and statistical uncertainties. Column four shows the probabilities, represented by the $p_0$ values, that the observed numbers of events are compatible with the background-only hypothesis (the $p_0$ values are obtained with pseudo-experiments). Columns five and six show respectively the expected and observed 95\% CL upper limit on non-SM events ($N_\text{non-SM}$), and columns seven and eight show respectively the 95\% CL upper limit on the visible signal cross-section ($\sigma_\text{vis} = \sigma_\text{prod} \times A \times \epsilon = N_\text{non-SM}/{\cal{L}}$). In the case where $N_\text{expected}$ exceeds $N_\text{observed}$, $p_0$ is set to $\geq 0.5$. \label{tab:results:discovery}}
\end{center}
\end{table*}

Model-independent upper limits on non-SM contributions are derived separately for each analysis, using the SR1 signal region for the total-jet-mass analysis. A set of generic signal models, each of which contributes only to the individual signal region, is assumed and no experimental or theoretical signal systematic uncertainties are assigned other than the luminosity uncertainty. A fit is performed in the signal regions to determine the maximum number of signal events which would still be consistent with the background estimate. The resulting limits on the number of non-SM events and on the visible signal cross-section are shown in the rightmost columns of \tabref{results:discovery}. The visible signal cross-section ($\sigma_\text{vis}$) is defined as the product of acceptance ($A$), reconstruction efficiency ($\epsilon$) and production cross-section ($\sigma_\text{prod}$); it is obtained by dividing the upper limit on the number of non-SM events by the integrated luminosity. The results of these fits are provided in \tabref{results:discovery}.

\begin{figure}[!htb]
\centering
\includegraphics[width=0.45\textwidth]{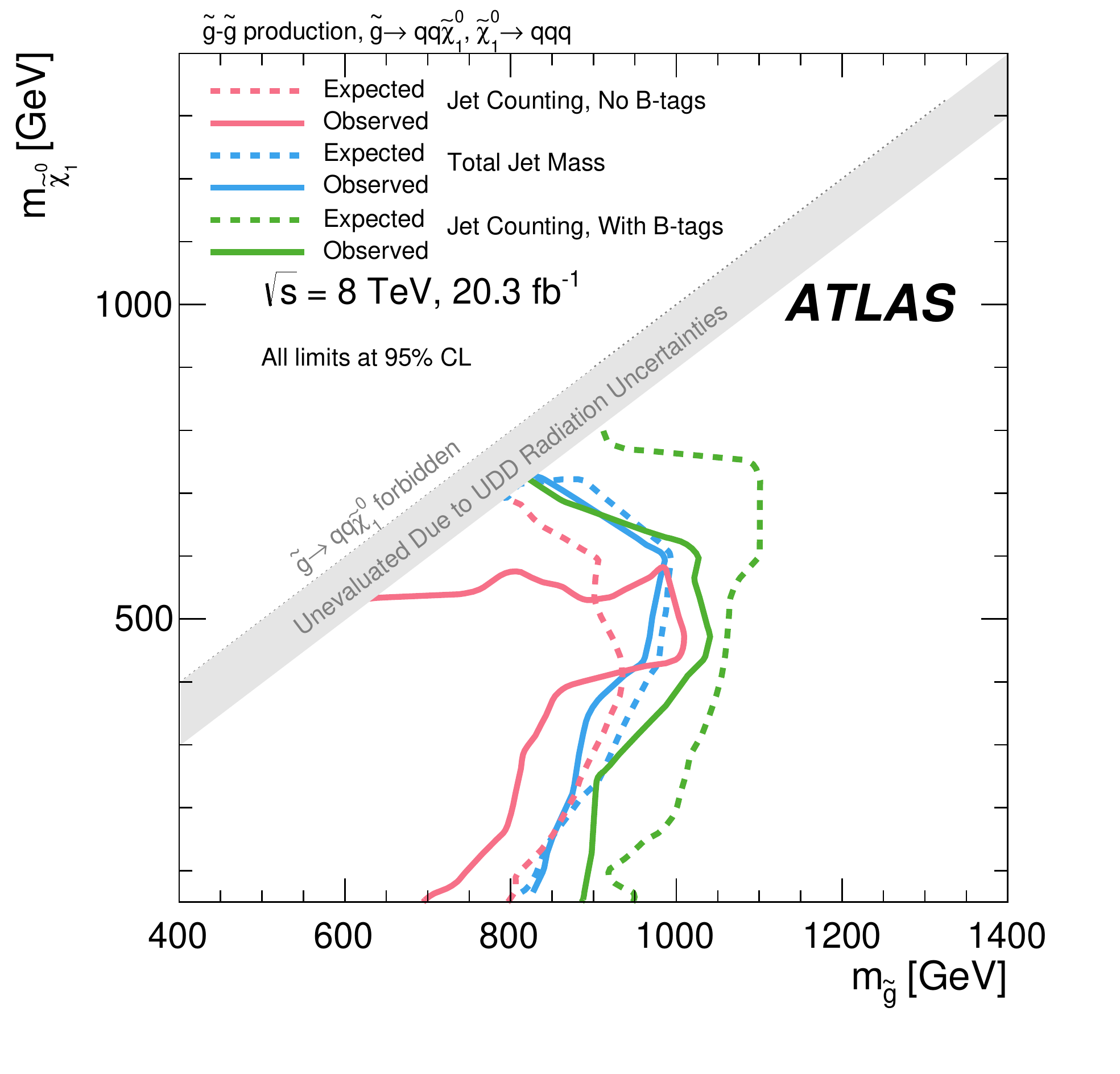}
\caption{Expected and observed exclusion limits in the (\mgluino, \mninoone) plane for the ten-quark model for the jet-counting analysis (with and without $b$-tagged jets) and the total-jet-mass analysis.} \label{fig:overlaidcontours}
\end{figure}

The interpretations of the results of the jet-counting and total-jet-mass analyses are displayed together in \figref{overlaidcontours} for the ten-quark model. This figure allows for the direct comparison of the results of the various analyses. Without $b$-tagging requirements, the jet-counting analysis sets slightly lower expected limits than the total-jet-mass analysis. With $b$-tagging requirements, the limits are stronger for the jet-counting analysis. The observed limits from the total-jet-mass analysis and jet-counting analysis with $b$-tagging requirements are also comparable.

\FloatBarrier

\section{Conclusions}
\label{sec:conclusions}

A search is presented for heavy particles decaying into complex multi-jet final states using an integrated luminosity of \intlumi of \sqseight \pp collisions with the ATLAS detector at the LHC. Two strategies are used for both background estimation and signal discrimination. An inclusive data-driven analysis using the total-jet-mass with a template method for background estimation is performed as well as a jet-counting analysis that includes exclusive heavy-flavor signal regions and provides limits on different branching ratios for the benchmark SUSY RPV UDD decays. For the ten-quark model, results from both analyses are presented with comparable conclusions. When the jet-counting analysis includes sensitivity to heavy flavor given by $b$-tagging requirements, mass exclusions are further increased.

Exclusion limits at the 95\% CL are set extending up to $\mgluino=917$~GeV in the case of pair-produced gluino decays to six light quarks and up to $\mgluino=1$~TeV in the case of cascade decays to ten quarks for moderate $\mgluino-\mninoone$ mass splittings. Limits are also set on different branching ratios by accounting for all possible decay modes allowed by the \lamppijk\ couplings in full generality in the context of $R$-parity-violating supersymmetry. These results represent the first direct limits on many of the models considered as well as the most stringent direct limits to date on those models previously considered by other analyses.

\section*{Acknowledgments}
\label{sec:Ack}


We thank CERN for the very successful operation of the LHC, as well as the
support staff from our institutions without whom ATLAS could not be
operated efficiently.

We acknowledge the support of ANPCyT, Argentina; YerPhI, Armenia; ARC,
Australia; BMWFW and FWF, Austria; ANAS, Azerbaijan; SSTC, Belarus; CNPq and FAPESP,
Brazil; NSERC, NRC and CFI, Canada; CERN; CONICYT, Chile; CAS, MOST and NSFC,
China; COLCIENCIAS, Colombia; MSMT CR, MPO CR and VSC CR, Czech Republic;
DNRF, DNSRC and Lundbeck Foundation, Denmark; EPLANET, ERC and NSRF, European Union;
IN2P3-CNRS, CEA-DSM/IRFU, France; GNSF, Georgia; BMBF, DFG, HGF, MPG and AvH
Foundation, Germany; GSRT and NSRF, Greece; RGC, Hong Kong SAR, China; ISF, MINERVA, GIF, I-CORE and Benoziyo Center, Israel; INFN, Italy; MEXT and JSPS, Japan; CNRST, Morocco; FOM and NWO, Netherlands; BRF and RCN, Norway; MNiSW and NCN, Poland; GRICES and FCT, Portugal; MNE/IFA, Romania; MES of Russia and NRC KI, Russian Federation; JINR; MSTD,
Serbia; MSSR, Slovakia; ARRS and MIZ\v{S}, Slovenia; DST/NRF, South Africa;
MINECO, Spain; SRC and Wallenberg Foundation, Sweden; SER, SNSF and Cantons of
Bern and Geneva, Switzerland; NSC, Taiwan; TAEK, Turkey; STFC, the Royal
Society and Leverhulme Trust, United Kingdom; DOE and NSF, United States of
America.

The crucial computing support from all WLCG partners is acknowledged
gratefully, in particular from CERN and the ATLAS Tier-1 facilities at
TRIUMF (Canada), NDGF (Denmark, Norway, Sweden), CC-IN2P3 (France),
KIT/GridKA (Germany), INFN-CNAF (Italy), NL-T1 (Netherlands), PIC (Spain),
ASGC (Taiwan), RAL (UK) and BNL (USA) and in the Tier-2 facilities
worldwide.



\bibliographystyle{../../../common/atlasBibStyleWithTitle}
\bibliography{../../../common/RPVGluino}

\providecommand{\href}[2]{#2}\begingroup\raggedright\begin{thebibliography}{10}

\bibitem{Miyazawa:1966}
H.~Miyazawa, {\em {Baryon Number Changing Currents}},
\href{http://dx.doi.org/10.1143/PTP.36.1266}{Prog. Theor. Phys. {\bfseries 36
  (6)} (1966) 1266--1276}.

\bibitem{Ramond:1971gb}
P.~Ramond, {\em {Dual Theory for Free Fermions}},
\href{http://dx.doi.org/10.1103/PhysRevD.3.2415}{Phys. Rev. {\bfseries D 3}
  (1971) 2415}.

\bibitem{Golfand:1971iw}
Y.~A. Gol'fand and E.~P. Likhtman, {\em {Extension of the Algebra of Poincare
  Group Generators and Violation of p Invariance}}, JETP Lett. {\bfseries 13}
  (1971) 323--326.
[Pisma Zh.Eksp.Teor.Fiz.13:452-455,1971].

\bibitem{Neveu:1971rx}
A.~Neveu and J.~H. Schwarz, {\em {Factorizable dual model of pions}},
\href{http://dx.doi.org/10.1016/0550-3213(71)90448-2}{Nucl. Phys. {\bfseries B
  31} (1971) 86}.

\bibitem{Neveu:1971iv}
A.~Neveu and J.~H. Schwarz, {\em {Quark Model of Dual Pions}},
\href{http://dx.doi.org/10.1103/PhysRevD.4.1109}{Phys. Rev. {\bfseries D 4}
  (1971) 1109}.

\bibitem{Gervais:1971ji}
J.~Gervais and B.~Sakita, {\em {Field theory interpretation of supergauges in
  dual models}},
\href{http://dx.doi.org/10.1016/0550-3213(71)90351-8}{Nucl. Phys. {\bfseries B
  34} (1971) 632}.

\bibitem{Volkov:1973ix}
D.~V. Volkov and V.~P. Akulov, {\em {Is the Neutrino a Goldstone Particle?}},
\href{http://dx.doi.org/10.1016/0370-2693(73)90490-5}{Phys. Lett. {\bfseries B
  46} (1973) 109}.

\bibitem{Wess:1973kz}
J.~Wess and B.~Zumino, {\em {A Lagrangian Model Invariant Under Supergauge
  Transformations}},
\href{http://dx.doi.org/10.1016/0370-2693(74)90578-4}{Phys. Lett. {\bfseries B
  49} (1974) 52}.

\bibitem{Wess:1974tw}
J.~Wess and B.~Zumino, {\em {Supergauge Transformations in Four-Dimensions}},
\href{http://dx.doi.org/10.1016/0550-3213(74)90355-1}{Nucl. Phys. {\bfseries B
  70} (1974) 39}.

\bibitem{Dimopoulos:1981zb}
S.~Dimopoulos and H.~Georgi, {\em {Softly Broken Supersymmetry and SU(5)}},
\href{http://dx.doi.org/10.1016/0550-3213(81)90522-8}{Nucl. Phys. {\bfseries B
  193} (1981) 150}.

\bibitem{Witten:1981nf}
E.~Witten, {\em {Dynamical Breaking of Supersymmetry}},
\href{http://dx.doi.org/10.1016/0550-3213(81)90006-7}{Nucl. Phys. {\bfseries B
  188} (1981) 513}.

\bibitem{Dine:1981za}
M.~Dine, W.~Fischler,  and M.~Srednicki, {\em {Supersymmetric Technicolor}},
\href{http://dx.doi.org/10.1016/0550-3213(81)90582-4}{Nucl. Phys. {\bfseries B
  189} (1981) 575--593}.

\bibitem{Dimopoulos:1981au}
S.~Dimopoulos and S.~Raby, {\em {Supercolor}},
\href{http://dx.doi.org/10.1016/0550-3213(81)90430-2}{Nucl. Phys. {\bfseries B
  192} (1981) 353}.

\bibitem{Sakai:1981gr}
N.~Sakai, {\em {Naturalness in Supersymmetric Guts}},
\href{http://dx.doi.org/10.1007/BF01573998}{Zeit. Phys. {\bfseries C 11} (1981)
  153}.

\bibitem{Kaul:1981hi}
R.~Kaul and P.~Majumdar, {\em {Cancellation of quadratically divergent mass
  corrections in globally supersymmetric spontaneously broken gauge theories}},
\href{http://dx.doi.org/10.1016/0550-3213(82)90565-X}{Nucl. Phys. {\bfseries B
  199} (1982) 36}.

\bibitem{Goldberg:1983nd}
H.~Goldberg, {\em {Constraint on the photino mass from cosmology}},
\href{http://dx.doi.org/10.1103/PhysRevLett.50.1419}{Phys. Rev. Lett.
  {\bfseries 50} (1983) 1419}.

\bibitem{Ellis:1983ew}
J.~Ellis, J.~Hagelin, D.~Nanopoulos, K.~Olive,  and M.~Srednicki, {\em
  {Supersymmetric relics from the big bang}},
\href{http://dx.doi.org/10.1016/0550-3213(84)90461-9}{Nucl. Phys. {\bfseries B
  238} (1984) 453}.

\bibitem{SUSYZeroLep2011}
{ATLAS} Collaboration, {\em {Search for squarks and gluinos using final states
  with jets and missing transverse momentum with the ATLAS detector in
  $\sqrt{s} = 7$~TeV proton-proton collisions}},
  \href{http://dx.doi.org/10.1016/j.physletb.2012.02.051}{Phys. Lett.
  {\bfseries B 710} (2012) 67},
\href{http://arxiv.org/abs/1109.6572}{{\ttfamily arXiv:1109.6572 [hep-ex]}}.

\bibitem{SUSYOneLep2011}
{ATLAS} Collaboration, {\em {Search for supersymmetry in final states with
  jets, missing transverse momentum and one isolated lepton in $\sqrt{s} =
  7$~TeV $pp$ collisions using 1 $fb^{-1}$ of ATLAS data}},
  \href{http://dx.doi.org/10.1103/PhysRevD.85.012006}{Phys. Rev. {\bfseries D
  85} (2012) 012006},
\href{http://arxiv.org/abs/1109.6606}{{\ttfamily arXiv:1109.6606 [hep-ex]}}.

\bibitem{SUSYTwoLep2011}
{ATLAS} Collaboration, {\em {Searches for supersymmetry with the ATLAS detector
  using final states with two leptons and missing transverse momentum in
  $\sqrt{s} = 7$~TeV proton-proton collisions}},
  \href{http://dx.doi.org/10.1016/j.physletb.2012.01.076}{Phys. Lett.
  {\bfseries B 709} (2012) 137},
\href{http://arxiv.org/abs/1110.6189}{{\ttfamily arXiv:1110.6189 [hep-ex]}}.

\bibitem{SUSYJetMult2011}
{ATLAS} Collaboration, {\em {Search for new phenomena in final states with
  large jet multiplicities and missing transverse momentum using $\sqrt{s} =
  7$~TeV $pp$ collisions with the ATLAS detector}},
  \href{http://dx.doi.org/10.1007/JHEP11(2011)099}{JHEP {\bfseries 1111} (2011)
  099},
\href{http://arxiv.org/abs/1110.2299}{{\ttfamily arXiv:1110.2299 [hep-ex]}}.

\bibitem{Arvanitaki:2013yja}
A.~Arvanitaki, M.~Baryakhtar, X.~Huang, K.~van Tilburg,  and G.~Villadoro, {\em
  {The Last Vestiges of Naturalness}},
  \href{http://dx.doi.org/10.1007/JHEP03(2014)022}{JHEP {\bfseries 1403} (2014)
  022},
\href{http://arxiv.org/abs/1309.3568}{{\ttfamily arXiv:1309.3568 [hep-ph]}}.

\bibitem{SUSYRPVemu2010}
{ATLAS} Collaboration, {\em {Search for a heavy particle decaying into an
  electron and a muon with the ATLAS detector in $\sqrt{s}=7$~TeV $pp$
  collisions at the LHC}}, Phys. Rev. Lett. {\bfseries 106} (2011) 251801,
\href{http://arxiv.org/abs/1103.5559}{{\ttfamily arXiv:1103.5559 [hep-ex]}}.

\bibitem{SUSYRPVemu20111fb}
{ATLAS} Collaboration, {\em {Search for a heavy neutral particle decaying into
  an electron and a muon using 1~fb$^{-1}$ of ATLAS data}}, Eur. Phys. J.
  {\bfseries C 71} (2011) 1809,
\href{http://arxiv.org/abs/1109.3089}{{\ttfamily arXiv:1109.3089 [hep-ex]}}.

\bibitem{SUSYRPVemu20112fb}
{ATLAS} Collaboration, {\em {Search for lepton flavour violation in the emu
  continuum with the ATLAS detector in $\sqrt{s} = 7$~TeV $pp$ collisions at
  the LHC}},
\href{http://arxiv.org/abs/1205.0725}{{\ttfamily arXiv:1205.0725 [hep-ex]}}.

\bibitem{SUSYRhadron2010}
{ATLAS} Collaboration, {\em {Search for stable hadronising squarks and gluinos
  with the ATLAS experiment at the LHC}},
  \href{http://dx.doi.org/10.1016/j.physletb.2011.05.010}{Phys. Lett.
  {\bfseries B 701} (2011) 1--19},
\href{http://arxiv.org/abs/1103.1984}{{\ttfamily arXiv:1103.1984 [hep-ex]}}.

\bibitem{SUSYRPVLL2010}
{ATLAS} Collaboration, {\em {Search for Heavy Long-Lived Charged Particles with
  the ATLAS detector in $pp$ collisions at $\sqrt{s} = 7$~TeV}}, Phys.Lett.
  {\bfseries B 703} (2011) 428,
\href{http://arxiv.org/abs/1106.4495}{{\ttfamily arXiv:1106.4495 [hep-ex]}}.

\bibitem{SUSYRPVDV2010}
{ATLAS} Collaboration, {\em {Search for displaced vertices arising from decays
  of new heavy particles in 7~TeV $pp$ collisions at ATLAS}}, Phys.Lett.
  {\bfseries B 707} (2012) 478,
\href{http://arxiv.org/abs/1109.2242}{{\ttfamily arXiv:1109.2242 [hep-ex]}}.

\bibitem{SUSYStopGluino2010}
{ATLAS} Collaboration, {\em {Search for decays of stopped, long-lived particles
  from 7~TeV $pp$ collisions with the ATLAS detector}}, Eur. Phys. J.
  {\bfseries C 72} (2012) 1965,
\href{http://arxiv.org/abs/1201.5595}{{\ttfamily arXiv:1201.5595 [hep-ex]}}.

\bibitem{RPVGluino7TeV}
{ATLAS} Collaboration, {\em {Search for pair production of massive particles
  decaying into three quarks with the ATLAS detector in $\sqrt{s}=7$ TeV $pp$
  collisions at the LHC}},
  \href{http://dx.doi.org/10.1007/JHEP12(2012)086}{JHEP {\bfseries 1212} (2012)
  086},
\href{http://arxiv.org/abs/1210.4813}{{\ttfamily arXiv:1210.4813}}.

\bibitem{Dreiner:1998wm}
H.~K. Dreiner, {\em {An introduction to explicit R-parity violation}},
  \href{http://dx.doi.org/10.1007/BF02827485}{Pramana {\bfseries 51} (1998)
  123--133},
\href{http://arxiv.org/abs/hep-ph/9707435v2}{{\ttfamily arXiv:hep-ph/9707435v2
  [hep-ph]}}.

\bibitem{Allanach:2003eb}
B.~Allanach, A.~Dedes,  and H.~Dreiner, {\em {R parity violating minimal
  supergravity model}}, \href{http://dx.doi.org/10.1103/PhysRevD.69.115002,
  10.1103/PhysRevD.72.079902}{Phys.Rev. {\bfseries D69} (2004) 115002},
\href{http://arxiv.org/abs/hep-ph/0309196}{{\ttfamily arXiv:hep-ph/0309196
  [hep-ph]}}.

\bibitem{PhysRevD.47.279}
I.~Hinchliffe and T.~Kaeding, {\em $B$- and $L$-violating couplings in the
  minimal supersymmetric standard model},
  \href{http://dx.doi.org/10.1103/PhysRevD.47.279}{Phys. Rev. D {\bfseries 47}
  (1993) 279--284}.

\bibitem{Allanach:1999ic}
B.~Allanach, A.~Dedes,  and H.~K. Dreiner, {\em {Bounds on R-parity violating
  couplings at the weak scale and at the GUT scale}},
  \href{http://dx.doi.org/10.1103/PhysRevD.60.075014}{Phys.Rev. {\bfseries D60}
  (1999) 075014},
\href{http://arxiv.org/abs/hep-ph/9906209}{{\ttfamily arXiv:hep-ph/9906209
  [hep-ph]}}.

\bibitem{Sher:1994sp}
M.~Sher and J.~Goity, {\em {Bounds on Delta B = 1 couplings in the
  supersymmetric standard model}},
\href{http://arxiv.org/abs/hep-ph/9503472}{{\ttfamily arXiv:hep-ph/9503472
  [hep-ph]}}.

\bibitem{Zwirner1983103}
F.~Zwirner, {\em {Observable $\Delta B = 2$ transitions without nucleon decay
  in a minimal supersymmetric extension of the standard model}},
  \href{http://dx.doi.org/10.1016/0370-2693(83)90230-7}{Physics Letters B
  {\bfseries 132} no.~1, (1983) 103}.

\bibitem{Bhattacharyya:1997vv}
G.~Bhattacharyya, {\em {A Brief review of R-parity violating couplings}},
\href{http://arxiv.org/abs/hep-ph/9709395}{{\ttfamily arXiv:hep-ph/9709395
  [hep-ph]}}.

\bibitem{Aaltonen:2011sg}
{CDF} Collaboration,, T.~Aaltonen {et~al.}, {\em {First Search for Multijet
  Resonances in $\sqrt{s} = 1.96$ TeV $ p\bar{p}$ Collisions}},
  \href{http://dx.doi.org/10.1103/PhysRevLett.107.042001}{Phys. Rev. Lett.
  {\bfseries 107} (2011) 042001},
\href{http://arxiv.org/abs/1105.2815}{{\ttfamily arXiv:1105.2815 [hep-ex]}}.

\bibitem{Chatrchyan:2013gia}
{CMS} Collaboration, {\em {Searches for light- and heavy-flavour three-jet
  resonances in pp collisions at $\sqrt{s} = 8$ TeV}},
  \href{http://dx.doi.org/10.1016/j.physletb.2014.01.049}{Phys.Lett. {\bfseries
  B730} (2014) 193--214},
\href{http://arxiv.org/abs/1311.1799}{{\ttfamily arXiv:1311.1799 [hep-ex]}}.

\bibitem{detPaper}
{ATLAS Collaboration}, {\em {The ATLAS Experiment at the CERN Large Hadron
  Collider}},
\href{http://dx.doi.org/10.1088/1748-0221/3/08/S08003}{JINST {\bfseries 3}
  (2008) S08003}.

\bibitem{Aad:2011dr}
{ATLAS} Collaboration, {\em {Luminosity determination in $pp$ collisions at
  $\sqrt{s}=7$~TeV using the {ATLAS} detector at the LHC}},
  \href{http://dx.doi.org/10.1140/epjc/s10052-011-1630-5}{Eur. Phys. J.
  {\bfseries C 71} (2011) 1630},
  \href{http://arxiv.org/abs/1101.2185}{{\ttfamily arXiv:1101.2185 [hep-ex]}}.

\bibitem{ATLAS-CONF-2011-116}
{ATLAS} Collaboration, {\em {Luminosity Determination in $pp$ Collisions at
  $\sqrt{s}=7$~TeV using the ATLAS Detector in 2011}},
  \href{http://cdsweb.cern.ch/record/1376384}{ATLAS-CONF-2011-116}.
  \url{http://cdsweb.cern.ch/record/1376384}.

\bibitem{Cacciari:2008gp}
M.~Cacciari, G.~P. Salam,  and G.~Soyez, {\em {The anti-$k_{t}$ jet clustering
  algorithm}}, \href{http://dx.doi.org/10.1088/1126-6708/2008/04/063}{JHEP
  {\bfseries 04} (2008) 063},
\href{http://arxiv.org/abs/0802.1189}{{\ttfamily arXiv:0802.1189 [hep-ph]}}.

\bibitem{Herwigpp}
M.~Bahr {et~al.}, {\em Herwig++ physics and manual},
  \href{http://dx.doi.org/10.1140/epjc/s10052-008-0798-9}{Eur. Phys. J.
  {\bfseries C 58} (2008) 639--707},
  \href{http://arxiv.org/abs/0803.0883}{{\ttfamily arXiv:0803.0883 [hep-ph]}}.

\bibitem{PDF-CTEQ}
J.~Pumplin {et~al.}, {\em {New generation of parton distributions with
  uncertainties from global QCD analysis}},
  \href{http://dx.doi.org/10.1088/1126-6708/2002/07/012}{JHEP {\bfseries 07}
  (2002) 012},
\href{http://arxiv.org/abs/0201195}{{\ttfamily arXiv:0201195 [hep-ph]}}.

\bibitem{cteq6l1}
P.~M. Nadolsky {et~al.}, {\em Implications of CTEQ global analysis for collider
  observables}, \href{http://dx.doi.org/10.1103/PhysRevD.78.013004}{Phys. Rev.
  {\bfseries D 78} (2008) 013004},
\href{http://arxiv.org/abs/0802.0007}{{\ttfamily arXiv:0802.0007 [hep-ph]}}.

\bibitem{Beenakker:1996ch}
W.~Beenakker, R.~Hopker, M.~Spira,  and P.~Zerwas, {\em {Squark and gluino
  production at hadron colliders}},
  \href{http://dx.doi.org/10.1016/S0550-3213(97)00084-9}{Nucl. Phys. {\bfseries
  B 492} (1997) 51},
\href{http://arxiv.org/abs/hep-ph/9610490}{{\ttfamily arXiv:hep-ph/9610490
  [hep-ph]}}.

\bibitem{Kulesza:2008jb}
A.~Kulesza and L.~Motyka, {\em {Threshold resummation for squark-antisquark and
  gluino-pair production at the LHC}},
  \href{http://dx.doi.org/10.1103/PhysRevLett.102.111802}{Phys. Rev. Lett.
  {\bfseries 102} (2009) 111802},
\href{http://arxiv.org/abs/0807.2405}{{\ttfamily arXiv:0807.2405 [hep-ph]}}.

\bibitem{Kulesza:2009kq}
A.~Kulesza and L.~Motyka, {\em {Soft gluon resummation for the production of
  gluino-gluino and squark-antisquark pairs at the LHC}},
  \href{http://dx.doi.org/10.1103/PhysRevD.80.095004}{Phys. Rev. {\bfseries D
  80} (2009) 095004},
\href{http://arxiv.org/abs/0905.4749}{{\ttfamily arXiv:0905.4749 [hep-ph]}}.

\bibitem{Beenakker:2009ha}
W.~Beenakker {et~al.}, {\em {Soft-gluon resummation for squark and gluino
  hadroproduction}},
  \href{http://dx.doi.org/10.1088/1126-6708/2009/12/041}{JHEP {\bfseries 0912}
  (2009) 041},
\href{http://arxiv.org/abs/0909.4418}{{\ttfamily arXiv:0909.4418 [hep-ph]}}.

\bibitem{Beenakker:2011fu}
W.~Beenakker {et~al.}, {\em {Squark and gluino hadroproduction}},
  \href{http://dx.doi.org/10.1142/S0217751X11053560}{Int. J. Mod. Phys.
  {\bfseries A 26} (2011) 2637},
\href{http://arxiv.org/abs/1105.1110}{{\ttfamily arXiv:1105.1110 [hep-ph]}}.

\bibitem{MC11c}
{ATLAS} Collaboration, {\em ATLAS tunes for Pythia6 and Pythia8 for MC11},
  \href{http://cdsweb.cern.ch/record/1363300}{ATLAS-PHYS-PUB-2011-009}.
  \url{http://cdsweb.cern.ch/record/1363300}.

\bibitem{pythia}
T.~Sjostrand, S.~Mrenna,  and P.~Z. Skands, {\em {PYTHIA 6.4 physics and
  manual}}, \href{http://dx.doi.org/10.1088/1126-6708/2006/05/026}{JHEP
  {\bfseries 0605} (2006) 026},
\href{http://arxiv.org/abs/0603175}{{\ttfamily arXiv:0603175 [hep-ph]}}.

\bibitem{Gleisberg:2008ta}
T.~Gleisberg {et~al.}, {\em {Event generation with SHERPA 1.1}},
  \href{http://dx.doi.org/10.1088/1126-6708/2009/02/007}{JHEP {\bfseries 02}
  (2009) 007},
\href{http://arxiv.org/abs/0811.4622}{{\ttfamily arXiv:0811.4622 [hep-ph]}}.

\bibitem{simulation}
{ATLAS} Collaboration, {\em {The {ATLAS} simulation infrastructure}},
  \href{http://dx.doi.org/10.1140/epjc/s10052-010-1429-9}{Eur. Phys. J.
  {\bfseries C 70} (2010) 823},
  \href{http://arxiv.org/abs/1005.4568}{{\ttfamily arXiv:1005.4568
  [physics.ins-det]}}.

\bibitem{Geant4}
{GEANT4} Collaboration,, S.~Agostinelli {et~al.}, {\em {GEANT4: A simulation
  toolkit}},
\href{http://dx.doi.org/10.1016/S0168-9002(03)01368-8}{Nucl. Instrum. Meth.
  {\bfseries A 506} (2003) 250}.

\bibitem{JetCleaning2011}
{ATLAS Collaboration}, {\em {Selection of jets produced in proton-proton
  collisions with the ATLAS detector using 2011 data}},
  \href{http://cdsweb.cern.ch/record/1430034}{ATLAS-CONF-2012-020}.
  \url{http://cdsweb.cern.ch/record/1430034}.

\bibitem{1748-0221-9-07-P07024}
{ATLAS} Collaboration, {\em Monitoring and data quality assessment of the ATLAS
  liquid argon calorimeter},
  \href{http://dx.doi.org/10.1088/1748-0221/9/07/P07024}{Journal of
  Instrumentation {\bfseries 9} no.~07, (2014) P07024},
  \href{http://arxiv.org/abs/1405.3768}{{\ttfamily arXiv:1405.3768 [hep-ex]}}.

\bibitem{TopoClusters}
W.~Lampl {et~al.}, {\em Calorimeter clustering algorithms: description and
  performance},
  \href{http://cdsweb.cern.ch/record/1099735}{ATL-LARG-PUB-2008-002}.
  \url{http://cdsweb.cern.ch/record/1099735}.

\bibitem{Aad:2014bia}
{ATLAS} Collaboration, {\em {Jet energy measurement and its systematic
  uncertainty in proton-proton collisions at $\sqrt{s}=7$ TeV with the ATLAS
  detector}},
\href{http://arxiv.org/abs/1406.0076}{{\ttfamily arXiv:1406.0076 [hep-ex]}}.

\bibitem{mv1}
{ATLAS} Collaboration, {\em Calibration of the performance of b-tagging for c
  and light-flavour jets in the 2012 ATLAS data},
  \href{http://cds.cern.ch/record/1741020}{ATLAS-CONF-2014-046}.
  \url{http://cds.cern.ch/record/1741020}.

\bibitem{Krohn2010}
D.~Krohn, J.~Thaler,  and L.-T. Wang, {\em {Jet trimming}},
  \href{http://dx.doi.org/10.1007/JHEP02(2010)084}{JHEP {\bfseries 2010} (2010)
  20}, \href{http://arxiv.org/abs/0912.1342}{{\ttfamily arXiv:0912.1342
  [hep-ph]}}.

\bibitem{Ellis1993}
S.~D. Ellis and D.~E. Soper, {\em {Successive combination jet algorithm for
  hadron collisions}}, \href{http://dx.doi.org/10.1103/PhysRevD.48.3160}{Phys.
  Rev. {\bfseries D 48} (1993) 3160},
\href{http://arxiv.org/abs/hep-ph/9305266}{{\ttfamily arXiv:hep-ph/9305266
  [hep-ph]}}.

\bibitem{Catani1993}
S.~Catani, Y.~L. Dokshitzer, M.~Seymour,  and B.~Webber, {\em {Longitudinally
  invariant $K_t$ clustering algorithms for hadron hadron collisions}},
\href{http://dx.doi.org/10.1016/0550-3213(93)90166-M}{Nucl. Phys. {\bfseries B
  406} (1993) 187}.

\bibitem{Aad:2013gja}
{ATLAS} Collaboration, {\em {Performance of jet substructure techniques for
  large-$R$ jets in proton-proton collisions at $\sqrt{s}$ = 7 TeV using the
  ATLAS detector}}, \href{http://dx.doi.org/10.1007/JHEP09(2013)076}{JHEP
  {\bfseries 1309} (2013) 076},
\href{http://arxiv.org/abs/1306.4945}{{\ttfamily arXiv:1306.4945 [hep-ex]}}.

\bibitem{Aad:2011tqa}
{ATLAS} Collaboration, {\em {Measurement of multi-jet cross sections in
  proton-proton collisions at a 7 TeV center-of-mass energy}},
  \href{http://dx.doi.org/10.1140/epjc/s10052-011-1763-6}{Eur. Phys. J.
  {\bfseries C 71} (2011) 1763},
\href{http://arxiv.org/abs/1107.2092}{{\ttfamily arXiv:1107.2092 [hep-ex]}}.

\bibitem{Hook2012}
A.~Hook, E.~Izaguirre, M.~Lisanti,  and J.~G. Wacker, {\em {High Multiplicity
  Searches at the LHC Using Jet Masses}},
  \href{http://dx.doi.org/10.1103/PhysRevD.85.055029}{JHEP {\bfseries 3} (2012)
  9},
\href{http://arxiv.org/abs/1202.0558}{{\ttfamily arXiv:1202.0558}}.

\bibitem{Hedri:2013pvl}
S.~El~Hedri, A.~Hook, M.~Jankowiak,  and J.~G. Wacker, {\em {Learning How to
  Count: A High Multiplicity Search for the LHC}},
  \href{http://dx.doi.org/10.1007/JHEP08(2013)136}{JHEP {\bfseries 1308} (2013)
  136},
\href{http://arxiv.org/abs/1302.1870}{{\ttfamily arXiv:1302.1870 [hep-ph]}}.

\bibitem{Cohen:2014epa}
T.~Cohen, M.~Jankowiak, M.~Lisanti, H.~K. Lou,  and J.~G. Wacker, {\em {Jet
  Substructure Templates: Data-driven QCD Backgrounds for Fat Jet Searches}},
  \href{http://dx.doi.org/10.1007/JHEP05(2014)005}{JHEP {\bfseries 1405} (2014)
  005},
\href{http://arxiv.org/abs/1402.0516}{{\ttfamily arXiv:1402.0516 [hep-ph]}}.

\bibitem{Aad:2013wta}
{ATLAS} Collaboration, {\em {Search for new phenomena in final states with
  large jet multiplicities and missing transverse momentum at $\sqrt{s}=8$~TeV
  proton-proton collisions using the ATLAS experiment}},
  \href{http://dx.doi.org/10.1007/JHEP10(2013)130}{JHEP {\bfseries 1310} (2013)
  130},
\href{http://arxiv.org/abs/1308.1841}{{\ttfamily arXiv:1308.1841 [hep-ex]}}.

\bibitem{Thaler:2010tr}
J.~Thaler and K.~Van~Tilburg, {\em {Identifying Boosted Objects with
  $N$-subjettiness}}, \href{http://dx.doi.org/10.1007/JHEP03(2011)015}{JHEP
  {\bfseries 1103} (2011) 015},
  \href{http://arxiv.org/abs/1011.2268}{{\ttfamily arXiv:1011.2268 [hep-ph]}}.

\bibitem{Thaler:2011gf}
J.~Thaler and K.~Van~Tilburg, {\em {Maximizing Boosted Top Identification by
  Minimizing $N$-subjettiness (2011)}},
\href{http://arxiv.org/abs/1108.2701}{{\ttfamily arXiv:1108.2701 [hep-ph]}}.

\bibitem{Gallicchio:2011xc}
J.~Gallicchio and M.~D. Schwartz, {\em {Pure Samples of Quark and Gluon Jets at
  the LHC}}, \href{http://dx.doi.org/10.1007/JHEP10(2011)103}{JHEP {\bfseries
  1110} (2011) 103},
\href{http://arxiv.org/abs/1104.1175}{{\ttfamily arXiv:1104.1175 [hep-ph]}}.

\bibitem{Aad:2014gea}
{ATLAS} Collaboration,, G.~Aad {et~al.}, {\em {Light-quark and gluon jet
  discrimination in $pp$ collisions at $\sqrt{s}=7\mathrm {\ TeV}$ with the
  ATLAS detector}},
  \href{http://dx.doi.org/10.1140/epjc/s10052-014-3023-z}{Eur.Phys.J.
  {\bfseries C74} no.~8, (2014) 3023},
\href{http://arxiv.org/abs/1405.6583}{{\ttfamily arXiv:1405.6583 [hep-ex]}}.

\bibitem{HistFitter}
A.~L. Read, {\em {Presentation of search results: The CL(s) technique}},
\href{http://dx.doi.org/10.1088/0954-3899/28/10/313}{J. Phys. {\bfseries G28}
  (2002) 2693--2704}.

\bibitem{Kramer:2012bx}
M.~Kramer {et~al.}, {\em {Supersymmetry production cross sections in pp
  collisions at $\sqrt{s} = 7$~TeV}},
\href{http://arxiv.org/abs/1206.2892}{{\ttfamily arXiv:1206.2892 [hep-ph]}}.

\end{thebibliography}\endgroup







\clearpage
\onecolumngrid
\begin{flushleft}
{\Large The ATLAS Collaboration}

\bigskip

G.~Aad$^{\rm 85}$,
B.~Abbott$^{\rm 113}$,
J.~Abdallah$^{\rm 152}$,
S.~Abdel~Khalek$^{\rm 117}$,
O.~Abdinov$^{\rm 11}$,
R.~Aben$^{\rm 107}$,
B.~Abi$^{\rm 114}$,
M.~Abolins$^{\rm 90}$,
O.S.~AbouZeid$^{\rm 159}$,
H.~Abramowicz$^{\rm 154}$
% ...
\bigskip
\\
% ...
$^{11}$ Institute of Physics, Azerbaijan Academy of Sciences, Baku, Azerbaijan\\
% ...
$^{85}$ CPPM, Aix-Marseille Universit{\'e} and CNRS/IN2P3, Marseille, France\\
% ...
$^{90}$ Department of Physics and Astronomy, Michigan State University, East Lansing MI, United States of America\\
% ...
$^{107}$ Nikhef National Institute for Subatomic Physics and University of Amsterdam, Amsterdam, Netherlands\\
% ...
$^{113}$ Homer L. Dodge Department of Physics and Astronomy, University of Oklahoma, Norman OK, United States of America\\
$^{114}$ Department of Physics, Oklahoma State University, Stillwater OK, United States of America\\
% ...
$^{117}$ LAL, Universit{\'e} Paris-Sud and CNRS/IN2P3, Orsay, France\\
% ...
$^{152}$ Institute of Physics, Academia Sinica, Taipei, Taiwan\\
% ...
$^{154}$ Raymond and Beverly Sackler School of Physics and Astronomy, Tel Aviv University, Tel Aviv, Israel\\
% ...
$^{159}$ Department of Physics, University of Toronto, Toronto ON, Canada\\
% ...
\end{flushleft}


\end{document}